\documentclass[12pt]{article}

\usepackage{graphics,epsfig}

\def\dalemb#1#2{{\vbox{\hrule height .#2pt
        \hbox{\vrule width.#2pt height#1pt \kern#1pt
                \vrule width.#2pt}
        \hrule height.#2pt}}}

\let\a=\alpha \let\b=\beta \let\g=\gamma \let\d=\delta \let\e=\epsilon
\let\z=\zeta  \let\th=\theta  \let\k=\kappa
\let\l=\lambda \let\m=\mu  \let\x=\xi \let\p=\pi %\let\r=\rho
\let\s=\sigma \let\t=\tau   \let\c=\chi 
\let\vp=\varphi \let\vep=\varepsilon
\let\w=\omega       \let\D=\Delta \let\Th=\Theta \let\L=\Lambda
\let\X=\Xi \let\P=\Pi \let\S=\Sigma  \let\Y=\Psi
\let\C=\Chi \let\W=\Omega
\let\la=\label \let\ci=\cite 
  
\def\nn{\nonumber} \def\bd{\begin{document}} \def\ed{\end{document}}
\def\ds{\documentstyle} \let\fr=\frac \let\bl=\bigl \let\br=\bigr
\let\Br=\Bigr \let\Bl=\Bigl
\let\bm=\bibitem
\let\na=\nabla
\def\tU{{\widetilde U}}
\let\pa=\partial \let\ov=\overline
\def\ie{{\it i.e.\ }}
\newcommand{\be}{\begin{equation}}
\newcommand{\ee}{\end{equation}}
\def\ba{\begin{array}}
\def\ea{\end{array}}
\def\ft#1#2{{\textstyle{{\scriptstyle #1}\over {\scriptstyle #2}}}}
\def\fft#1#2{{#1 \over #2}}
\def\F#1#2{{ F_{#1}^{(#2)} }}
\def\cF#1#2{{ {\cal F}_{#1}^{(#2)} }}

\def\={\, =\, }
\def\+{\, +\, }
\def\-{\, -\, }

\def\R{{\bf R}}
\def\sst#1{{\scriptscriptstyle #1}}
\def\oneone{\rlap 1\mkern4mu{\rm l}}
\def\e7{E_{7(+7)}}
\def\td{\tilde}
\def\wtd{\widetilde}
\def\im{{\rm i}}
\newcommand{\ho}[1]{$\, ^{#1}$}
\newcommand{\hoch}[1]{$\, ^{#1}$}
\newcommand{\bea}{\begin{eqnarray}}
\newcommand{\eea}{\end{eqnarray}}
\newcommand{\ra}{\rightarrow}
\newcommand{\lra}{\longrightarrow}
\newcommand{\Lra}{\Leftrightarrow}
\newcommand{\ap}{\alpha^\prime}
\newcommand{\bp}{\tilde \beta^\prime}
\newcommand{\cB}{{\cal B}}
\newcommand{\cO}{{\cal O}}
\newcommand{\vecx}{\vec{x}}
\newcommand{\vecy}{\vec{y}}
\newcommand{\vecp}{\vec{p}}
\newcommand{\vecq}{\vec{q}}
\newcommand{\tr}{{\rm tr} }
\newcommand{\Tr}{{\rm Tr} }

\newcommand{\cL}{{\cal L}}
\newcommand{\cA}{{\cal A}}
\newcommand{\cD}{{\cal D}}
\def\sst#1{{\scriptscriptstyle #1}}
\def\0{{\sst{(0)}}}
\def\1{{\sst{(1)}}}
\def\2{{\sst{(2)}}}
\def\3{{\sst{(3)}}}
\def\4{{\sst{(4)}}}
\def\5{{\sst{(5)}}}
\def\6{{\sst{(6)}}}
\def\7{{\sst{(7)}}}
\def\8{{\sst{(8)}}}
\def\ve{\varepsilon}
\def\vf{\varphi}
\def\F{\Phi}
\def\wg{\wedge}

\def \foot {\footnote}
\def \bi{\bibitem}

\def \tr {{\rm tr}}
\def \ha {{1 \over 2}}
\def \td {\tilde}
\def \ci{\cite}
\def \N {{\mathcal N}}
\def \ww {\Omega}
\def \const {{\rm const}}
\def \ss {\sum_{i=1}^3 }
\def \t {\tau}
\def\S{{\mathcal S} }
\def \nn {\nu}
\def \XX {{\rm X}}

\def \lra {\leftrightarrow}
\def \vom {{\bar \omega}}
\def \E {{\mathcal  E}} \def \J {{\mathcal  J}}
\def \YY {{\rm Y}}

\def \d {\del}
\def \rJ {{J}}
\def \sms {sigma models\ }
\def \sm {sigma model\ }
\def \L {\Lambda}
\def \gl {\ell}
\def \tr {{\rm tr\ }}
\def\z{\zeta}
\def\zi{\zeta_1}
\def\zii{\zeta_2}
\def\K{\mbox{K}}
\def\eE{\mbox{E}}   \def \vt {\vartheta}
\def \vr {\varrho}
\def \wup {w}

\def\dg{\dagger}
\def\a{\alpha}
\def\b{\beta}
\def\e{\varepsilon}
\def\p{\phi}
\def\ap{\alpha^\prime}
\def\I{{\cal I}}

\def\R{{\bf R}}
\def\Z{{\bf Z}}
\def\C{{\bf C}}
\def\P{{\bf P}}
\def\xb{{\bar X}}
\def\Tr{{\rm  Tr}}
\def\tr{{\rm  tr}}

\def \del{\partial}
\def \a {\alpha}
\def \aa {{\a'}}
\def\g{\gamma}
\def\s{\sigma}
\def\z{\zeta}
\def\zi{\zeta_1}
\def\zii{\zeta_2}
\def\ov{\over}

\def\I{{\cal I}}
\def\J{{\mathcal J}}
\def \ok {{1\ov \k}}
\def\LL{{\mathcal L }}
\def \jL {{J}}
\def \om {\omega}
\def \cL {{\mathcal L}} \def \cH {{\mathcal H}}
\def\E{{\mathcal E}}
\def\w{\omega}
\def\b{\beta}
\def\l{\lambda}
\def\eps{\epsilon}
\def\vep{\varepsilon}
\def \De {{\mathcal D}}

 \def \cV {{\cal V}}
\def  \Jt {  {J}_{\rm tot}    }

\def \k {\kappa}
\def\foot{\footnote}
\def \four{{\textstyle {1\ov 4}}}
 \def \third { \textstyle {1\ov 3
}}
\def\det{\hbox{det}}
\def \ci {\cite}

\def \foot {\footnote}
\def \bi{\bibitem}

\def \tr {{\rm tr}}
\def \ha {{1 \over 2}}
\def \tid {\tilde}
\def \vv {{\rm v}}
\def \tl {{\tilde \l}}
\def \XX {{\rm X}}
\def \ta {{\tilde \a}}
\def \fo { {1\ov 4}}
\def \ep {\epsilon}
\def \inti {{\int^{2\pi}_0 {d \sigma \ov 2 \pi}}}

\def \d {\partial}
\def \K {{\rm S}}
\def \el {\ell}
\def \Tr {{\rm Tr}}
\def \P {\Phi}
\def \l  {\lambda}
\def \tl {{\tilde \l}}
\def \bl {{\tilde \l}}
\def \const {{\rm const}}
\def \V {v}

\def \bv {v^*}
\def \vv {{\rm v}}
\def \LL {{\mathcal L}}
\newcommand{\PV}[1]{P_{\!\!_{V_{#1}}}}

\def \bL {\ell}
\def \M {{\mathcal M}}
\def \N {{\mathcal N}}
\def \S {{\rm S}}
\def \vn {\vec n}
\def \tl {\td \l}
\def \td {\tilde}
\def \Prod {\Pi}
\def \O {{\mathcal O}}
\def \Q {{\rm  Q}}
\def \D {\Delta}
\def \N {{\mathcal N}}
\def\tN{{\tilde N}}

\def \m {\mu}
\def \vs {\vec \s}
\def \ie {i.e.}

\def \cD {{\cal D}}

\def  \le  {\l_{\rm eff}}

\def \rS {{\rm S}}
\def\as{{\a}}
\newcommand{\bra}[1]{\mbox{$\langle #1 |$}}
\newcommand{\ket}[1]{\mbox{$| #1 \rangle$}}

\newcommand{\auth}{AUTHORS}

\def\thb{\bar{\theta}}
\def\Thb{\bar{\Theta}}
\def\barp{\bar{p}}
\def\barq{\bar{q}}
\def\barc{\bar{c}}
\def\bard{\bar{d}}
\def\e{\epsilon}

\def \bi{\bibitem}
\def \la {\label}

\def \l {\lambda}
\def\foot{\footnote}
\def \tl  {{\tilde \l}}
\def \sql {{\sqrt \l}}
\def \adss {$AdS_5 \times S^5$\ }
\newcommand{\rf}[1]{(\ref{#1})}
\def \ov {\over}

\def\th{\theta}
\def\Th{\Theta}
\def\vth{\vartheta}
\def\btheta{{\bar\theta}}
\def\ttheta{{{\tilde\theta}}}
\def\bttheta{{{\bar\ttheta}}}
\def\vth{\vartheta}

\def\ra{\rightarrow}
\def\N{{\cal N}}
\def\F{{\cal F}}
\def\uM{\underline{M}}
\def\uN{\underline{N}}
\def\uP{\underline{P}}
\def\cc{\circ}
\def\eqv{\equiv}

\def\ni{\noindent}
\def \ha{{1\ov 2}}
\def \bw {{\rm w}}

\def\r{{\rm r}}

\def \cT {{\cal T}}
\def \no {\nonumber}

\def\a{{\rm\bf a}}
\def\b{{\rm\bf b}}
\def\c{{\rm\bf c}}

\def\Y{{\rm Y}}
\def\X{{\rm X}}
\def\tY{\tilde{\rm Y}}
\def\tX{\tilde{\rm X}}
\def\dY{\dot{\rm Y}}
\def\dX{\dot{\rm X}}

\def \J {\mathcal{J}}
\def \del {\partial}

\def \bps {{\bar \psi}}
\def \sqbl {\sqrt{\bar \lambda}}

\def\dF{\dot{F}}
\def\dG{\dot{G}}
\def\df{\dot{f}}
\def \E {{\cal E}}

\def \S {{\cal S}}
\def \J {{\cal J}}

\def\ms{\mathcal{S}}
\def\mj{\mathcal{J}}
\def\soj{\fr{\ms}{\mj}}
\def \R {{\bf R}}
\def \om {\omega}
\def \tH {\widetilde H}
\def \bE {\bar E}
\def \x {{\cal X}}

\def \hV {{\hat V}}

 \def \bb {\bar \beta}
\def \W {{\cal E}}

\def \bi{\bibitem}
\def \la {\label}

\def \l {\lambda}
\def\foot{\footnote}
\def \tl  {{\tilde \l}}
\def \sql {{\sqrt \l}}
\def \sqtl {{\sqrt {\tilde \l}}}

\def \HH {{\rm E}}
\def \cS {{\cal S}}
\def \cL {{\cal L}}

\def \adss {$AdS_5 \times S^5$\ }

\def \D {\Delta}
\def \thet {\theta}
 \def \t {\tau}
 \def \p {\phi}
 \def \r {\rho}
 \def \rN {{\rm N}}
 \def\tw{{\tilde w}}
 \def\hJ{{J}}
 \def\hw{{w}}
 \def\hl{{\lambda}}
 \def\hth{{\theta}}
 \def\NN{{\cal N}}
 \def \bv {{ \bar w}}
\def \vn {{\vec n}}
\newcommand{\sfrac}[2]{{\textstyle\frac{#1}{#2}}}
\def \bl {{ \bar \lambda}}
\def \bp {{\bar p}}
\def \bu {{\bar u}}
\def \sha {\sfrac{1}{2}}
\def \w {\omega}
\def \ov {\over}
\def \vl { \vec \ell}
\def \varpi {{\rm w}}
\def \OO {{\cal O}}

\def\pic #1#2{\hbox{\lower#1pt\hbox{~\mbox{\epsfxsize=20truemm \epsffile{#2}}}}}

\def\pic #1#2#3{\hbox{\lower#1pt\hbox{~\mbox{\includegraphics[scale=#3]{#2}}}}}

\begin{document}
\overfullrule=0pt
\parskip=2pt
\parindent=12pt
\headheight=0in \headsep=0in \topmargin=0in \oddsidemargin=0in

\vspace{ -3cm} \thispagestyle{empty} \vspace{-1cm}
\begin{flushright}
Imperial-TP-AT-6-2

\end{flushright}

\begin{center}

{\Large\bf  Asymptotic  Bethe Ansatz S-matrix \\
\vspace{0.15cm}
 and
Landau-Lifshitz  type
%\vspace{0.15cm}
 effective 2-d actions
%\vspace{0.3cm}
   }

 \vspace{.5cm} { R. Roiban$^{a,}$\footnote{radu@phys.psu.edu},
  A. Tirziu$^{b,}$\footnote{tirziu@mps.ohio-state.edu}
 and A.A.
 Tseytlin$^{c,b,}$\footnote{Also at
 Lebedev  Institute, Moscow.
  %tseytlin@mps.ohio-state.edu
 }}\\
 \vskip 0.3cm

{\em $^{a}$Department of Physics, The Pennsylvania  State University,\\
University Park, PA 16802 , USA\\
$^{b}$Department of Physics, The Ohio State University,\\
Columbus, OH 43210, USA\\
\vskip 0.08cm $^{c}$  Blackett Laboratory, Imperial College,
London SW7 2AZ, U.K. }

\end{center}

 \begin{abstract}
 %%%%%%%%%%%%%%%%%%%%%%%%%%%%%%%%%
Motivated by  the desire   to relate   Bethe  ansatz equations
for anomalous  dimensions found on the
gauge theory side of the AdS/CFT correspondence to
superstring theory on $AdS_5 \times S^5$   we explore a connection
between the asymptotic S-matrix that enters the Bethe ansatz
and an effective two-dimensional quantum field theory. The latter  generalizes
the standard ``non-relativistic''
Landau-Lifshitz (LL) model  describing  low-energy  modes   of
 ferromagnetic  Heisenberg spin chain
and should be  related to a  limit
of superstring effective action.
We find the exact  form of the  quartic interaction terms in the
generalized LL type action  whose quantum
 S-matrix matches the low-energy limit
of the asymptotic  S-matrix of the spin chain of Beisert, Dippel and Staudacher (BDS).
This generalises to all orders in the `t~Hooft coupling $\l$
an  earlier  computation  of Klose and Zarembo
of  the S-matrix of the standard LL   model.
We  also  consider a generalization to the
case when  the spin chain S-matrix contains an
extra ``string'' phase and determine the exact form of the LL
4-vertex  corresponding to the low-energy limit
of the ansatz of Arutyunov, Frolov and Staudacher (AFS).
We explain  the   relation between the resulting
``non-relativistic'' non-local action and the second-derivative
string sigma model.  We comment
on modifications introduced by strong-coupling
corrections to the AFS phase.
We mostly discuss   the $SU(2)$ sector but also
present generalizations to the  $SL(2)$  and $SU(1|1)$ sectors,
confirming universality of the dressing phase contribution
by matching the low-energy limit  of the AFS-type  spin chain S-matrix
with  tree-level string-theory S-matrix.
\end{abstract}
\newpage

\renewcommand{\theequation}{1.\arabic{equation}}
 \setcounter{equation}{0}

\setcounter{equation}{0} \setcounter{footnote}{0}
\setcounter{section}{0}

%%%%%%%%%%%%%%%%%%%%%%%%%%%%%%%%%%%%%%%%%%%%%%%%%%%%%%%%%%%%%%%%
\section{Introduction}
%%%%%%%%%%%%%%%%%%%%%%%%%%%%%%%%%%%%%%%%%%%%%%%%%%%%%%%%%%%%%%%%%%%%5

To demonstrate the  AdS/CFT duality one is to establish
a direct relation between the spectrum of  the $N=4$ SYM
gauge-theory dilatation
operator and the spectrum of quantum string energies in \adss.
There are strong indications  that both spectra are described by solutions
of certain spin chain-type Bethe Ansatz.
In the simplest bosonic sector of the  gauge theory, the $SU(2)$ sector,
the
%%R
%
%In the simplest $SU(2)$ sector on the gauge theory side
%the corresponding
%
spin chain is a long-range
  extension of the ferromagnetic XXX$_{1/2}$
model \ci{mz,bks,bds,s,bs2}. Its Hamiltonian is  known explicitly up to three
loops; beyond this order the spin chain is defined by the  Bethe ansatz
%
%with the corresponding Bethe ansatz   having
%the form
\ci{bs2,bds,afs,s}
 \be \la{ook}
 e^{ip_k L}=\prod_{j\neq k}^{M} S(p_k,p_j; \l) \ , \ \ \ \ \ \ \  \ee
 \be \la{sss}
 S(p_k,p_j;\l)= S_1 (p_k,p_j;\l)\ e^{i \theta(p_k,p_j;\l)}  \ ,\ \ \ \  \
 \ \ \ \  \  S_1  =  { u_k-u_j + i \ov u_k-u_j -i} \ .
\ee
%%R
Here $p_j$ ($j=1,...,M$) are  momenta of  excitations
which at one loop reduce to those diagonalizing the XXX$_{1/2}$ monodromy matrix
%elementary excitations
(i.e. magnons) and $u_j$ are their  rapidities related to $p_j$ by \ci{bds}
\bea
u_j= u(p_j;\l) \ , \ \ \ \ \ \ \ \ \ \ \ \ \
u(p;\l) \equiv  \sfrac{1}{2}\cot{\sfrac{p}{2}}\  \sqrt{1+  \sfrac{\l}{\pi^2}
\sin^2\sfrac{p}{2}} \ ,
 \la{ooo}  \eea
where $\l$ is the 't Hooft coupling.
 $S(p_k,p_j;\l)$  is
a phase shift \ci{s} due to magnon scattering
which one may try to interpret  as a  two-particle  scattering matrix
of an integrable two dimensional field theory \ci{s,manpol,kz,kaz}
%%R
whose fundamental excitations correspond to the spin chain magnons.
%%R
The momenta $p_j=p_j (\l, L,M) $ satisfying \rf{ook}  are also subject
to the quantization condition $\sum_{k=1}^M p_k =2\pi m$ encoding the
fact of  cyclicity of the trace of the corresponding  gauge-theory operators.
%
% One is supposed to find $p_j=p_j (\l, L,M) $ satisfying
%\rf{ook}   with the condition that total  momentum
%$\sum_{k=1}^M p_k  $
%is  quantized due to the
%periodicity of the spin chain  implied by the
%cyclicity of the trace in the gauge-theory operators.
Then, the energy of the spin-chain state
or the anomalous dimension of the corresponding operator
is given by
\be
E=   \sum^M_{j=1} \Big( \sqrt{1+  \sfrac{\l}{\pi^2} \sin^2\sfrac{p_j}{2}} -1\Big)\ .
\la{jy}  \ee
The factor
$S_1 $  in  \rf{sss}  is the standard Heisenberg model phase shift
which  enters also  the asymptotic (large $L$)
BDS gauge theory Bethe ansatz \ci{bds}. An
extra phase $\theta$ (common  to all sectors \ci{beis,beiklos})
is expected to be present in the
exact  ansatz which, according to the AdS/CFT correspondence,
 should match  the conjectured  Bethe ansatz on the string theory  side
\ci{afs,afz}.
% (cf. also \ci{afz}).
The precise structure of this phase (which
at weak coupling should start from 3-loop $\l^2$ terms \ci{afs,cal,ss} and at strong
coupling
 should include quantum string $1\ov \sql$ corrections \ci{bt,szz,sz})
  is a key  open problem  at the moment \ci{beis,janik,hl,af,fk}.
In addition to showing that $S$ with correct dressing phase $\theta$
does come out of the \adss string theory of \ci{mt}, one is also to provide a string-theory
derivation of the dispersion relation  \rf{ooo},\rf{jy}
containing the ``discreteness'' factor  $\sin^2\sfrac{p}{2}$;
 important steps in the  latter direction
  were  recently made  in \ci{beis,mald,dor}.\foot{
On the gauge-theory side,
the history of derivation of the square root  formula
 $\sqrt{1+  \sfrac{\l}{\pi^2} \sin^2\frac{p}{2}}$
 for ``magnon'' energy  starts  (in the small $p$ limit, $\sin\frac{p}{2} \to
   \frac{p}{2}$)  with
 \ci{bmn}.  All-order arguments for  the validity of the formula with full
$\sin^2\frac{p}{2}$  were given  in
 \ci{gmr} (see  eq. (62) there)
 % contained full $\sin^2$ under the square root)
 and in \ci{saz} (where
  $\sin^2\frac{p}{2}$  appeared at intermediate steps  of the derivation
  of the BMN relation).
 More recently,
 \rf{jy}  was derived \ci{bcv} using a  matrix model obtained by s-wave truncation
 of SYM theory on $S^3$. An interesting geometrical picture found in \ci{bcv}
 apprears to provide a  link to
  a related discussion  on the string side in \ci{mald}.
}

\bigskip

%In order
To understand  how   the Bethe ansatz
\rf{ook}  may be related to string theory one may  try to
directly associate to it a two-dimensional
%world-sheet
action describing magnon  interactions.
At leading (1-loop) order in $\lambda$ the
effective 2d action describing the ``low-energy'' part of the
spectrum of the ferromagnetic XXX$_{1/2}$ model is
the non-relativistic Landau-Lifshitz (LL)  action.
It can be found by taking the continuum limit
in  the coherent state path integral
representation for the Heisenberg model  \ci{fradkin,kru}.
 The  S-matrix of
the LL model on an infinite line   does
match \ci{kz}
the leading term in the small-momentum expansion
(i.e. $u_j\rightarrow \ha \cot { p_j \ov 2} \to  { 1 \ov p_j}$)
%of the Heisenberg model
of the $S_1$--factor in \rf{sss}.
%(with $u_j= \ha \cot { p_j \ov 2} \to  { 1 \ov p_j}$) .
The same  LL  action appears as a ``fast-string''   limit of the
 classical string action on $R \times S^3$
\ci{kru,krt,kmmz}.
One can also reconstruct higher order in $\l$  terms in
a generalized effective  LL action
by matching the energies of the Bethe ansatz  states
with their field theory  counterparts \ci{krt,rt,tsecarg,mtt,mtt1,t}.

\bigskip

Our interest in this  generalized LL action is
due to the fact  that it may serve as a bridge between
 quantum string theory and the generalized Bethe ansatz \rf{ook}.
 %that at weak coupling
%and large length should reduce to the asymptotic BDS ansatz.
We expect that the large $L=J$ limit of a quantum effective string action will
be related to  an effective LL action that reproduces the spin chain
S-matrix.   %%R
\foot{It is important to emphasize that  quantum
corrections computed  by quantizing    the large $J$ limit of a classical action
need not necessarily be  the same as the large $J$ limit of corrections found from the
quantum effective action.
}

Here  we will not  address in detail
the relation to  quantum string theory, concentrating
as a first step on the correspondence  between the scattering phase  entering
the spin chain Bethe ansatz and  the  generalized
 LL model   that reproduces it as its  S-matrix.
 We shall demonstrate that a low-momentum form
 of the BDS S-matrix $S_1$ in \rf{sss}
   is the same as  the  quantum S-matrix
  for   a  LL type action with a particular quartic interaction term,
  thus generalizing to all orders in ${\lambda}$
the  S-matrix relation  \ci{kz} between the
  Heisenberg model   and the standard LL action.
%     to all orders in $\l$.
  The fact that (a  limit of)  the  BDS S-matrix
  can be  interpreted as a  quantum  field theory
  S-matrix  is    non-trivial,
 indicating  the existence of a two dimensional field theory
description behind the asymptotic gauge theory  spin chain.

We shall also show that including the AFS \ci{afs} phase in \rf{sss}
leads in a similar way of  matching the S-matrices
to a  non-relativistic field theory model with quartic interaction
vertex  that
 matches exactly the one
 extracted (using the approach of  \ci{krt})
  from the classical string action on $R \times S^3$.
This  may   not be too surprising, given  that the AFS Bethe ansatz was
obtained by discretizing  \ci{afs} the  classical $R \times S^3$
string Bethe equations of \ci{kmmz},
but this relation may help  to relate the quantum deformation \ci{bt,hl,af}
 of the AFS phase to  world sheet quantum  corrections in a more direct
 fashion.
%%new
In the same spirit we shall discuss the non-relativistic limit of
the AFS-type scattering matrix  proposed in \cite{bs2} for the
$SL(2)$ sector and find that it coincides with the tree-level
scattering matrix of classical string theory on $AdS_3\times
S^1$. This lends strong support to the idea that the dressing phase
relating the ``gauge'' and ``string'' Bethe ans\"atze is universal \ci{beis}
for all
%(bosonic)
 sectors of the theory.

%i was thinking about the meaning  of that  a-- > 0 limit:
%it seems it is related to conformal limit that people take in  sigma
%model S-matrices
%-- they end up with large J   but  also with exploiting asymptotic
% freedom (like Kazakov,
% or conf inv in Polchinski-Mann) -- I am just curious if indeed we
%can get full BDS S-matrix from field theory
%without a-- > 0 ... but we know that full string theory at the end of
%the day should give an S-matrix of similar type...
%that raises a question if there is a ``string-like''  action behind
%BDS ansatz (definiely, there are two  very similar LL actions for BDS
%and AFS so why can't they have all order extensions)
%These random comments are supposed to indicate my search for
%motivating of what were doing

\bigskip

This paper is  organized as follows.

In section 2 we shall first review the  structure of the
generalized LL action for  the $SU(2)$ sector, first in the $SO(3)$
invariant form and then in the complex scalar form found by expanding near the vacuum
state. We shall also discuss  the definition of the
 theory on an infinite line (as required for computation of  S-matrix),
the role of  2d UV cutoff and its relation to the spin chain.
In section 3  we shall illustrate  how to compute the tree-level and 1-loop
corrections to the 2-particle S-matrix for the generalized LL model
containing the  all-order
kinetic term and few higher-derivative interaction terms. We shall
follow mostly the same methods as used at the leading order in $\l$ in
 \ci{kz}.

In section 4 we shall start  with the spin chain  scattering phase  in \rf{sss}
and find  its low-energy  limit in which  one keeps only the leading in
momentum term at each order in expansion in $\l$.
We shall consider separately
the BDS and AFS ans\"atze and in the latter case
we emphasize the new features introduced
by the presence of non-trivial corrections to the phase in \rf{sss}.
%(this limit  has close
%similarity to   the BMN limit).
% may also be interpreted as a kind of continuum limit
%in which 2d short-distance cutoff of spin chain is taken to zero.
Then in section 5 we shall reconstruct  the exact (all-order in $\l$)
quartic vertex in the
generalized LL action  and show  that the resulting  quantum field theory
S-matrix matches exactly the  low-energy limit of the BDS spin chain
scattering phase.

In section 6 we shall comment on a generalization to larger
(compact) sectors containing $SU(2)$ sector.
In section 7 we shall discuss a relation between a
non-relativistic LL type action
  reconstructed  from the S-matrix of the AFS ansatz
%nnew
   and string theory action  on $R \times S^3$, on $AdS_3\times S^1$ and the
   fermionic action \cite{aaf} obtained by
truncation  of the full superstring
action \ci{mt}  to two fermionic fields  corresponding to
 $SU(1|1)$ sector.
   We shall show that the corresponding
   tree-level string S-matrices  matches the  low-energy, strong coupling
%nnew
   limit of the AFS-type S-matrix in  the  $SU(2)$,$SL(2)$ and
   $SU(1|1)$ sectors    respectively.
    %  all rank one sectors.
%   in the $SU(2)$ and $SL(2)$ sectors.
 We shall also explain   that a specific non-local structure of the
 quartic interaction  term in the LL action has its origin in  the
elimination of
  the negative-energy modes when passing from a second-derivative
   to a non-relativistic first-derivative action.

Section 8 will contain some  concluding remarks.
In Appendix A we shall present the results for the quartic
interaction vertex   in the
 LL actions corresponding to
 the $SU(1|1)$ and $SL(2)$ spin chain sectors described by the
BDS-type Bethe  ansatz.
 %and comment also on the case of the ``string'' deformation of the
 %BDS  S-matrix by an extra phase.
  In Appendix B we shall give  some details of the small
  momentum expansion of the  leading quantum correction \ci{bt,hl}
   to the AFS phase.

 \iffalse
  EDIT

\fi

%%%%%%%%%%%%%%%%%%%%%%%%%%%%%%%%%%%%%%%%%%%%%%%%%%%%%%%%%%%%%%%%%%%%%%%%%%%%%%%%%%%%%%%%%%%
\renewcommand{\theequation}{2.\arabic{equation}}
 \setcounter{equation}{0}

\section{ General  structure of the effective\\
 Landau-Lifshitz type  action }

The LL type  action we will be interested in appears in the
 description  the low-energy modes of the ferromagnetic  $SU(2)$
gauge theory spin chain.
Its derivation from the spin chain Hamiltonian involves several steps
 \ci{krt,tsecarg}.
First,
%one rewrites
the quantum-mechanical path integral is expressed in terms of
spin coherent states parametrized by  a unit 3-vector $\vn_a$
at each site $a=1, ... , J$.\foot{$\vn$ represents two  ``phase-space''  variables  of
the  ``classical spin'' $ U^* \vec \sigma U = \vn $.
On the  string side  $\vn$ corresponds to the
two  transverse modes of a ``fast'' string on $R \times S^3$.}
  The resulting (discrete) action
%getting the action
contains a WZ-type  (Berry phase) term \ci{fradkin} linear in the
time derivative  of $\vn_a$   and a Hamiltonian part
$\sum^J_{a=1} [ \l ( n_{a+1} - n_a)^2 +  O(\l^2)  ] $.
 One then considers the large   $J$
 region  and takes the  continuum limit  by truncating away
all but
%assuming that one is interested only in
the
 low-energy spin wave excitations of the periodic
 chain;   only the leading lowest-derivative terms are kept
 at each order in $\l$. It turns out then that
%Then
$\l$
%happens to
 combines with powers of $J$
 into an  effective parameter
 $\tl = { \l \ov J^2}$  which manifests  the existence (at least in the first
 few orders of expansion in $\l$) of a  scaling   BMN-type limit.
 Furthermore, $J$ then appears in front of the action, implying that
 for fixed $\tl$ the large $J$ limit
 is the same as the semiclassical limit, with  $1/J$
 corrections  playing the role of  quantum corrections to the classical LL model.

%%%%%%%%%%%%%%%%%%%%%%%%%%%%%%%%%%%%%%%%%%%%%%%%%%%%%%%%%%%%%%%%%%%%
 \subsection{$O(3)$ invariant $\vn$-field   action}
 %%%%%%%%%%%%%%%%%%%%%%%%%%%%%%%%%%%%%%%%%%%%%%%%%%%%%%%%%%%%%%%%%%%%%%%%%%%

The resulting action  has the following  structure
 ($\del_0 = \del_t, \  \partial_{1}  \equiv \del_\sigma$, \ $\vn^2=1$)
\begin{equation}\label{aal}
\cS=J\int dt \int_{0}^{2\pi}\frac{d\sigma}{2\pi}\ \cL\,,
\ \ \ \ \ \ \ \ \ \  \cL=  \vec C (n) \cdot \del_0 \vn - \cH(\del_1 n) \ ,
\end{equation}
 \be \la{lol}
 \cH= \cH_2 +  \sum^\infty_{k=2}  \cH_{2k} \ , \ \ \ \ \
 \cH_{2k} \sim   (\sfrac{\l}{  J^2})^k  (  \del_1^{2k}  n^{4}
 + ... +  \del_1^{2k}  n^{2k} ) \ ,  \ee
 where $J$ is the total spin equal to the  spin chain length $L$
 and  $\vec C(n) $
is the same as a monopole potential on $S^2$, i.e.
$d C= \epsilon^{ijk} n_i d n_j \wedge d n_k$.
  The general form of the ``kinetic'' part  $\cH_2$ can be found \ci{rt,krt}
 from the continuum limit of the coherent state
 expectation value of the leading spin-spin part of the gauge-theory
 dilatation operator \ci{bks,beisert}
  (assuming consistency with the BMN limit which is  also
 implied in \rf{ook},\rf{sss})
 \be   \la{qua}
 \cH_2 = \ \frac{1}{4}\vn\
\big(\sqrt{1-\tilde{\lambda}\partial_{1}^2}-1\big)\ \vn\ ,\ \ \ \ \ \ \ \ \
\tl \equiv { \l \ov J^2}     \ . \ee
 The ``two-loop'' \ci{krt},  ``three-loop'' \ci{mtt,mtt1}
 and ``four-loop'' \ci{t}  terms in $\cH$  are
 \be \la{two}
 \cH_4= \sfrac{1}{32} a_1  \tl^2  (\partial_{1}\vn)^4 \  ,
 \ee
 \be \la{tro}
 \cH_6= \sfrac{1}{64}\tl^3 \left[ b_1 (\partial_{1}\vn)^2
(\partial_{1}^2 \vn )^2\ +\  b_2  (\partial_{1}\vn
\partial_{1}^2 \vn)^2\  + \  b_3  (\partial_{1}\vn)^6\right]    \ , \ee
\bea
\cH_8&=&
\tl^4 \bigg[ c_{1}  (\partial_{1}^2 \vec{n})^4  +c_{2}
(\partial_{1} \vec{n})^2(\partial_{1}^2 \vec{n} \partial_{1}^4
\vec{n}) + c_{3}
(\partial_{1}\vec{n}\partial_{1}^5\vec{n})(\partial_{1}\vec{n})^2
+  c_{4}  (\partial_{1}^3 \vec{n})^2(\partial_{1}\vec{n})^2\nonumber\\
&& + \
 c_{5}  (\partial_{1}\vec{n})^4(\partial_{1}^2 \vec{n})^2+
c_{6}  (\partial_{1}\vec{n}\partial_{1}^2
\vec{n})^2(\partial_{1}\vec{n})^2+  c_{7}  (\partial_{1}\vec{n})^8\bigg]
\la{jou}
  \ .
\eea
 The known  coefficients consistent with the
 leading terms in the gauge theory dilatation operator
 \ci{bks,beisert} and thus with the
 BDS ansatz
 (i.e. \rf{ook} with £$S=S_1$) are \ci{krt,mtt1,t}
 \be  \la{coef}
 a_1= {3 \ov 4} \ ; \ \ \ \ \ \ \ \ \ \
 b_1 = -{7 \ov 4} \ , \ \ \
 b_2 = -{23 \ov 2} \ , \ \ \
 b_3 = {3 \ov 4} \ ,  \ee
 \be
 c_5 =\frac{111}{4096}\ , \quad \quad
c_7 =-\frac{267}{32768}\ , \quad \quad
c_1-  c_2 + c_3 + c_4 =-\frac{59}{2048} \ .
\label{s1}
\end{equation}
 While the coefficients
 $a_1$ and $b_1$ appear to be  non-renormalized
 when going from small  to large $\l$ region, the
 coefficients $b_2,b_3$ and at least $c_5$ and $c_7$
 are,  in fact,
 %(interpolating)
%actually
functions of $\l$ \ci{bt,mtt1}. They
 have unequal   values at $\l\to 0$ and $\l \to \infty$, i.e.
 they  are found to be  different  from the BDS coefficients
 in \rf{coef} when one starts from the ``string'' AFS Bethe ansatz
  (which includes a non-trivial phase $\theta$ \ci{afs}
  in $S$ in \rf{ook}).
   The ``string'' values that agree with the classical string
  theory predictions are \ci{krt,mtt1,t}
  \be \la{stri}
  b_2 = -{25 \ov 2}\   , \ \ \ \ \ \ \ \ \
 b_3 = {13 \ov 16}\ ,  \ \ \ \ \ \ \ \ \ \
   c_5 =\frac{119}{4096}\ , \ \ \ \ \ \ \ \ \ \
c_7 =-\frac{323}{32768}   \ .
\ee

 %%%%%%%%%%%%%%%%%%%%%%%%%%%%%%%%%%%%%%%%%%%%%%%%%%%%%%%%%%%%%%%%%%%%%%
 \subsection{Complex scalar form of the action}
 %%%%%%%%%%%%%%%%%%%%%%%%%%%%%%%%%%%%%%%%%%%%%%%%%%%%%%%%%%%%%%%%%%%%%%%%%%%
One  may solve the
constraint $\vn^2=1$, i.e. $n_3= \sqrt{ 1 - n_s n_s}$\ ($s=1,2$)
and express the action \rf{aal}
 in terms of the two independent ``magnon'' fields
$n_s$ whose fluctuations describe deviations from the
ferromagnetic vacuum $\vn= (0,0,1)$ representing the gauge-theory BPS state tr$Z^J$.
As already mentioned,
since  $J$  appears in front of the action \rf{aal}
defined  on a circle of  radius 1, the large $J$ expansion
for fixed $\tl$ and fixed length of the string
%
% keeping  $\tl$   and the length of the circle fixed
%while expanding in  large $J$  corresponds to quantum corrections in
%the LL model.
%
represents quantum  loop expansion of the LL model. Keeping also the
excitation number of  a magnon state fixed,  these $1/J$ quantum
corrections  to the energies of the LL states then  match  finite-size
corrections computed directly from the Bethe ansatz
 \ci{btz,mtt,mtt1}.

Our aim here will be  to compute the magnon S-matrix from the LL model
and to compare it to the spin chain
scattering phase $S$ in \rf{sss}. For this purpose
%
%To compute the magnon scattering matrix from the LL model
%and to compare it to the BA  S-matrix in \rf{sss}, which
% will be our aim here,   one should  consider
%
 a different limit is appropriate,
 in which the LL model is defined on an infinite line \ci{kz}.
This can be accomplished by taking $J \to \infty$
while keeping  the 't~Hooft coupling $\lambda$ and the magnon
momenta fixed.\foot{Similar limit was  considered
in \ci{manpol} and  in connection with the
antiferromagnetic state  of spin chain \ci{rss,za,at}.
Recently it  was emphasized also  in  \ci{mald,dor}.}
%of elementary excitations.
As follows from the structure of \rf{aal},\rf{lol},
rescaling the spatial coordinate
 \begin{equation} \la{re}
x=\frac{J}{2\pi}\sigma\ , \ \ \ \ \ \ \ \s \in (0, 2 \pi) \ , \
%\quad t t \frac{1}{h}, \quad
%f,g\rightarrow \sqrt{J}g, \sqrt{J}g
\end{equation}
and  making the field redefinition \ci{mtt}
\be \la{nz}
 n_{s}= 2\sqrt{1-z^2} \ z_{s} \ , \ \ \ \ \ \ \ \ \ \ \ \phi\equiv  z_1 + i z_2
 \  , \ \
\ee
we can rewrite the LL action  \rf{aal} as a ``first-order'' action for a
complex scalar  ``magnon'' field $\phi$
%(below $\phi'= \del_x \phi$)
\begin{equation}
\cS=\int dt \int_{0}^{J}dx\ \bigg\{
%\sfrac{i}{2}(\phi^{*} \dot{\phi}-\phi \dot{\phi}^{*})-
  \phi^* \bigg[i \del_t - (\sqrt{1- \bl  \partial_x^2}-1)\bigg]\phi
  \   - \  V(\p,\p^*)
 \bigg\} , \label{quad}  \ee
\be \la{lamm}  \bl \equiv { \l \ov ( 2 \pi)^2}  \ . \ee
Here $V$  contains terms of all orders in powers of
$\phi$ and  its spatial derivatives   and depends only on $\l$ and not on $J$:
\be  \la{vve}
V= V_4 + V_6 + ... \ ,\ \ \ \ \ \ \ \ \ \
V_{2n} \sim \sum_{k=1}^\infty  \bl^k \del_x^{2k} (\p^* \p)^n  \ .\ee
The  dependence on $J$ is now only in the length of the spatial direction and thus
$J\to \infty$ corresponds  to  a theory on an  infinite line
(provided we also scale the quantum numbers  $m_k$ of modes on a circle so that momenta
$p_k = { 2 \pi m_k \ov J}$   stay fixed in the limit).

 Explicitly, the leading
 quartic interaction term $V_4$ originating from the first three  terms
 in $\cH$ in \rf{qua},\rf{two}  has the form  ($\phi'= \del_x \phi$)
\begin{eqnarray}
V_4&=&  |\phi|^2
\sqrt{1-\bl\partial_{1}^2}\ |\phi|^2-\frac{1}{2} |\phi|^2\big(  \phi^*
\sqrt{1-\bl\partial_{1}^2}\
\phi +c.c.\big)     +\ha a_1 \bl^2  |\phi'|^4\nonumber\\
&+&    { 1 \ov 16}  {\bl^3}  \bigg[2(2b_1+ b_2)|\phi'|^2 |\phi''|^2
+     b_2  (\phi''^{2}\phi'^{*2}+ c.c.) \bigg]  +  O(\bl^4) \ ,
\la{piu}
\end{eqnarray}
or, expanded in $\bl$ to ``4-loop'' order,
\begin{eqnarray}
V_4&=&\frac{\bl}{4}
(\phi^{*2}\phi'^2+c.c)\nonumber\\
&-&\frac{\bl^2}{8}\bigg[\frac{1}{2}|\phi|^2(\phi''''\phi^{*}+c.c.)+
4|\phi|^2(\phi'''\phi'^{*}+c.c.)+6|\phi''|^2|\phi|^2- 4 a_1 |\phi'|^4\bigg] \no\\
&-&\frac{\bl^3}{4}\bigg[\frac{1}{8}|\phi|^2(\phi^{(6)}\phi^{*}+c.c.)+
\frac{3}{2}|\phi|^2(\phi^{(5)}\phi'^{*}+c.c.)+
\frac{15}{4}|\phi|^2(\phi^{(4)}\phi''^{*}+c.c.)\nonumber\\
&+& 5|\phi|^2
|\phi'''|^2    -    \ha  (2b_1+ b_2)|\phi'|^2 |\phi''|^2
-    { 1 \ov 4}   b_2  (\phi''^{2}\phi'^{*2}+ c.c.)    \bigg]+  O(\bl^4)  \ . \la{kj}
\end{eqnarray}
The action \rf{quad}
has manifest $U(1)$ symmetry  and   ``hidden'' $O(3)$ symmetry
(which was explicit in \rf{aal}).
Since we expect this action  to describe an integrable field theory,
the quartic interaction term may effectively determine all higher order terms
(modulo field redefinitions):
the S-matrix should factorize and thus should be  obtainable from
bubble  graphs with quartic interactions only, just as in the leading-order LL
action case discussed  in \ci{kz}.

 \bigskip

 %%%%%%%%%%%%%%%%%%%%%%%%%%%%%%%%%%%%%%%%%%%%%%%%
 \subsection{Infinite line limit and small momentum expansion}
 %%%%%%%%%%%%%%%%%%%%%%%%%%%%%%%%%%%%%%%%%%%%%%%%%%%%

% Below
In section 3  we shall first consider the   tree-level
 2-particle S-matrix for the action \rf{quad},\rf{kj} and then also compute
the first few terms in its
%quantum
loop expansion.
% quantum loop contributions.
We shall find  that the results for the choice of coefficients
 in \rf{coef},\rf{s1}   match
 the low-momentum limit of the  small $\l$ expansion of the BDS S-matrix in \rf{sss}.
  Then in section 4 we shall  consider the opposite problem of reconstructing
  higher-order terms in \rf{piu},\rf{kj}   by  starting with a low-momentum limit
  of the full BDS  S-matrix.

  To prepare for this discussion, it is
   important to clarify the nature of limits we will be taking  and also the
  role of the 2d field theory
   cutoff in this context.
 To consider the S-matrix, we  should  ignore the periodicity condition
 in the spatial coordinate and define the  field theory on an infinite line.
 Formally, it   may seem that this can be achieved by sending  $J$ in \rf{quad}
 to infinity but this ignores the presence of a hidden UV  scale in the problem.
 An indication of a need  for a spatial scale
 can be seen, e.g,   from the fact that
 $x$  in \rf{quad} does not have the standard
 length  dimension ($\bl$ and $J$  should be   dimensionless).

 Let us go back to the spin chain picture
   and consider the limit  in which the number of
sites
%points
 $J$ is sent  to infinity while the  periodicity condition is not imposed.
 In that case we get  an infinite  1d lattice whose  spacing
  may be  denoted  as
 $a$.
  For a
   finite number of points $J$  of  a  periodic chain of length  $L$
   the  step of the lattice is  $a= { L \ov J}$.
 The limit we are interested in is when both $L$ and $J$ are sent to infinity with
 $a$ kept finite.
  More precisely, it is the dimensionless product
 $a p$ where $p$ is a one-dimensional  momentum
 (with  canonical mass dimension) that  should be
 kept finite.
  The momenta of magnons on a circle  are $p_k = { 2 \pi  n_k \ov L} =
  { 2 \pi  n_k \ov a J }$
  and they remain finite provided $n_k$ is also
   scaled to infinity together with $J$.

 Next, if  we  take  a  continuum limit  $a \to 0$
 of the  spin chain Hamiltonian on an
 infinite lattice (using that $\vn_{x+a} - \vn_x =
  a \vn' + \ha  a^2 \vn'' + ...$, etc.)
 the result will differ from \rf{quad} by a rescaling $x \to a^{-1}  x$. Then
 $\del_x$ will be replaced by $a \del_x $
  and there will be a factor of $  1 \ov a $
 in front of the action (coming from the integration measure).
  Higher derivative terms will be suppressed by
 higher powers  of $a$; most of them   can be ignored assuming  that
 one  keeps only the leading in $a$ term at each order of expansion in $\l$.
 Note that the
  presence of the UV cutoff factor $ {1 \ov a}$ in front of the action
is natural
%%R
 on power counting grounds:   the standard loop expansion of the leading-order
 LL
 action  contains
 linear  UV divergences \ci{btz,kruznew}. One  may  choose to  ignore all power divergences
 using, e.g.,  the zeta-function or
  dimensional regularization
   prescription as in \ci{kz}.\foot{Such a  prescription that ignores all power divergences
 appears to be  necessary in order to match
 the BDS S-matrix (see section 5). It is also consistent with  the
 expected conformal invariance
 of the dual string theory,  predictions of which we   should eventually match by
 starting with a properly
 modified AFS ansatz.}
 Then $a$ will play the role  of an effective coupling or an  effective
  2d  Planck constant  that
 counts loop order.

 If we ignore all  power divergences  then the field-theory
   S-matrix will  involve  only dimensionless products
    ($a p,\ a p'$)  of the scale $a$ and
     momenta. In  the continuum limit, it is natural
      to expect that it will match
     the spin-chain S-matrix
     only in the region when momenta are small compared  to the cutoff.
     Indeed, in the small momentum expansion
%is supposed to mean that
both $p$ and $p'$ are small
   compared to the cutoff scale $ a^{-1}$, i.e. $ap \to 0, \ ap' \to 0$ but
      their ratio $p/p'$ is fixed.
   Taking this  limit can be formally implemented    by scaling $a$ to zero while
     assuming  that $\bl  (a p)^2$
   is kept finite.
   %when $ap$  is sent  to zero.
    This  does not necessarily   mean that $\l$
   is taken to be large: this
    means   only that one  wants to  keeps  the leading in  $ap\to 0$ expansion
     term at each order in expansion in  $\l$, i.e. the limit  of small $ap$ is taken
     {\it before} the limit of small $\l$.

\bigskip

  The momenta $p_i$ in the spin chain expressions \rf{sss},\rf{ooo}
 are dimensionless,  corresponding  to the choice of $a=1$,
 i.e. of unit step of the lattice.
 Then   $p_i$ in  \rf{ooo} should
  stand for $a p_i$  if we want $p_i$   to have  canonical mass dimension.
 Taking $a \to 0$   corresponds to uniformly scaling all momenta  to zero,
  so that
 \be    u =\sfrac{1}{2}\cot{\sfrac{ap}{2}} \sqrt{1+
  \sfrac{\l}{\pi^2}
\sin^2\sfrac{ap}{2}}\  \to \  ( { 1 \ov a p } + ...)
\sqrt{1+  \sfrac{\l}{\pi^2} [ (ap )^2  +
...]}
\ee  We shall discuss such  an expansion
 of the spin-chain S-matrix \rf{sss}
in section 4.

Let us mention also  the analogy of this limit of the spin chain
S-matrix
with the BMN-type scaling
 limit  in the Bethe ansatz equations \rf{ook}.
 Suppose   we take the large length $L=J \gg 1$   limit  in \rf{ook}
 by  rescaling at the same time the momenta
 so that the l.h.s part of \rf{ook} stays finite,
 $p_k= { \bp_k \ov J}$,
 i.e. $\bp_k$ will be  finite in the limit.
 Then $u(p)$ in \rf{ooo} that enters the scattering phase
  \rf{sss}  will become
  \be\la{kiu}
   u = J \bar u +  ... \ , \ \ \ \ \ \ \ \ \ \
\bar u =  { 1 \ov \bp}  \sqrt{ 1 + \sfrac{ \l}{ ( 2 \pi)^2  J^2} \bp^2 } \ee
and thus
\be \la{iu}S_1 (p',p) \  \ \to   \ \  \hat S_1 (p',p)= { \bu (p') - \bu (p) + i J^{-1}  \ov
\bu (p') - \bu (p) - i J^{-1} }  \ . \ee
%%R
There is then a  direct analogy with the discussion above with the role of $a$
played by $J^{-1}$,  assuming that we   keep only the  leading term in the $J^{-1}$
expansion at each order in the small $\l$ expansion.
%
%Assuming that one keeps leading term in $1/J$
%  at each order in expansion in $\l$ which is
%This is  formally the same as
%  keeping   $\tl \equiv { \l \ov   J^2}$ fixed while taking $J $ to be large
% one finds direct analogy  with the above
% discussion  where the role of $a$  is played by $1/J$.
This is  formally the same as
  keeping   $\tl \equiv { \l \ov   J^2}$ fixed while taking $J $ to be large.
 Expanding $\hat S_1(p',p)$ in  powers of $ J^{-1}$ will be  analogous to
 the small momentum expansion or quantum
 loop expansion in the corresponding
 effective field theory. The Bethe ansatz equations \rf{ook}
 make sense of course only
 for the theory on a circle,
 implying that at leading order  in $J^{-1}$ one has $e^{ i \bp_k} =1$,
 i.e. $\bp_k = 2\pi n_k + O(J^{-1})$.\foot{In \ci{kz}  the logic was to
 start with the LL model on a  line, derive the
 corresponding quantum
 S-matrix
 $1/p_k-1/p_j +i  \ov  1/p_k-1/p_j-i $, and then use it in the Bethe ansatz equations
like \rf{ook}  with $e^{i p_k J}$ in the l.h.s.
 The main observation was that the resulting  Bethe  ansatz is the same
 as the  limit of
  the Heisenberg model  Bethe  ansatz  in which
   the  (dimensionless)
 momenta $p_j$ are  taken to be small compared to 1. Indeed,
the resulting solutions for low-energy modes found from the two Bethe ansatze
are then the same in the large $J$ limit
(up to order $1/J^2$ terms).
}

 \bigskip

 One may wonder  if it is possible to extend the matching between
the two-dimensional field theory S-matrix and the spin chain S-matrix by keeping
all the higher-derivative terms in the kinetic term
($\vn_{x+a} - \vn_x =  2 \sinh { a \del_x \ov 2} \ \vn_{x+{ a \ov 2}}$)
but still replacing the lattice sum by an infinite integral
and $\vn_x(t)$ by a continuous field $\vn(t,x)$
(with $J$  assumed to be  taken to infinity so that the
 theory is defined on an infinite line).
 In this case
the kinetic term in \rf{qua} or in \rf{quad} will be  replaced by its
``discreet''  counterpart \ci{mtt1}:
\be \la{exx}
i \del_t -  \bigg(  \sqrt{ 1 - 4 \bl \sinh^2 \sfrac{ a \del_x}{ 2} }
-1 \bigg)   \ .
\ee
%%R
The corresponding dispersion relation is the same as
 for the spin-chain magnons:
 $\omega =   \sqrt{ 1 + 4 \bl \sin^2 { a p \ov 2} } -1$.
% with the corresponding dispersion relation  being the same as
% for the spin-chain magnons:
% $\omega =   \sqrt{ 1 + 4 \bl \sin^2 { a p \ov 2} } -1$.
 Moreover, the  range of momenta is restricted to
 $( - { \pi \ov a},{ \pi \ov a})$, so that
 the loop integrals  should be
  automatically finite for a finite
 cutoff $a$. One may then try to fix the
  quartic interaction in the corresponding analog of \rf{quad}
 so that to match the spin chain S-matrix beyond the
  small $a$ or low-momentum limit.
 We will not attempt to do this here. One conceptual issue is that if one does not use the
 small $a$ expansion, it is not clear how  to reinterpret  the BDS S-matrix as a sum of
 bubble
%loop
 graphs in  field theory, following the  LL example  of \ci{kz}.
 One possibility is that the resulting action may  be  considered as a
 quantum effective action,
 whose {\it tree level}  S-matrix should then match
 the exact spin chain S-matrix in.\foot{This interpretation may be useful in order make
 contact with string theory:
 presumably, such action  may be derived  by taking large $J$ limit
  in the quantum string effective
 action for a string moving on $S^3$ part of \adss, just like the
 classical LL model
  followed from the classical string
 action \ci{kru,krt}. We shall return to the discussion of
 related  issues in section 7.}

%%%%%%%%%%%%%%%%%%%%%%%%%%%%%%%%%%%%%%%%%
%%%%%%%%%%%%%%%%%%%%%%%%%%%%%%%%%%%%%%%%%

\renewcommand{\theequation}{3.\arabic{equation}}
 \setcounter{equation}{0}

 \section{Field theory S-matrix}

 The quadratic part of the action \rf{quad} resembles
the action  for the positive-energy part  of
 a massive relativistic scalar field  in two dimensions.   Indeed,
 the classical solutions in the free-field limit are\foot{We are using a
  different normalization of creation operators than in \cite{kz}
  and thus some subsequent formulae differ by factors of $2\pi$.}
 \begin{equation}\la{fii}
\phi(x,t)=\int  {dp \ov \sqrt{2 \pi}}   \ a_{p} \ e^{-i\omega_p  t+ipx}, \quad\quad\quad
\phi^{*}(x,t)=\int  {dp \ov \sqrt{2 \pi}} \ a_{p}^{*} \ e^{i\omega_p t-ipx}
\end{equation}
where\foot{If one  rescales the time coordinate and thus $\omega_p$ by
$\bl$    then $\omega_{p}=  \sqrt{  p^2  + m^2}- m^2 ,\
\ m^2 \equiv { 1/ \bl}$. This normalization corresponds to
extracting one power of $\bl$ from the spin chain energy,
so that the Heisenberg  model energy
does not have an overall $\bl$ factor.}
\begin{equation} \la{dii}
\omega_{p}=  e(p)-1  \ , \ \ \ \ \ \ \
\ \ \ \ \ \ \ e(p) \equiv \sqrt{1+\bl  p^2}\ .
\end{equation}
In the quantum theory $[a_p, a^\dagger_{p'}]= \delta(p-p')$.
The interaction term $V$ in \rf{quad}, however, depends
only on spatial derivatives implying that the S-matrix is not
 expected to be relativistic-invariant.
 % even in a restricted  sense.
A possible approach to finding  a similar action from string theory  is
to solve for
half of modes  at the classical level \ci{kru,krt} or
effectively
 to integrate them out at the quantum level (see section 7).

To compute the S-matrix  corresponding to \rf{quad} we
 follow the same steps as in case of the leading-order LL action in
\ci{kz}.
The crucial simplifying point is that the propagator
 can be chosen as the  retarded one, $D(x,t)\sim \theta(t)$,
 i.e.
 \be\la{prop}
 D(\omega,p)=\frac{i}{\omega- \omega_p +i\epsilon}   \ . \ee
 This implies that the two-body
S-matrix is a sum of bubble diagrams  with $V_4$ in \rf{vve}
as vertices.
Let us consider the 2-body
scattering process with the initial state
$|pp'\rangle=a_{p}^{\dagger}a_{p'}^{\dagger}|0\rangle$
(initial particles being  2 magnons with
momenta $p,p'$),   and  the final state
as
$|kk'\rangle=a_{k}^{\dagger}a_{k'}^{\dagger}|0\rangle$. The two-body
scattering matrix is
\be \la{jjjk}
 \langle kk'|\hat S|pp'\rangle=\langle k k'|T e^{-i\int dt dx V_{4}}|p p'\rangle  \ .
\end{equation}
As usual, the  translational invariance of the action
implies momentum conservation, i.e. that  $\langle kk'|\hat S |pp'\rangle$ is proportional  to
$\delta^{(2)}(k^{\mu}+k'^{\mu}-p^{\mu}-p'^{\mu})$.
 In two dimensions  the
energy and momentum conservation allow the two particles to only
exchange their momenta, so that the
 energy-momentum conservation delta-function becomes
%\foot{The
%different normalization of $\delta_+=\langle kk'|pp'\rangle$
%compared to \cite{kz} is related to the different normalization of the
%creation operators \rf{fii}.}
\be
\label{delta}
\delta(\omega_{p}+\omega_{p'}-\omega_{k}-\omega_{k'})\delta(p+p'-k-k')=
  K(p,p') \   \delta_{+} (p,p',k, k') \ , \ \ \ee
\be
\delta_{+}\equiv \delta(p-k)\delta(p'-k')+\delta(p-k')\delta(p'-k) \ ,\ \
\ \ \ \ \ \
K(p,p') = { 1 \ov   {d \omega_p \ov dp} -  {d \omega_{p'} \ov dp'}  }  \ .
\label{lta}
\ee
Using \rf{dii}, we get
\bea
&& \ \ \ \ \ \ \ \ \ \ \ \ \ \ \ \ \ \
K(p,p') =      \   { \bl{}^{-1} e(p)\ e(p') \ov  p\ e(p') - p'\ e(p) }   \label{ta}\\
 &&\ =     \frac{\bl{}^{-1} }{p-p'}\bigg[1+ \ha \bl (p^2+p'^2+pp')  +
\frac{1}{8} \bl^2  (p^3p'+pp'^3+3p^2p'^2-p^4-p'^4)  + O(\bl^3)
 \bigg]
 \nonumber
\eea
One finds that
\begin{equation}\la{sev}
\langle kk'|\hat S|pp'\rangle\ =\ S(p',p)\ \delta_{+} (p,p',k,k') \ ,
%\ \ \ \ \ \ \ \ \ \
%S(p',p)= 1 + i \cT(p,p')
\end{equation}
where  the ``kinematic'' factor $K(p,p')$ is
 included into the 2-body S-matrix  $S(p',p)$.

%%%%%%%%%%%%%%%%%%%%%%%%%%%%%%%%%%%%%%%%%%%%%%%%%%%%%%%%%%%%%%%%%
\subsection{Leading-order  tree-level term
 }
%%%%%%%%%%%%%%%%%%%%%%%%%%%%%%%%%%%%%%%%%%%%%%%%%%%%%

Starting with the
interaction term
in \rf{piu}, \rf{kj}  and computing the leading-tree-level
contribution
of  the quartic vertex  we obtain the following expression
\bea
\pic{18}{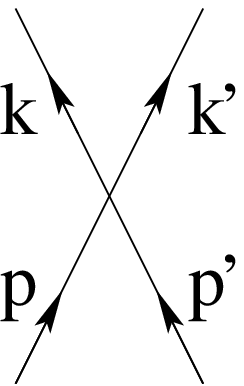}{0.45} &=&- i \langle kk'|V_{4}|pp'\rangle \cr
 &=& - i \bigg[\sqrt{1+\bl (p'-k')^2}+
\sqrt{1+\bl (p'-k)^2}+\sqrt{1+\bl  (p-k')^2}\nonumber\\
&&+ \sqrt{1+\bl  (p-k)^2}-\sqrt{1+\bl  k'^2}-
\sqrt{1+\bl  k^2}-\sqrt{1+\bl  p'^2}- \sqrt{1+\bl  p^2}\bigg]
\nonumber\\
&&-2 i a_1  \bl^2 p
p' k k' \label{vertex} \\
&&- {i \ov 8}  \bl^3p p' k k' \bigg[ (2 b_1 + b_2 )\ (p+p')(k+k') -
2 b_2  \ (p p'+k k')\bigg]+O(\bl^4),    \nonumber
\eea
%%R
%
% original paragraph: to restore remove \iffalse and \fi
%
\iffalse
\begin{figure}[ht]
\centerline{\includegraphics[scale=0.2]{vertex}}\caption{}\nonumber
\end{figure}
\begin{eqnarray}
- i \langle kk'|V_{4}|pp'\rangle = &-& i \bigg[\sqrt{1+\bl (p'-k')^2}+
\sqrt{1+\bl (p'-k)^2}+\sqrt{1+\bl  (p-k')^2}\nonumber\\
&+& \sqrt{1+\bl  (p-k)^2}-\sqrt{1+\bl  k'^2}-
\sqrt{1+\bl  k^2}-\sqrt{1+\bl  p'^2}- \sqrt{1+\bl  p^2}\bigg]
\nonumber\\
&-&2 i a_1  \bl^2 p
p' k k' \label{vertex} \\
&-& {i \ov 8}  \bl^3p p' k k' \bigg[ (2 b_1 + b_2 )\ (p+p')(k+k') -
2 b_2  \ (p p'+k k')\bigg]+O(\bl^4)   \nonumber
\end{eqnarray}
\fi
%
%
where for generality we kept the exact
 form of  the  square root terms  in \rf{piu}.
Expanding in $\lambda$ we get
%%R
%
% original equation: to restore remove \iffalse and \fi
%
\iffalse
\begin{eqnarray}
- i \langle kk'|V_{4}|pp'\rangle &=&i\bl (pp'+kk')+\frac{i \bl^2}{8}\bigg[p^4+p'^4+k^4+k'^4
-4(k+k')(p^3+p'^3) \nonumber    \\
&-&4 (k^3+k'^3)(p+p') +6(k^2+k'^2)(p^2+p'^2)- 16 a_1
\ kk'pp'\bigg]  \label{v}\\
&+&\frac{i \bl^3}{4}\bigg\{-\frac{1}{4}(p^6+p'^6+k^6+k'^6)+
\frac{3}{2}\bigg[(k+k')(p^5+p'^5)+(p+p')(k^5+k'^5)\bigg]\nonumber\\
&-&\frac{15}{4}\bigg[(k^2+k'^2)(p^4+p'^4)+(p^2+p'^2)(k^4+k'^4)\bigg]+
5(p^3+p'^3)(k^3+k'^3)\nonumber\\
&-& \ha ( 2 b_1 + b_2)\ p p' k k' (p+p')(k+k')  +   b_2\ p p' k k' (p p'+k k') \bigg\}
+ O( \bl^4)  \no
\end{eqnarray}
\fi
%
%
\begin{eqnarray}
- i \langle kk'|V_{4}|pp'\rangle &=&i\bl (pp'+kk')+\frac{i \bl^2}{8}\bigg[p^4+p'^4+k^4+k'^4
-4(k+k')(p^3+p'^3) \nonumber    \\
&-&4 (k^3+k'^3)(p+p') +6(k^2+k'^2)(p^2+p'^2)- 16 a_1
\ kk'pp'\bigg]  \label{v}\\
&+&\frac{i \bl^3}{4}\bigg\{-\frac{1}{4}(p^6+p'^6+k^6+k'^6)+5(p^3+p'^3)(k^3+k'^3)
\nonumber\\
&&\!\!
+\frac{3}{2}~\,\bigg[(k+k')(p^5+p'^5)+(p+p')(k^5+k'^5)\bigg]\nonumber\\
&&\!\!
-\frac{15}{4}\bigg[(k^2+k'^2)(p^4+p'^4)+(p^2+p'^2)(k^4+k'^4)\bigg]\nonumber\\
&&\!\!
- \ha ( 2 b_1 + b_2)\ p p' k k' (p+p')(k+k')  +   b_2\ p p' k k' (p p'+k k') \bigg\}
+ O( \bl^4)  \no
\end{eqnarray}
%The external lines just give factors of $1$ in momentum space.
Taking into account the relations between $p,p'$ and $k,k'$ implied by momentum
conservation \rf{delta},\rf{lta}  we find  the following contribution
of the 4-point vertex
\begin{eqnarray}
&& \!\!\!\!\!\!\!
- i \langle kk'|V_{4}|pp'\rangle =    2i \bl pp'-i \bl^2 \bigg[   pp'(p^2+p'^2)    -
 (  {3 \ov 2} - 2a_1 )\  p^2 p'^2 \bigg]
      \la{hh} \\
&& \!\!\!\!\!\!\!
+\frac{i }{4}\bl^3 pp'\bigg[3(p^4+p'^4)- \ha (15 +2b_1
+  b_2)\  pp'(p^2+p'^2)+(10-2 b_1 + b_2 )\ p^2 p'^2\bigg]+ O( \bl^4) \no
\end{eqnarray}
Multiplying this by the kinematic factor in the delta-function (\ref{ta})
we obtain the leading terms in the
tree-level 2-particle S-matrix
\be
S(p',p)= 1 + S_{\rm tree}(p',p) + ...\ ,
     \la{jol}\ee
 corresponding to the action
 %\foot{For comparison with
% the Bethe ansatz  results, here we rescaled the time coordinate in the  LL action
% so that $\bar{\lambda}$ does not appear at $1$-loop.}
  \rf{quad},\rf{piu}
\begin{eqnarray}
&& S_{\rm tree}(p',p) = \frac{2 i pp'}{p-p'}\bigg\{ 1 +   \bl pp' (- { 1 \ov 4 }  +  a_1)
\no\\
&&-\frac{1}{8} \bl^2  pp' \bigg[  (17 + 2 b_1  + b_2 )(p^2+p'^2)
+    \ha (-9 + 2 b_1  - b_2 ) p p'\bigg]    + O( \bl^3) \bigg\} \ .
\la{jkl}
\end{eqnarray}
The comparison  with the gauge Bethe ansatz  allows one to fix the 2-loop
coefficient $a_1$
and two  of the three 3-loop coefficients -- $b_1$ and $b_2$. Indeed, for the values in
\rf{coef} we get
\begin{equation}
 S^{(g)}_{\rm tree}= \frac{2ipp'}{p-p'}+  \bl \frac{i p^2p'^2}{p-p'}
-  \bl^2 \frac{ip^2p'^2(p^2+ p'^2-pp')}{4(p-p')}+O(\bl^3) \ , \la{kop}
\end{equation}
and this is the same S-matrix  that comes out of the expansion of
$S_1$ in \rf{sss},\rf{ooo} in small momenta $p_k=p',\ p_j=p$
at each order in expansion in $\l$ (see \rf{BDS} below).
 We thus generalize the
result of  \ci{kz} that the
 leading ``1-loop''  term $\frac{2ipp'}{p-p'}$  in \rf{jkl}
 which is the tree-level
 S-matrix of the standard
  LL model is the same as the small momentum expansion of the
 phase shift $S_1$ \rf{sss} of the Heisenberg spin chain model
  to the next two orders in
 small $\l$  expansion.

In the case of the LL model that originates from string theory and matches the predictions
of the ``string'' Bethe ansatz  that includes the particular AFS
\ci{afs}   dressing phase $\theta$
in \rf{sss}, the coefficients in the equation \rf{lol} are given
by \rf{stri} \ci{krt,mtt1} (i.e. $a_1,b_1$ are the same while $b_2$ is smaller by 1)
one finds instead
\begin{equation}
 S^{(s)}_{\rm tree}  =   \frac{2ipp'}{p-p'}+   \bl \frac{i p^2p'^2}{p-p'}-
  \bl^2  \frac{ ip^2p'^2(p^2+p'^2)}{8(p-p')}+O(\bl^3) \ . \la{op}
\end{equation}
As expected, the ``gauge''  \rf{kop} and ``string'' \rf{op}
 S-matrices differ
starting at 3-loop order.
The expression \rf{op} follows indeed  from the small $\l$ expansion of the
phase shift  factor of the AFS ansatz (see \rf{AFS}).

We have  also repeated the above computation  including the $\l^4$ terms in \rf{jou}
 and checked that the resulting S-matrix for the values of ``4-loop'' coefficients
 in \rf{s1},\rf{stri}  is again in agreement with with the next-order
 $O(\bl^3)$ term in \rf{kop} in the small momentum
 expansion of the BDS and AFS   S-matrices  \rf{sss},\rf{ooo}
 given below in \rf{BDS} and \rf{AFS}.

%%%%%%%%%%%%%%%%%%%%%%%%%%%%%%%%%%%%%%%%%%%%%%%%%%%%%%%%%%%%%%%%%
\subsection{1-loop correction:  order $\l$ term}
%%%%%%%%%%%%%%%%%%%%%%%%%%%%%%%%%%%%%%%%%%%%%%%%%%%%%%%%%%%%%%%%%%%%%%

Let us now consider the 1-loop correction to the above tree-level
S-matrix \rf{jkl}  following
the  same steps as in the leading-order ``1-loop'' LL model case
in  \ci{kz}.
One may try  to compute  subleading in small momentum expansion term at
 each order in small $\lambda$ expansion.
 Here we shall consider the leading correction to the
 first two $O(\l^0)$ and $O(\l^1)$ terms in \rf{jkl}.
While we will be expanding in $\l$, it is useful to keep
the $\l$-dependent corrections to propagator before expanding in $\l$.\foot{This
takes into account the diagrams with extra insertions   of
the 4-derivative term in the kinetic part of \rf{quad}
into the internal lines.} At this order, the correction to the
scattering matrix is  (as mentioned before, kinematical constraints
require that $p=k$ or $p=k'$; we consider explicitly only the diagram
with $p=k, \ p'=k'$)
%The $1$-loop graph in Figure 2 is
%(we consider the diagram with $p=k, \ p'=k'$)
\begin{eqnarray}
&&
\pic{15}{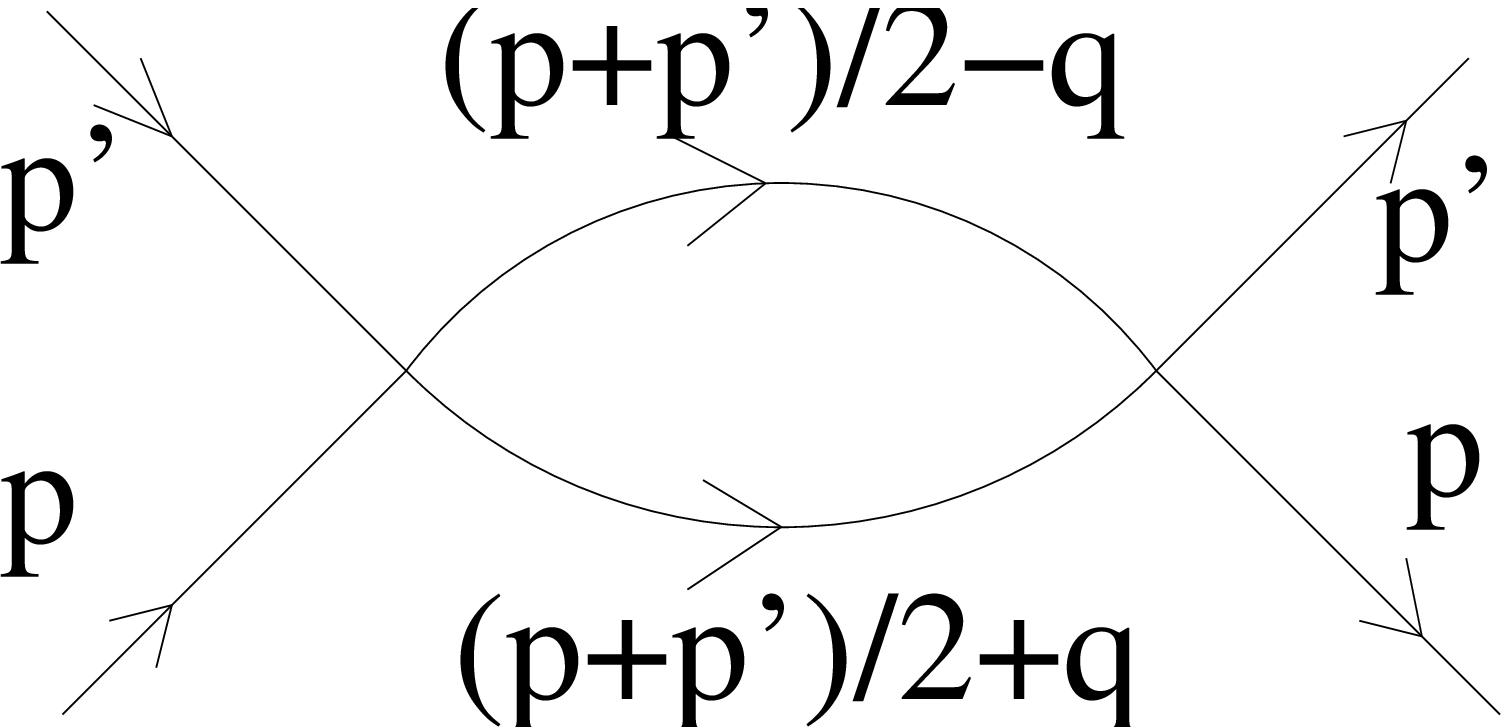}{.18}      \no \\
&&~~~~
=\int
\frac{d\omega dq}{(2\pi)^2} \  D(\sfrac{p+p'}{ 2}+  q) \  D(\sfrac{p+p'}{ 2}-  q )
\bigg[V(p, p', \sfrac{p+p'}{ 2}+  q,
\sfrac{p+p'}{2}-  q)\bigg]^2
\la{dia}
\end{eqnarray}
%%R
% original paragraph: to restore remove \iffalse and \fi
%
\iffalse
\begin{figure}[ht]
\centerline{\includegraphics[scale=0.25]{1loop}}\caption{}
\end{figure}
\be
  \int
\frac{d\omega dq}{(2\pi)^2} \  D(\sfrac{p+p'}{ 2}+  q) \  D(\sfrac{p+p'}{ 2}-  q )
\bigg[V(p, p', \sfrac{p+p'}{ 2}+  q,
\sfrac{p+p'}{2}-  q)\bigg]^2     \ , \la{dia}
\ee
\fi
%
%
\bea
&&D(\sfrac{p+p'}{ 2}+  q)= \frac{i}{  \ha ( \w_p + \w_{p'} ) + \w   -
\w_{ \ha ( p+p')+  q}
 +i\epsilon  }
\la{kp}
\\
 &&=
\frac{i}{\w + \frac{\bl}{4} (p^2+p'^2) -\frac{\bl^2}{16}(p^4+p'^4)-
\frac{\bl}{2}\big(\frac{p+p'}{2}+q\big)^2+\frac{\bl^2}{8}\big(\frac{p+p'}{2}+
q\big)^4 +  O(\bl^3)  +i\epsilon}\no
\eea
\bea
&&V=i\bl \bigg[pp'+\frac{(p+p')^2}{4}-q^2\bigg] \la{ver} \\
&&+\ \frac{i\bl^2}{8}\bigg[2q^4+
3q^2(p-p')^2-\frac{7}{8}(p^4+p'^4)-\frac{9}{2}pp'(p^2+p'^2)-\frac{21}{4}p^2p'^2\bigg]
+ O(\bl^3) \no
\eea
where $\w$ is the energy of the virtual  particle
with momentum $q$  and
we used \rf{kj} with $a_1= { 3 \ov 4}$.
The integral over $\omega$ is easily   done, and  expanding the
propagator in $\lambda$ we  find that the $O(\lambda^{0})$
contribution to the scattering amplitude \rf{sev}  is
(ignoring $\delta_{+}$ factor) \ci{kz}
\begin{equation} \la{hih}
-2\frac{p^2p'^2}{(p-p')^2} \ .
\end{equation}
Here we divided  by a symmetry factor of $2$ and included the leading
 $ \frac{\bl{}^{-1}}{p-p'}\delta_{+}$ term  from the kinematic factor
   \rf{ta}.

 At the next order in $\l$ there are two
contributions:
\begin{eqnarray}
(i)\ =&-&\frac{i \bl^2}{4}\int \frac{dq}{2\pi}\
\frac{  pp'+\frac{(p+p')^2}{4}-q^2    }{q^2-\frac{(p-p')^2}{4}}
\bigg[2q^4+3q^2(p-p')^2
-\frac{7}{8}(p^4+p'^4)\nonumber\\
&-&\frac{9}{2}pp'(p^2+p'^2)-\frac{21}{4}p^2p'^2\bigg]
\end{eqnarray}
and
\begin{eqnarray}
(ii)\ = \frac{i \bl^2}{8}\int  \frac{d q}{2\pi}\
\frac{\big(pp'+\frac{(p+p')^2}{4}-q^2\big)^2}{\big(q^2-\frac{(p-p')^2}{4}\big)^2}
\bigg[p^4+p'^4-\big(\frac{p+p'}{2}+q\big)^4-\big(\frac{p+p'}{2}-q\big)^4\bigg]
\end{eqnarray}
Evaluating the two integrals  ignoring  the power divergences by
using the dimensional
regularization prescription
$\int dq \ q^{\alpha}=0$ ($\alpha=0,1,2...$) as in \ci{kz},  we obtain
\begin{equation}
(i)=4 \bl^2 \frac{p^2p'^2(p^2+p'^2)}{p-p'}\ , \quad \quad\quad
(ii)=-2  \bl^2  \frac{p^2p'^2(p^2+p'^2+pp')}{p-p'} \ .
\end{equation}
Adding them together (while dividing by a symmetry factor of $2$)
%\begin{equation}
%h\frac{p^2p'^2}{p-p'}(p^2+p'^2-pp')
%\end{equation}
and including the delta-function factor in (\ref{ta}) we
finally obtain the next to leading order term in the small
momentum expansion of the 1-loop contribution  to the  S-matrix of the
the generalized LL model  \rf{quad}. Adding this 1-loop correction to
the tree-level expression  \rf{hih} we get
\begin{equation} \la{ress}
S_{\rm tree+ 1-loop} (p,p')= -2\frac{p^2p'^2}{(p-p')^2}-2\bl \frac{p^3p'^3}{(p-p')^2}+O(\bl^2)
\ . \end{equation}
This expression  agrees  with  the next-order term in the  expansion in momenta
of the S-matrix in the BDS and AFS
Bethe ansatze in \rf{BDS} and \rf{AFS}.
%(\ref{BDSSmatrix},\ref{AFSSmatrix}).
%This is an
%extension of the work done in \cite{KZ} to higher loops in
%$\lambda$ and matching with the BDS S-matrix.

Furthermore, we can  follow \ci{kz} and
consider higher-loop ``bubble'' graphs  and show that their contributions
form a geometric series  as in the case of the leading order LL
action.
%simplest leading-order LL
%action  case.
We shall postpone the details of this until  section 5, where we will construct
the all-order scattering matrix.
%the discussion of the general case below.

\renewcommand{\theequation}{4.\arabic{equation}}
 \setcounter{equation}{0}

%%%%%%%%%%%%%%%%%%%%%%%%%%%%%%%%%%%%%%%%%%%%%%%%%%%%%%%%%%%%%%%%%%%%%%%%%%%%%%%%%%%%%%%%%%%
\section{Small momentum expansion of S-matrix of  BDS and AFS   Bethe
ans\"atze \label{lowmomlim}}
%%%%%%%%%%%%%

Let us now determine  explicitly the low-energy
form  of the spin chain S-matrix  in \rf{sss}.

\subsection{BDS    case }
%%%%%%%%%%%%%%%%%%%%%%%%%%%%%%%%%%%%%%%%%%%%%%%%%%%%%%%%

Starting  with the S-matrix of the BDS ansatz, i.e. with $S_1$ in \rf{sss},\rf{ooo}
\be \la{bds}
S_{ \rm BDS} (p',p) = \frac{u(p')-u(p)+i}{u(p')-u(p)-i} \ ,
\ee
\be  \la{fgh}
u(p)=\frac{1}{2}\cot \frac{p}{2}\ \sqrt{1+  4 \bl \sin^2 \frac{p}{2}} \ , \ \ \ \ \ \ \ \ \ \ \
\ \ \ \ \
\bl = { \l \ov (2 \pi)^2}  \ ,
\end{equation}
we may expand it in small momenta, and then also  expand in $\l$
\begin{eqnarray}
S_{\rm BDS}(p',p)&=&1+\frac{2ipp'}{p-p'}-2\frac{p^2p'^2}{(p-p')^2}-
\frac{i p^2p'^2(p^2+10 p p'+p'^2)}{6(p-p')^3}+O(p^4)\nonumber\\
&+&\bl \bigg[ \frac{i p^2p'^2}{p-p'} -\frac{2 p^3p'^3}{(p-p')^2}
- \frac{ip^2p'^2(p^4+16p^2p'^2+p'^4)}{6(p-p')^3}+O(p^6)\bigg]\nonumber\\
&-&\bl^2 \bigg[\frac{ip^2p'^2(p^2-pp'+p'^2)}{4(p-p')}
-\frac{p^3p'^3}{2} \label{BDS}\\
& -& \frac{ip^2p'^2(p^6-3p^5p'+12p^4p'^2-32
p^3p'^3+12p^2p'^4-3p
p'^5+p'^6)}{16(p-p')^3}+O(p^8)\bigg]\nonumber\\
&+&\bl^3\bigg[\frac{i p^2 p'^2(p^4-p^3 p'+p^2
p'^2-pp'^3+p'^4)}{8(p-p')}+O(p^8)\bigg]+O(\bl^4)\ . \nonumber
\end{eqnarray}
To compare with the S-matrix of the
 effective  Landau-Lifshitz model  \rf{aal},\rf{qua}
 with the dispersion relation \rf{dii}
 we are to consider a particular resummation of part of the terms
 in the expansion \rf{BDS}.
 %AAT
 We shall first  keep only the leading term in small momentum expansion
 at each order in $\l$ and then resum
%up
 the series in $\l$. That will determine the tree-level S-matrix of the
 corresponding LL model. We can then keep also certain subleading terms
 in small momentum expansion  that will
 combine into  the geometric series (as in the leading-order LL model case in \ci{kz}).
 In this way  we get
 \begin{equation}
S_{\rm BDS} \ \to \  \td S_{\rm BDS}(p',p)
=1+ \frac{2i pp'}{p \ e(p')-p' \ e(p)}-\frac{2 p^2 p'^2}{[p
\ e(p')-p' \ e(p)]^2}+ \  \ldots, \label{bdv}
\ee
$$ \ e(p) \equiv
\sqrt{ 1 + \bl p^2}      $$
where $\frac{2i pp'}{p \, e(p')-p' \, e(p)}$
represents all leading terms in $p$ at each order in $\l$.

 The reason for this particular structure can be  understood by noting that
 taking $p$ small and keeping only the leading in $p$
 term in the expansion of the $\l \sin^2 {p \ov 2} $ term in $u(p)$
 in \rf{bds} gives
 \be \la{pu}
 u(p) \to  {1 \ov p} \sqrt{ 1 + \bl p^2}   \ ,
 \ee
 \be    \la{puu}
 S_{\rm BDS}=
 \frac{u(p')-u(p)+i}{u(p')-u(p)-i}\ \ \ \ \  \to \ \ \ \ \ \td S_{\rm BDS} =
 \frac{1 + \frac{i pp'}{p  e(p')-p'  e(p)} }{ 1 - \frac{i
  pp'}{p  e(p')-p'  e(p)} }
  \  .  \ee
  Then \rf{bdv} follows upon expansion in small momenta with $e(p)$ kept fixed.
  This  limit is thus { formally} equivalent to taking $p\to 0$ while keeping
  $\l p^2$ fixed. As was mentioned in section 2.3  this limit
   is  reminiscent of the BMN-type scaling  limit with small $p$ expansion
   corresponding to $1/J$ expansion
%with
(note that  the structure
 of the LL action   \rf{aal} is indeed consistent with this scaling limit).

  %which was consistent with  the LL action   \rf{aal}.
  %(though here the momenta are assumed to be continuous and
  %there is no explicit $J$ parameter).
\bigskip

 While the first two  terms  \rf{bdv}
 in the small momentum expansion  of
   $\td S_{\rm BDS}$ \rf{puu} at fixed $e(p)$   are the
   same as in $S_{\rm BDS}$, the  higher order terms
   %(indicated by ellipsis \rf{bdv})
are  different.
%from those coming from the
% expansion in powers of   $\frac{i pp'}{p  e(p')-p'  e(p)}$.
    This is clear already at the leading order in $\l$, i.e.
    at the level of the Heisenberg model vs the standard LL model. We have
      % replacing $u= \ha \cot {p\ov 2}$ by $ { 1 \ov p}$ in
    from    \rf{puu}
    leads to
    \be \la{klo}
    \td S_{\rm BDS} (\l\to 0)  = \frac{1 + \frac{i pp'}{p  -p'} }{ 1 - \frac{i pp'}{p  -p'} }
    = 1+\frac{2ipp'}{p-p'}-\frac{2p^2p'^2}{(p-p')^2}
    - \frac{2i p^3p'^3 }{(p-p')^3}+O(p^4) \ , \ee
    where the  order $p^3$ term is different from the similar one in the
    first line of \rf{BDS} by
  \be \la{loc}
 \frac{i p^2p'^2(p^2+10 p p'+p'^2)}{6(p-p')^3}
 -   \frac{2i p^3p'^3 }{(p-p')^3} = \frac{i p^2p'^2 }{6(p-p')}
  \ee
%%R
%
 This difference has its origin in the small momentum limit, which replaces
%has to  do with the difference between
 $ \ha \cot {p\ov 2}$  with  $ { 1 \ov p}$ on the effective field theory
  side. Interestingly,  after extracting the kinematic factor \rf{ta}
proportional to $\frac{1}{p-p'}$
%(which is associated to the change of
%variables in the delta-function enforcing energy conservation)
this difference is represented  by a local vertex which may thus
  be interpreted as a contribution  of a local counterterm
 one  may  add to the leading-order LL  action.\foot{One choice for such counterterm
 is $\vn'^4$, which will have one  less power of $\l$ compared to  the  2-loop
 term in \rf{two}.
  A more  natural alternative would be the term
 $\vn''^2$ coming out of subleading term in expansion of $\sinh^2$ term in \rf{exx}.
 Strangely, the required coefficient of such  counterterm (5/6) happens
 to be different from the one following from \rf{exx}.
 }

%Including the next order in dispersion relation we find that the
%expansion in (\ref{BDSSmatrix}) can be obtained from the exact
%expression
%\begin{eqnarray}
%S(p',p)&=&1+\frac{2i
%pp'}{p\bigg(1-\frac{p'^2}{12}\bigg)\sqrt{1+hp'^2\bigg(1-\frac{p'^2}{12}\bigg)}-
%p'\bigg(1-\frac{p^2}{12}\bigg)\sqrt{1+hp^2\bigg(1-\frac{p^2}{12}\bigg)}}\nonumber\\
%&-&\frac{2p^2p'^2}{[p \ e(p')-p' \ e(p)]^2}-\frac{2i p^3 p'^3}{[p
%\  e(p)-p' \ e(p)]^3} \label{BDSvertex1}
%\end{eqnarray}

\bigskip

\subsection{AFS  case }

In the case of the ``string'' Bethe ansatz of AFS \cite{afs} the
scattering matrix \rf{sss}   contains a particular dressing phase:
\be \la{kolp}
S_{\rm AFS}= S_{\rm BDS} \ e^{i \theta_{\rm AFS}} \ , \ \ \ \ \ \ \ \
\theta_{\rm AFS}   =  2
\sum_{r=2}^{\infty}\left(\frac{\bl}{4}\right)^{r}
\bigg[q_{r+1}(p)q_{r}(p')-q_{r+1}(p')q_{r}(p)  \bigg] \ ,
\ee
where
\begin{equation}\la{kilp}
q_{r}(p)=\frac{2 \sin ((r-1) \frac{p}{2})}{r-1}\bigg(\frac{\sqrt{1 + 4 \bl \sin^2
\frac{p}{2}}-1}{  \bl \sin
\frac{p}{2}}\bigg)^{r-1}
\end{equation}
To match quantum string theory results, the
 phase  in \rf{sss} must receive modifications at
  subleading orders in  strong  coupling
expansion  \ci{bt}.

The general expression for the phase is
   given by a double sum of the charges $q_r$
\ci{beiklos,bt,beis} with
coefficients having a nontrivial \ci{bt,hl,af,fk}
dependence on $\l$:
\be\la{doa}
\theta (p',p; \l)  =  2
\sum_{r=2}^{\infty}\sum^\infty_{s=r+1} c_{rs} (\l)  \left(\frac{\bl}{4}\right)^{r+ s - 1 \ov 2 }
\bigg[q_{s}(p)q_{r}(p')-q_{s}(p')q_{r}(p)  \bigg] \ . \ee
Here
\be \la{cvc} c_{rs}(\l) = \delta_{s,r+1} + { 1 \ov \sqbl}   a_{rs} + { 1 \ov (\sqbl)^2}   b_{rs} + ... \
,  \ee
and $a_{rs}= { 4 \ov \pi}
 { (r-1)(s-1) \ov (r-1)^2 - (s-1)^2}$ for $r+s=$odd
and zero otherwise  \ci{hl}.
%, and it is likely that its weak-coupling  expansion
%should  also be modified.
%\foot{The general expression
%for the phase  is   given by a double sum of the charges
%\ci{beiklos,bt,beis} with the
%coefficients $c_{rs}$ having a nontrivial \ci{bt,hl,af,fk}
%dependence on $\l$. These corrections appear to be irrelevant in
%the limit of large $\l$   for fixed $p,p'$.}
%v2
The relation  between  the general Bethe ansatz \rf{ook}
with the phase \rf{doa}  and the AFS ansatz  should be understood
as a statement
that the coefficients $c_{rs}$ at large $\lambda$  reduce to
$\delta_{s,r+1}$.\foot{To say
that  AFS ansatz is a strong coupling limit of the general  string ansatz
is not precise as $\lambda$ enters  not only in $c_{rs}$ but also
in the expressions for $u_j$ and $q_r$.}

 Let us  first  ignore  the subleading  terms in  \rf{cvc}
  and
 consider  small momentum expansion  of the  AFS  S-matrix
 \rf{kolp} in the same way as
we did above in the BDS case.
Expanding  in small momenta and then in  $\lambda$ we
obtain
\begin{eqnarray}
S_{\rm AFS}(p',p)&=&1+\frac{2ipp'}{p-p'}-\frac{2p^2p'^2}{(p-p')^2}-
\frac{ip^2p'^2(p^2+10 p
p'+p'^2)}{6(p-p')^3}+O(p^4)\nonumber\\
&+&\bl \bigg[ \frac{i p^2p'^2}{p-p'} -\frac{2p^3p'^3}{(p-p')^2}
- \frac{ip^2p'^2(p^4+16p^2p'^2+p'^4)}{6(p-p')^3}+O(p^6)\bigg]\nonumber\\
&-&\bl^2\bigg[ \frac{ip^2p'^2(p^2+p'^2)}{8(p-p')}-\frac{p^3p'^3}{4}
+O(p^7)\bigg]  \cr
&+&\bl^3\bigg[ \frac{ip^2p'^2(p^4+p'^4)}{16(p-p')}+O(p^8)\bigg] -
\bl^4\bigg[ \frac{5\,ip^2p'^2(p^6+p'^6)}{128(p-p')}+O(p^{10})\bigg]\cr
&+&\bl^5\bigg[ \frac{7\,ip^2p'^2(p^8+p'^8)}{256(p-p')}+O(p^{12})\bigg]
%-  \bl^6\bigg[ \frac{21\,ip^2p'^2(p^10+p'^10)}{1024(p-p')}+O(p^{14})\bigg]
+O(\bl^6)
\la{AFS}\end{eqnarray}
This expression   is different from the expansion in \rf{BDS}
starting with the 3-loop $\bl^2$ terms.

%AAT
As  in the BDS case,  we  may  collect all the leading-order terms  in small
momentum  at each order in $\lambda$, and then   sum up the expansion
in $\l$. The result  may  be again interpreted as a  tree-level
S-matrix of an effective field theory.

Given that the ``string'' Bethe ansatz  was constructed  by starting with the strong-coupling
region,
here it may be more appropriate to  view this low-energy limit
 as\foot{Here $p$ stands, of course,  for both $p$ and $p'$.}
\be \la{ssdd}
p \to 0 \ , \ \ \ \ \ \ \ \ \ \   \ \ \  \l p^2={\rm  fixed}\ ,
\ee
 i.e. as $p \to 0$ with  $\l \sim p^{-2}  \to \infty$.
Since $\l$ is then effectively taken to be large
this  suggests that in this limit  quantum string $1 \ov \sql$ corrections
to the phase in \rf{cvc}   may be ignored.
 Indeed, as we shall find  in section 7.2,  this low-energy, strong coupling
  limit of the AFS S-matrix
 is in perfect agreement with  the classical
  S-matrix of the LL type model
 originating in  a ``non-relativistic'' limit from the string  sigma model on
 $R \times S^3$.
%At the end, this  is not too surprising, given that the
% AFS  ansatz was found \ci{afs}  by  ``discretizing''  the
% integral equation  \ci{kmmz} describing  ``fast'' string motion on $R \times S^3$.

 Taking the limit $p \to 0$ with $\l p^2$=fixed  in \rf{kolp},\rf{kilp}
 as in \rf{pu},\rf{puu} we get
 \be \la{lii}
 q_r (p) \ \to \ p  \bigg[  {e(p) -1   \ov \ha \bl p }\bigg]^{r-1} \ ,\ee
 \be \la{pha}
 \theta_{\rm AFS} \ \to \  \td  \theta_{\rm AFS}=
 \bigg(p' [ e(p)-1]  - p [ e(p')-1]\bigg) \sum_{r=2}^\infty
   \bigg({[e(p) -1][e(p') -1]   \ov  \bl p p' } \bigg)^{r-1}\ . \ee
   Thus
 \be \la{tii}
 \td  \theta_{\rm AFS}= \bigg(p' [ e(p)-1]  - p [ e(p')-1]\bigg)
 { [e(p) -1][e(p') -1]   \ov  \bl p p' - [e(p) -1][e(p') -1]     } \, \ee
 so that
 \be \la{kpu} S_{\rm AFS}\  \to \  \td  S_{\rm AFS}= \td  S_{\rm BDS }  \ e^{i  \td  \theta_{\rm AFS}} \ ,
 \ee
 \be \la{tida}
  \td  S_{\rm AFS}(p',p) =
 \frac{1 + \frac{i pp'}{p  e(p')-p'  e(p)} }{ 1 - \frac{i
  pp'}{p  e(p')-p'  e(p)} } \  {\rm exp} \bigg[ i {(p' [ e(p)-1]  - p [ e(p')-1])
 [e(p) -1][e(p') -1]   \ov  \bl p p' - [e(p) -1][e(p') -1]     } \bigg]
 \  .  \ee
 Note that  like  $ \frac{i pp'}{p  e(p')-p'  e(p)}$ in \rf{puu} the phase
 $ \theta_{\rm AFS}$ scales linearly with momentum  (at fixed  $\l p^2$)
  so that  the leading term in the small momentum expansion is then
  \be\la{jpp}
   \td  S_{\rm AFS} =1 + (\td  S_{\rm AFS})_{\rm tree} +  ...\  , \ee
  \be \la{lea}
 (\td  S_{\rm AFS})_{\rm tree}
 = \frac{2i pp'}{p  e(p')-p'  e(p)} +   {i(p' [ e(p)-1]  - p [ e(p')-1])
 [e(p) -1][e(p') -1]   \ov  \bl p p' - [e(p) -1][e(p') -1]  } \ . \ee
 The expansion of $(\td  S_{\rm AFS})_{\rm tree}$ in powers of $\bl$ then
 reproduces all the leading in small momentum terms at each order in $\l$ in \rf{AFS}.
 An equivalent form of \rf{lea}
 %$(\td  S_{\rm AFS})_{\rm tree}$
 %(that can be found, e.g.,  by using that $\sqrt{ \bl} p=\sqrt{ e^2(p)-1}$, etc.)
 is
 \be \la{afss}
 (\td  S_{\rm AFS})_{\rm tree}= \frac{2i F(p,p') }{p \ e(p')-p' \ e(p)} \ ,
 \ee
 where
\bea \la{jpl}
F(p,p')&=&\  pp' + \ha  [p \ e(p')-p' \ e(p)] \ \td  \theta_{\rm AFS}\\
&=&\
\bl^{-1} \bigg(  \bl pp' - [ e(p)-1]  [ e(p')-1]  \bigg)
\bigg[ 1 + { 1 \ov 4}  \big( \bl pp' -  [ e(p)-1]  [ e(p')-1]  \big) \bigg] \no
\eea
%We  have given explicitly  only  the leading
%in small momentum term in the expansion (the analog
%of the first non-trivial term in \rf{bdv}).
%ATT
By analogy with the BDS case  one could  expect that
the expression in \rf{tida} may be possible to put into a
``ratio'' form similar to \rf{puu}, i.e. that
the subleading terms in \rf{jpp} should form
 geometric series
 \be  \la{faa}
 \td  S  = \frac{1  +  \frac{iF(p,p')}{p  e(p')-p'  e(p) }}
 { 1  -  \frac{iF(p,p')}{p  e(p')-p'  e(p) }    }   \ . \ee
  This, however, does not follow from \rf{tida}.
 Moreover, the expression in \rf{tida} or in \rf{jpp}
 cannot be trusted beyond the leading term
$(\td  S_{\rm AFS})_{\rm tree}$
    which scales  as first  power of   momentum.

  The reason is that  the corrections to the phase \rf{doa},\rf{cvc}
    which we  ignored  produce extra terms  in the exponent in \rf{tida}
     that scale  as
    quadratic and higher power of momenta.
 Indeed, in our low energy limit \rf{ssdd}  $q_r$ in \rf{lii}  scales as $p^r$
 and so the leading (AFS) term in the phase \rf{doa},\rf{cvc}
 scales linearly with $p$. The subleading  terms ignored
 in the AFS approximation \rf{tida} then scale as higher powers $p^2,p^3,...$
 and thus potentially contribute to the terms indicated by  ellipsis
%$+...$ terms
in \rf{jpp}.

%AAA
 It  could  happen  that for a special choice of the coefficients in
\rf{cvc} we could indeed end up with  \rf{faa}.
It is possible to test this
 conjecture   at the level of the first subleading term in \rf{faa}
 using the explicit value of the coefficients $a_{rs}$  in
 \rf{cvc}. One finds (see Appendix B) that the  coefficient of the corresponding
 order $p^2$ correction to the phase depends on odd powers  of
 our fixed parameter $\sql\ p$,
 \begin{eqnarray}\la{jlle}
\delta {\tilde S}_{\rm AFS}=
-\frac{i}{3 \pi } {\bar\lambda}^{3/2} p^2 p'^2 (p-p') + ...
\end{eqnarray}
i.e. it explicitly involves $\sql$, while
 $F$ in \rf{jpl} contains only integer powers of $\l$.
 This non-analyticity  resulting from the first quantum  correction
 to the phase  has  of course the  same origin  as  the one found
  in \ci{bt,sz}. It implies that the first subleading correction in
\rf{jpp} may not
%does not
  agree with the conjecture \rf{faa}.\foot{
  It appears, however,  that a more definitive statement requires knowing all
higher order corrections. While the  particular $\sqrt{\lambda}$ dependence
of the first correction to $\theta$ leads to
$\sqrt{\lambda}$-dependent corrections to \rf{tida}, a  resummation
of the full series may  change this dependence.}
%NEW
Thinking of the LL action as a low-energy approximation to an  effective
quantum 2d string action (which contains string $\alpha' \sim {1 \ov \sql}$
corrections) one may be able to reproduce \rf{jlle}  and similar corrections starting with
\rf{quad},\rf{kj} where $b_2$ and other higher coefficients
in \rf{lol} are actually functions of $\l$,
\be b_2(\l)  = b_{2s}  + {k_1 \ov \sqbl}  + ..., \ee
interpolating between the weak coupling \rf{coef} and strong-coupling
($b_{2s} = - { 25 \ov 2}$)  \rf{stri}
values \ci{bt,mtt1} (cf. \rf{jkl}).

 \bigskip

The  expression
for $  (\td S_{\rm AFS})_{\rm tree}$ \rf{lea} is   more complicated
than the corresponding one for $(\td S_{\rm BDS})_{\rm tree}$ \rf{puu}.
%%%R
%the BDS one in \rf{bdv},\rf{puu}.
%(which, curiously,
%is reproduced  from $  \td S_{\rm AFS}$ by
%keeping only the lowest in momenta term in \rf{jpl}, i.e.
%by $F(p,p') \to  pp'$).
 For that reason
in the next section   we  shall use the BDS case  to illustrate how
to reconstruct a LL type field theory model that
reproduces  such an S-matrix.
We shall return  to the AFS case in section 7.2 where we
will show that  \rf{afss} is precisely the S-matrix
corresponding to the   4-point vertex in the ``non-relativistic'' limit of the
classical string theory  on $R \times S^3$.

\renewcommand{\theequation}{5.\arabic{equation}}
 \setcounter{equation}{0}

%\section{Exact in $\lambda$ S-matrix computation from gauge LL}

\section{All-order  Landau-Lifshitz type
 action \\
  corresponding to BDS S-matrix}
%%%%%%%%%%%%%%%%%%%%%%%%%%%%%%%%%%%%%%%%%%%%%%%%%%%%%%%%%%%%%%%%%%%%%%%%%%%%%%%%%%%

We have seen that the
generalized LL action \rf{aal}-\rf{jou}  defined on an infinite line
leads to the S-matrix  which is the same as the
 leading terms in the small momentum expansion
of the magnon S-matrix that enters  the gauge-theory BDS  or ``string'' AFS
Bethe ans\"atze.
 The matching depends on the proper choice of the coefficients
 of the interaction terms in the LL action:
with the ``gauge theory'' choice  \rf{coef}, \rf{s1}   we found that
equations \rf{kop}, \rf{ress} match  the respective terms in \rf{BDS}, while
for the ``string theory'' choice  \rf{stri}   we found that
the
%third
``3-loop'' term in  \rf{op}
matches the corresponding  term in \rf{AFS}.
Equivalently, matching onto  BDS or AFS S-matrix could
be used to fix (part of) the coefficients ($a_1,b_1,b_2,c_5,c_7,...$)
in the LL action.

We may then turn the problem
around, i.e.  follow the standard field theory logic and try
to reconstruct the   low-energy effective field theory which
will be consistent with
a particular small-momentum  limit \rf{bdv},\rf{puu}
of the  BDS S-matrix to {\it all} orders in expansion in $\l$.
The same can be done
also in the AFS case
and thus   may be important for understanding of how
  a similar  S-matrix may be originating from quantum string theory.

The low-energy limit $\td S_{\rm BDS}$  \rf{puu} of the BDS S-matrix
implies the dispersion relation \rf{dii},
so that  we shall assume that the effective action corresponding to
 $\td S_{\rm BDS}$
has the structure \rf{quad}, where the interaction part $V$ is to be determined.

%%R
%
Quite generally, a 2-body S-matrix fixes  the on-shell value of the
quartic vertex in $V$. The assumption  that this
field theory is integrable (implying  factorization of the
multi-particle  S-matrix) determines the on-shell values of the
interaction terms with higher number of fields in terms of the quartic
one.\foot{In particular, that means that the value of the  coefficient $b_3$ in \rf{tro}
is fixed by the values of $b_1$,$b_2$, and
 similar relations should hold  at higher orders in derivative expansion.}
  Some of the relations constraining  them stem also  from the $SO(3)$
symmetry of the $\vn$-field  LL action which is spontaneously broken
to  $U(1)$ symmetry of the action constructed from the on-shell
vertex for the magnon fields $\phi, \phi^*$.
% This symmetry imposes
% nontrivial relations between quartic and higher-order interaction terms.

%
% Starting with  the 2-body S-matrix we are able only to
%determine the on-shell value of the quartic
%  vertex in $V$, but we may assume that integrability (implying
%  factorization of the  multi-particle  S-matrix)  should  fix the structure of the
%  higher-point interaction terms in terms of the quartic one.
%
%  Also, the action for $\phi, \phi^*$  we will thus find will have the $SO(3)$
%  symmetry of the $n$-field  LL action spontaneously  broken down to $U(1)$;
%  the condition of the hidden $SO(3)$ symmetry   should also  impose  relations
%  between quartic and higher-order interaction terms.
%

The above discussion
 suggests that the leading non-trivial term in \rf{bdv}
 \be \la{iii}
 (\td S_{\rm BDS})_{\rm tree} =  { 2 i pp' \ov   p\ e(p') - p'\ e(p) }
 \ee
  or in  $\td S_{\rm BDS}$ in \rf{puu}
 may  be interpreted as a tree-level field-theory S-matrix.
%
% To find the exact quartic vertex $V_4$  we note that the leading non-trivial term
% \be \la{iii}
% (\td S_{\rm BDS})_{\rm tree} =  { 2 i pp' \ov   p\ e(p') - p'\ e(p) }
% \ee
%  in \rf{bdv} or in  $\td S_{\rm BDS}$ in \rf{puu}
% should be interpreted as a tree-level S-matrix.
 A nontrivial consistency check that  $ \td S_{\rm BDS}$ can indeed
 be interpreted  as a quantum S-matrix of an interacting field theory of LL type
  is that higher powers  of $ {  i pp' \ov   p e(p') - p' e(p) }  $
        (coming from the expansion of $\td S_{\rm BDS}$ in \rf{puu}
 in powers of momenta with $e(p)$ kept fixed)
%should then be
are the same as  loop corrections to the S-matrix
 of a two-dimensional field theory with the interaction vertex
determined from  \rf{iii}. More precisely, these higher order terms should
represent the contributions  of  bubble graphs with several  insertions  of
 this   quartic vertex.

 This is what we are going to show below, thus
   generalizing  the relation \ci{kz}  between the  low-momentum
  (or ``continuum'')  limit of the Heisenberg chain  S-matrix  and
 the  S-matrix of the quantum LL model
  to the BDS case, i.e. to
  all orders in $\l$. As a result, we will  find an effective
   two-dimensional field theory
behind the low-energy limit of the BDS S-matrix.

%%%%%%%%%%%%%%%%%%%%%%%%%%%%%%%%%%%%%%%%%%%%%%%%%%%%%%%%%%%%%%%%%%%%%%
\subsection{Tree-level 4-point interaction vertex}
%%%%%%%%%%%%%%%%%%%%%%%%%%%%%%%%%%%%%%%%%%%%%%%%%%%%%%%%%%%%%%%%%%%%%%%%

  Dividing
  \rf{iii} by the exact kinematic factor coming from the momentum conservation delta
  function \rf{ta} we conclude that, up to the use of momentum
conservation constraints, the exact on-shell four-point vertex should be given by
  \be
  \cV_{\rm on-shell} (p,p') = i \bl  \frac{2 pp'  }{e(p) \ e(p')} \ .
  \label{ex}
\end{equation}
  Here the leading term in the expansion in $\l$ is indeed consistent with \rf{hh}.
  More precisely,  the quartic term
  in the effective action \rf{quad}   written in momentum representation
  with the four fields put on-shell has the form
\begin{equation} \la{ine}
\int dp dp' dk dk'  \ \cV(p,p';k,k')\  K(p,p')\delta_{+}(p,p',k,k')\  a_p a_{p'} a^*_k a^*_{k'}\ ,
\end{equation}
where $a_p$ is the Fourier transform of  the on-shell  field $\phi$
\rf{fii}.
 The
vertex $\cV(p,p';k,k')$ should be  symmetric under the
interchanges $p\leftrightarrow p'$, $k\leftrightarrow k'$, and
also under $(p,p')\leftrightarrow (k,k')$ to ensure  the reality
of the above expression.
To extend  the vertex off shell we shall
assume that the action has the same  structure as at lowest orders \rf{piu},\rf{kj}, i.e.
the interaction terms should involve only spatial  derivatives.
There are many possible off-shell extensions of \rf{ex}
consistent with symmetries of \rf{ine}; they will lead to actions differing
only by field
redefinitions. The simplest possible choice for the off-shell vertex  is
a $(p,p')\leftrightarrow (k,k')$ symmetrization of  \rf{ex}:
\begin{equation}
\cV(p,p';k,k') = \frac{i\bl pp'}{e(p) \ e(p')}+ \frac{i\bl kk'}{e(k) \ e(k')}\ .
\label{exo}
\end{equation}
Expanding this  in $\bl$ we obtain
\begin{equation}
\cV=i\bl (pp'+kk')-\frac{i}{2}\bl^2 \bigg[p p'(p^2+p'^2)+k
k'(k^2+k'^2)\bigg]+O(\bl^3)  \ . \label{vertex2}
\end{equation}
The $\bl^2$ term here   appears to be different from
the one in \rf{v}, though the two agree on-shell
(i.e. the $\bl^2$ terms in  \rf{hh}  and in \rf{ex}
are the same).  As we shall explain below, this is   a reflection of a
 different choice of  an off-shell extension.

In  coordinate space \rf{exo}
 corresponds to the following term in the  Lagrangian
in \rf{quad},\rf{vve}:
\begin{equation} \la{vvv}
V_4=\frac{1}{4} \bl \bigg[   \bigg(\phi^{*}\frac{\partial_{x}}{\sqrt{1-\bl
\partial_{x}^2}}\phi\bigg)^2 +c.c.\bigg]
\end{equation}
Again, the   order $\bl^2$ term in the expansion of  \rf{vvv}
is different from its counterpart in \rf{kj}.
It may seem puzzling   how the two actions  can be related
by a field redefinition given that the interaction terms do not
 involve time derivative while
the kinetic term does. The way how it works
  happens to be a peculiarity  of first-order two-dimensional field
theories.
%However, as expected from the above discussion
%their on-shell values are the same. This is a reflection of the
%fact that two different actions can give the same S-matrix. It
%is known fact that when this happens the two actions are related
%by field redefinition.
Consider a field
redefinition $\phi\rightarrow \phi+\delta \phi$ with
\begin{equation}
\delta\phi=-\frac{3}{2}\bl (\phi''^{*}\phi^2+2 \phi \phi'
\phi'^{*}+\phi^{*}\phi'^2)
\end{equation}
This produces the following  leading-order
change in the Lagrangian in \rf{quad}
\begin{equation}\la{kkk}
\delta \cL=-i(\dot{\phi}-\frac{i}{2} \bl \phi'')^{*}\delta
\phi+c.c.+ O(\phi^6) \ .
\end{equation}
The key observation is that the time-derivative term drops
out since it can be written (using integration by parts)
 as a  total derivative:
\begin{equation}
-i\dot{\phi}^{*}\delta \phi+c.c.=\frac{3 i \bar{\lambda}}{4}
\frac{d}{d t} \big(\phi'^2 \phi^{*2}-\phi^2 \phi'^{*2} \big) \ .
\end{equation}
The remaining  term $ \frac{\bl}{2}  \phi''^{*} \delta \phi+c.c.$
gives an extra $\bl^2$ contribution to the 4-vertex which is
precisely the difference between \rf{vvv}   and  \rf{kj}. As
expected, \rf{kkk} and thus this remaining spatial derivative term
vanishes on-shell, i.e. the two actions give the same tree-level
S-matrix.

One may wonder which is an   $SO(3)$ invariant action \rf{aal}
for the
unit vector $\vn$  which
 leads to the  action \rf{quad}   with the vertex given in \rf{vvv}.
 As follows   from \rf{piu},\rf{vertex},   the natural quadratic in $\vn$ term
 \rf{qua}   in \rf{aal} produces a  non-trivial contribution to the 4-point amplitude
 that should be subtracted from \rf{exo} in order to determine the contribution that comes
 solely from the ``interaction'' $\vn^4$ terms in this action. The resulting
 ``interaction'' vertex is then:
\begin{eqnarray}
\td \cV&=& \frac{i\bl p p'}{e(p) \ e(p')}+\frac{i\bl  k k'}{e(k) \ e(k')}
\\  &+&   i \bigg[e(p'-k')+e(p'-k)+e(p-k')+ e(p-k) -e(p)-e(p')-e(k)-e(k')\bigg]
\no  \end{eqnarray}
It is not immediately  clear
 how to  write down explicitly the all-order  $\vn^4$ term
 which  is
%%R
%would  be
consistent with  such a vertex
 and also generalize  the known terms in \rf{two},\rf{jou};
 one may need to use  some field redefinitions to  simplify it.

%AAT
As already mentioned, the
 off-shell form \rf{vvv}  of the interaction vertex \rf{exo}
is obviously not unique: in section 7.2 we shall present an alternative
to \rf{vvv} in which  non-local factors are distributed symmetrically between the 4 fields
in the vertex and which will be related to a simple scalar action
with 2-derivative kinetic term
whose non-relativistic limit is BDS LL model.

\bigskip

%%%%%%%%%%%%%%%%%%%%%%%%%%%%%%%%%%%%%%%%%%%%%%%%%%%%%%%%%%%%%%%%%%%%%%
\subsection{Loop corrections to field-theory S-matrix}
%%%%%%%%%%%%%%%%%%%%%%%%%%%%%%%%%%%%%%%%%%%%%%%%%%%%%%%%%%%%%%%%%%%%%%%%

Let us now consider quantum corrections to the 2-particle S-matrix
using the vertex \rf{exo}   and the same LL-model
propagator as in  \rf{quad},\rf{prop}, i.e.
$D(\omega,p)=\frac{i}{\omega- e(p) + 1 +i\epsilon}$.
Let us start  with the  1-loop  contribution, i.e. \rf{dia}  now with the vertex  (cf. \rf{ver})
\begin{eqnarray} \la{gghh}
U(p,p';q) \equiv \cV(p,p', \sfrac{p+p'}{2}+q, \sfrac{p+p'}{2}-q)
=\frac{i\bl pp'}{e(p)\ e(p')}+
\frac{i\bl \big(\frac{(p+p')^2}{4}-q^2\big)}{e\big(\frac{p+p'}{2}-
q\big)\ e\big(\frac{p+p'}{2}+q\big)}
\end{eqnarray}
The energy ($\omega$) integral in the one-loop graph receives contribution from
a single pole; it  yields
%
%Doing the integral over the virtual energy  $\omega$
% the contribution from the 1-loop  graph can be
%written as
\begin{eqnarray}
I_1=-i \int \frac{d q}{2\pi}\
\frac{[U(p,p';q)]^2 }{e(p)+e(p')-e\big(\frac{p+p'}{2}+q\big)-e\big(\frac{p+p'}{2}-q\big)+i\epsilon}
\label{opi}
\end{eqnarray}
Note  that  at large $q$
the vertex   $U(q)$  approaches a  constant value while
 the denominator in \rf{opi}  scales as $\sql q$  (the propagator
 of 1-st order theory scales linearly with inverse momentum)
 implying the absence of power divergences  but a potential presence of a logarithmic
 divergence (this is the same behaviour  as, e.g. in the Thirring model).
  Indeed, expanding the integrand at large $q$ we get\foot{Somewhat surprisingly, the coefficient of the logarithmic divergence happens to have
  a non-analytic dependence on $\l$ (cf. \ci{mtt1}).}
 \be \la{logg}
 I_1=-i \int \frac{d q}{2\pi q }  \ \bigg[ { 1 \ov 2 \sqrt{ \bl} }
 \bigg(\frac{\bl pp'}{e(p)\ e(p')} - 1 \bigg)^2  + O( { 1 \ov q}) \bigg]
 \ee
 This  discussion was under an  implicit assumption that $\l$ was kept finite
 while one integrated over the momentum.
 If instead we first expand the integrand in $\l$ and then do the integration over $q$
 separately at each order in $\l$
 we get power divergences but no  logarithmic divergences.
 Similar result was previously  found  in \ci{mtt1}
 in the   discussion of quantum corrections coming from quadratic in fluctuations terms
 in   the generalized LL model \rf{aal}, \rf{qua}.
  To match the BDS  S-matrix (which is essentially perturbative in $\l$)
  we need to adopt this  second prescription
  and also to drop all  power  divergences
  (using, e.g., the dimensional  regularization  as  in \ci{kz}
  as we already did   at low order in $\l$  in \rf{dia}).
  Equivalently, that means that we   should  omit this ``unphysical''
  logarithmic divergence of the integral in \rf{opi}, \rf{logg}.

\bigskip

To evaluate the finite part of the integral we note  that,
interestingly, the  denominator of the integrand in  \rf{opi}
 has two zeroes, regardless the value of the coupling constant $\bl$, at
\begin{eqnarray}
q^2 =  \frac{1}{4}(p-p')^2 \ .
\end{eqnarray}
 They
correspond to simple poles. Only one of them contributes to the
evaluation of the integral, independently of the choice of a contour.
The residue at the relevant pole yields
%The residue at one of the two poles
%gives (assuming $p  \ge p'$):
\begin{equation}\la{poll}
I_{1 \ \rm pole}=[U\big(p,p';\sfrac{p-p'}{2}\big)]^2 \ \frac{\bl e(p)\ e(p')}{p\ e(p')-p' \
e(p)} \ , \
\end{equation}
where from \rf{gghh},\rf{ex} we get
\be \la{ktk}
U\big(p,p';\sfrac{p-p'}{2}\big) = { 2  \bl i p p' \ov e(p) \ e(p') } =
\cV_{\rm on-shell} (p,p') \ .
\ee
In addition to the contribution of the
residue at the pole at finite distance,  there is also a contribution from the
contour at infinity in the complex $q$ plane.
To evaluate it  let us
set $q=R \ e^{i\psi}$ and  pull out a factor
of $1\ov e(q)$, while expanding the
 rest of the expression in large $R$.
We obtain then for the contour integral
\begin{equation}\la{puy}
\bl^2  \int_{\pi}^{2\pi}\frac{d\psi}{2\pi}\frac{R \
e^{i\psi}}{\sqrt{1+\bl  \ R^2 \
e^{2i\psi}}}\ \ \bigg[ \frac{1}{2}\bigg(\frac{\bl pp'}{e(p) \
e(p')} - 1 \bigg)^2+ \ O({f(\psi) \ov R})\bigg]
\end{equation}
The  $O({f(\psi) \ov R})$  terms contain
convergent integrals, so that after  taking the $R\rightarrow \infty$
limit they vanish.
The remaining term which should be formally added
to the pole contribution in \rf{poll} gives the
logarithmically  divergent contribution, i.e.
% which we have already found in \rf{logg},
%pole contribution in \rf{poll} we get
\begin{equation}\la{quy}
I_1 =[U\big(p,p';\sfrac{p-p'}{2}\big)]^2 \ \frac{\bl{}^{-1} e(p)\ e(p')}{p\ e(p')-p' \
e(p)}+\frac{i}{2\pi \sqrt{\bl} }\bigg(\frac{\bl  pp'}{e(p)
e(p')}-  1 \bigg)^2 \ \log (2 R  \sqrt{\bl})
\end{equation}
As we have already  discussed above,
this   logarithmic divergence, i.e. the contribution of the contour integral,
 should be
omitted assuming that  the BDS S-matrix
 and thus the corresponding LL model  should be
  understood perturbatively in $\l$.

Dividing by  the symmetry factor  $2$, and multiplying by the kinematic
factor in
\rf{ta} we thus find  the
following 1-loop contribution to the 2-particle field-theory S-matrix
\begin{equation}\la{onp}
 { 1 \ov 2 }  (I_1)_{ \rm fin}\   \frac{\bl{}^{-1} e(p)\ e(p')  }{p\ e(p')-p'\ e(p)}
  =     - \frac{2 p^2 p'^2}{\big[p\ e(p')-p'\ e(p)\big]^2}
\end{equation}
This matches exactly the second term in \rf{bdv},
i.e. the term in the small momentum expansion of
$\td S_{\rm BDS}$ in \rf{puu}.

One can extend  this computation to higher loop bubble graphs (Figure 1)
as in \ci{kz}.
\begin{figure}[ht]
\centerline{\includegraphics[scale=0.35]{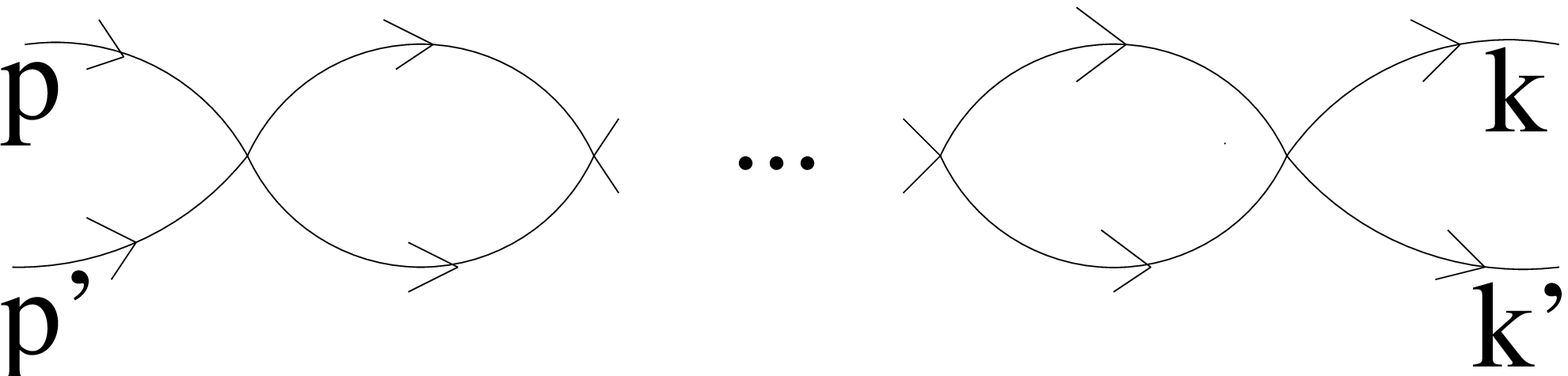}}\caption{}
\end{figure}
Omitting again logarithmic divergences (or all  power divergences
if one first expands in $\l$ as discussed above)
we finish with the following  $n$-loop  contribution to the 2-particle scattering
\begin{equation}\la{hhg}
(I_n)_{ \rm fin} = [U\big(p,p';\sfrac{p-p'}{2}\big)]^{n+1} \
\bigg[ \frac{  \bl{}^{-1}     e(p)\ e(p')}{p\ e(p')-p' \
e(p)}\bigg]^n
\ee
Dividing by the symmetry  factor $2^n$ and  multiplying
by the kinematic factor from \rf{ta} we finish with the following generalization of
\rf{onp}
\be
{ 1 \ov 2^n }  (I_n)_{ \rm fin}\   \frac{ \bl{}^{-1} e(p)\ e(p')  }{p\ e(p')-p'\ e(p)}
  =    2 \bigg[   \frac{i  p  p' }{ p\ e(p')-p' \ e(p)}\bigg]^{n+1}
\end{equation}
Summing up all these  bubble diagram  contributions  gives
(as at leading order in $\l$ \ci{kz})
a simple geometric series
and thus we finish  with
%$i{\cal
%T}$. Then, from the first equation \rf{jjjk} we obtain
the following
field-theory S-matrix
\begin{equation}\la{ll}
S_{\rm LL}(p',p)=\frac{p e(p')-p'e(p)+i pp'}{pe(p')-p'e(p)-ipp'} \ .
\end{equation}
This  is indeed exactly  the same as the low-energy limit of the BDS S-matrix, i.e.
$\td S_{\rm BDS}$ in \rf{puu}.
It   may be viewed as a  generalization to all orders in $\lambda$ of the
the standard (``one-loop'')  Landau-Lifshitz model S-matrix obtained in
\cite{kz}.

It is worth  stressing that it was not a priori clear that the result of this
calculation should yield (the low-energy limit of) the BDS S-matrix.
This  conclusion  rests on
  a  number of details, in particular, on
 the structure of the quadratic term
  as well as of the  quartic vertex in the LL action, making the
agreement nontrivial.
In the Appendix A we shall  generalize the  construction  of the
quartic vertex
in the BDS-related  LL action  to the
$SU(1|1)$ and $SL(2)$ sectors.

\bigskip
%AAT
The  above discussion may  be repeated also in the AFS case by starting
with the 4-vertex  consistent with
\rf{afss}  which we shall explicitly
determine  in section 7.2.
Loop corrections to the S-matrix
of such  LL model   produce again a geometric series
combining into \rf{faa}.\foot{Since
the  vertex corresponding to  \rf{afss} scales with momentum
in the same way as in the BDS case
here again we shall get a  formal logarithmic divergence of
1-loop integral which should be  discarded in the LL framework.
 Similar divergences should  be automatically cancelling only in
 the full superstring calculation where  both positive and negative-energy
modes will
 be propagating in
 loops and also contributions of other bosons and fermions will be
included.}
 However, the expression \rf{faa}
does not
%directly
%%R
naturally follow from the  AFS S-matrix \rf{tida}
(see comments at the end of section 4);  the  important issue
of the relation between the low-energy limit of the exact  string S-matrix
and the  quantum S-matrix  \rf{faa} of the  LL model with tree-level
AFS vertex
remains to be clarified.

%in the case  of a similar LL field theory reconstructed from the
%``string'' AFS Bethe ansatz \rf{afss} (see section 7). In that case
%one need not assume that one is to expand in powers of $\l$.

%Since the  full  dual superstring  theory computation  should   be finite,
%one may expect that such divergence  should get cancelled.
% for further comments
%  related to the finiteness of
% the calculations with an LL-type obtained from string theory.

\renewcommand{\theequation}{6.\arabic{equation}}
\setcounter{equation}{0}

\section{Comments on larger sectors}
%%%%%%%%%%%%%%%%%%%%%%%%%%%%%%%%%%%%%%%%%%%%%%%%%%%%%%%%%%

In the previous sections we have constructed a field theory
whose loop expansion reproduces the BDS S-matrix. A natural question
is whether a similar  field theory exists for larger sectors.
Here we shall make few comments on the case of sectors including the $SU(2)$ sector.
% and, if so,
%what are its properties.

\subsection{S-matrix of Landau-Lifshitz model  for the $SU(1|2)$
sector}

Let us consider, for example,   the $SU(1|2)$ sector
(containing  gauge theory operators  built out of 2 chiral scalars and 1 component of
gaugino \ci{bei2})
where
the leading-order  LL Lagrangian is given by
\cite{hl1,st1} (cf. \rf{aal}, \rf{qua})
\begin{equation}\la{gfd}
\cL=-i
U_{i}^{*}\partial_{0}U_{i}-i{\psi^*}D_{0}\psi-\frac{\tl }{2}\bigg[(1-{\psi^*}\psi)|
D_{1}U_{i}|^2+D_{1}^{*}{\psi^*}D_{1}\psi\bigg]\ ,
\end{equation}
where $D_{a}=\partial_{a}-iC_{a}$,
$C_{a}=-iU_{i}^{*}\partial_{a}U_{i}$ and $|U_{1}|^2+|U_{2}|^2=1$.
Expanding near the  vacuum configuration
$U_1=1, \ U_2=0$ (i.e. $\vec{n}=(0,0,1)$ for
$\vec{n}= U^*\vec  \sigma U$)  and $ \psi=0$
   and rescaling the spatial coordinate by $J$ as in \rf{re}
we obtain the following action
to quartic order in the fluctuation fields $\phi$ and $\psi$
which generalizes  \rf{quad} to the presence of a complex fermion field:
%quartic order in fluctuation Lagrangian
%\begin{equation}
%L=2
%\dot{f}g-H_{2}+{\psi^*}\bigg(-i\partial_{0}+\frac{\partial_{1}^2}{2}
%\bigg)\psi-
%(\dot{f}g-\dot{g}f){\psi^*}\psi-H_4+{\psi^*}\psi
%H_{2}+\bigg(\frac{i}{2}(f'g-fg')\psi
%\partial_{1}{\psi^*}+c.c.\bigg)
%\end{equation}
%where
%\begin{equation}
%H_2=\frac{1}{2} (f'^2+g'^2), \quad H_4=\frac{1}{4J}[2 (f f'+g
%g')^2-(f^2+g^2)(f'^2+g'^2)]
%\end{equation}
%Doing the rescalings as in the $SU(2)$ case we obtain
\be \la{fer} \cS = \int dt \int_{0}^{J}dx\ \bigg[\phi^{*}\ \big(i
\del_t + \sha \bl \del_x^2    \big)\  \phi - {\psi^*}\ \big(i
\del_t - \sha \bl \del_x^2    \big)\ \psi
  - V_4 (\phi, \phi^*, \psi, \psi^*)  \bigg] \ee
\be \la{inte} V_4 = \frac{\bl}{4}\bigg[ ( \phi^{*2} \phi''^2
+c.c.)  -  2 {\psi^*}\psi |\phi'|^2  + \big[(\phi' \phi^{*}-\phi
\phi'^{*}) {\psi^*}{}' \psi + c.c.\big] \bigg] +
\frac{i}{2}(\dot{\phi}\phi^{*}-\phi \dot{\phi}^{*}){\psi^*} \psi
\ . \ee
%NEW%
Here we followed the previously used notation for $\phi,\phi^*$ in
\rf{nz} and the notation for the fermions in \cite{hl1,st1}; to
make the signs in the respective kinetic terms in (\ref{fer}) the
same it is sufficient to interchange $\phi$ with $\phi^*$ or
$\psi$ with $\bar \psi$.

The time derivative dependent interaction term in \rf{inte}
can be
converted into a spatial derivative one by a field redefinition.
More precisely, following the logic outlined in equation (\ref{kkk})
combined with the transformation $\phi\rightarrow
\phi+\frac{1}{2}\phi \psi\bar{\psi}$ replaces the time derivatives in
\rf{inte} with spatial derivatives. The
%
%More precisely doing as in (\ref{kkk}) a field redefinition as
%$\phi\rightarrow \phi+\frac{1}{2}\phi \psi\bar{\psi}$ we end up
%with the above $V_4$ where time derivatives are replaced by space
%derivatives and the
resulting $V_4$ can be simplified to:
\be
\la{inte1} V_4 = \frac{\bl}{4}\bigg[ ( \phi^{*2} \phi''^2 +c.c.) +
2( {\psi^*}\psi'\phi\phi'^{*}+c.c.) \bigg] . \ee
The solutions to the free fermion equations of motion  may be chosen as
 \begin{equation}\la{fiif}
{\psi^*}(x,t)=\int \frac{dp}{\sqrt{2\pi}} \ b_{p} \ e^{-i\omega_p  t+ipx}, \quad\quad\quad
\psi(x,t)=\int \frac{dp}{\sqrt{2\pi}} \ b_{p}^{*} \ e^{i\omega_p t-ipx}
\end{equation}
with $\omega_p$ being the same as in \rf{dii}
and  $\{b_{p},b_{p'}^{*}\}=\delta(p-p')$
(this choice  assures the positivity of energy).
 Then the  fermionic
propagator is the same as the bosonic one  \rf{prop}.
%NEW
\foot{As already mentioned,
an alternative way to make the  analogy with the free bosonic
 theory manifest  is to
switch
%what we mean by magnon and conjugate magnon
${\psi^*}\leftrightarrow \psi$.
}

\begin{figure}[t]
\centerline{\includegraphics[scale=0.6]{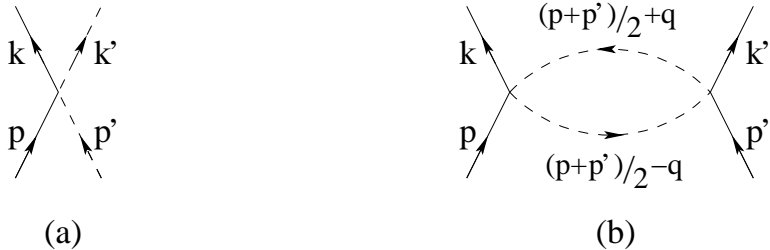}}\caption{ (a) Elastic
scattering of a bosonic and fermionic magnon; (b) potential fermionic
correction to the scattering of bosonic magnons  \label{es2m}}\nonumber
\end{figure}

In addition to the bosonic vertex \rf{kj} that we had in the $SU(2)$
sector, now we have also a $2$ boson--$2$ fermion vertex shown
in Figure \ref{es2m}(a), where the dashed line denotes the fermion.
Suppose we are interested in the bosonic sector of the
S-matrix where  the \emph{in} and \emph{out}
particles are bosons as in the $SU(2)$ case.
The tree-level scattering matrix
is then the same as in the $SU(2)$ sector, while
at  $1$-loop level we  could  get an additional contribution
from the fermionic loop   in Figure \ref{es2m}(b).
However, this contribution {\it vanishes}  since, like
 the bosonic propagator, the fermionic propagator is also
retarded (see also \ci{kz}). This argument extends also to higher loop
contributions.

 We conclude that the bosonic sector of
the  S-matrix of the  $SU(1|2)$ LL model
is exactly the same as that of the $SU(2)$ LL model.
This appears to be in agreement with the structure of the
corresponding Bethe ansatz S-matrix.

\bigskip

\subsection{Absence  of mixing of  magnon S-matrices}

The decoupling observed in the previous subsection for the all-loop
scattering of bosons and fermions is not restricted to the $SU(1|2)$
sector. It is possible to see that in the context of the  LL-type models,
all larger sectors containing $SU(2)$ have the same property: the
quantum LL
S-matrix  with external states from the $SU(2)$ sector is simply  that of
the quantum $SU(2)$  LL model.

There are two essential ingredients which lead to this type of
decoupling.

 First,  the scattering of
magnons is flavor-diagonal. In the case of either the ``gauge'' or ``string''
Bethe Ansatz the magnons are associated to simple roots of
$PSU(2,2|4)$ and scatter following its Dynkin diagram \ci{bs2}.  As a consequence,
 a field theory realising the magnon scattering should have  a very
particular form of  the four-field interaction
terms. Since the magnons are
associated to the nodes of the Dynkin diagram, we may assign  to
them abelian conserved charges. Each term in the magnon Lagrangian
is neutral with respect to these charges; thus each term must contain
an equal number of  magnon fields and their complex conjugates
(in particular, cubic vertices are prohibited).

Second, the magnons around the ferromagnetic ground
state of any spin chain are nonrelativistic. Therefore, it is always
possible to choose the vacuum of the field theory describing them in
such a way that it is annihilated by holomorphic fields. Hence
  all propagators  are retarded: $\langle
\phi^*(t,x)\phi(t',y)\rangle|_{t'<t}=0$.

%Together with the
%structure of the magnon scattering matrix,
It follows then that at the
tree-level there is no term in the effective field theory
Lagrangian which corresponds  to  annihilation of some type of
magnon and  pair-production of a different kind of magnon.
Instead, all terms describe elastic scattering (as in Figure
\ref{es2m}(a), showing the scattering of a boson and a fermion).
%We have therefore argued that, if magnons are nonrelativistic,
Also, loop contributions containing different sorts of magnons
than those on external legs vanish.
The conclusion is
that  magnon scattering in each unit rank sector
 does not receive corrections
from other sectors.
%%R
%

It is
worth mentioning that if  the magnons were  charged under more than one
Cartan generator, then cubic vertices would be allowed in the effective field
theory action,  and, in spite of  the propagators being retarded,
 there
could exist, e.g., a  nontrivial fermion contribution to the
scattering of bosons.

This pattern is obviously  different from what is found
in analogous string-theory  second-derivative
 sigma model computations, where
loop diagrams involving  all states  provide non-trivial contributions   to
diagrams with external states from the $SU(2)$ sector (and,  in fact,  to all
unit-rank sectors). These contributions are  crucial, in particular, for the
cancellation of  2d UV  divergences
%%R
%
and presumably for obtaining the complete dressing phase $\theta$.

\renewcommand{\theequation}{7.\arabic{equation}}
 \setcounter{equation}{0}

\section{The relation to  string theory}
%%%%%%%%%%%%%%%%%%%%%%%%%%%%%%%%%%%%%%%%%%%%%%%%%%%5

Our motivation for  reconstructing the field theory whose
scattering matrix reproduces the asymptotic S-matrix  of the
spin chain \rf{ook} is the expected  close
relation of this S-matrix
 with  string theory in \adss.
The field theory action we discussed above   is a Landau-Lifshitz-type
non-relativistic action which is first order in time derivative.
At the same time, the \adss string action
expanded near the  point-like string  moving along the
$S^5$ geodesic
has a two-dimensional
relativistically-invariant kinetic term \ci{mets,bmn}
and  the interaction terms which are not
 relativistically invariant  \ci{ft1,parn,cal,afz}.

To relate such a second-order  action for ``BMN magnons''
to   first-order LL type action \rf{quad}
for ``spin-chain magnons''
it is
necessary to eliminate
half of the  modes in the former action  -- the negative-energy modes.
% --
%
It is also necessary to eliminate  time-derivative  terms from
the interaction part.

A  general systematic  procedure (not assuming the expansion of $\vn$  near
the $(0,0,1)$ vacuum)  for relating the classical string action on $R \times S^3$
to generalized LL  action \rf{aal}--\rf{tro} was presented  in  \ci{krt}.
It was based on the ``fast-string'' expansion
(i.e. it assumed that the time derivatives of
``transverse'' string profile
$\vn$ are small compared to spatial ones)
and on performing   field redefinitions  to eliminate time derivatives
from
the interaction terms.
That determined the exact quadratic term in the action and also
few  leading coefficients $a_1,b_1,b_2,b_3$ in  \rf{two},\rf{tro} (see  \rf{coef},\rf{stri}).
The ``string'' values of these coefficients are  indeed consistent with
the string AFS-type  Bethe ansatz \ci{mtt1}.

While directly extending the approach of  \ci{krt} to determine the coefficients
of higher order terms in the corresponding effective  LL action
appears to be complicated,
expanding near the vacuum $\vn=(0,0,1)$ and
concentrating only on the quartic interaction vertex  one is able
to fix its exact form as we shall do  below in section 7.2.

%AAT

We shall find that  it matches exactly
the vertex  extracted from the tree-level part of the low-energy limit
of the AFS S-matrix \rf{afss}.
This may be expected, given  that the AFS Bethe ansatz was
obtained by discretizing  \ci{afs} the  classical $R \times S^3$
string Bethe equations of \ci{kmmz},
but it  gives a hope of more direct understanding of
the correspondence   between  quantum string corrections
and the structure of string S-matrix  (cf. also \ci{kaz}).

%
% to restore previous version remove \iffalse and \fi
%

\iffalse
One may take two opposite
standpoints regarding the string Bethe ansatz S-matrix.
One the one hand, it is possible to use the S-matrix approach
to determine the exact form of the 4-magnon interaction term
corresponding  to the string Bethe ansatz S-matrix; the result would
presumably be similar in structure with  $V_4$ \rf{vvv}
found from the BDS ansatz.
On the other hand, since the AFS scattering phase was constructed by
discretizing the classical string Bethe equations, one may contemplate
interpreting $(S_{\rm AFS}-1)$ as the tree-level vertex of a field theory.
The $\l$ expansion of one of these two vertices should match
the results found directly from string action to all orders in
derivative expansion.
\fi

Below  we will first  explain
 how one  can   relate the LL model with 1-st order kinetic term and
 the vertex
of the type of \rf{vvv},\rf{ex} or its AFS analog
%extracted from the BDS S-matrix
to an interacting
 2d action with a relativistic kinetic term.
We  shall then see that  there is indeed
a close connection  between the string sigma model  action on $R\times S^3$
and the LL action of the type  \rf{quad},\rf{vvv},   which
 generalizes
the leading-order classical correspondence
described in \ci{kru,krt} to all orders in $\l$
(but only for quartic interaction terms).
%
%
%
%We will then show that the quartic vertex extracted from the
%nonrelativistic limit of classical string action is the same
%as the low energy limit of $(S_{\rm AFS}-1)$ -- i.e. of the
%AFS S-matrix interpreted as arising from classical scattering.
%Curiously, the truncation of the string vertex to leading momentum
%terms generates the BDS vertex \rf{vvv}.

%%%%%%%%%%%%%%%%%%%%%%%%%%%%%%%%%%%%%%%%%%%%%%%%%%%%%%%%%%%%%%%%%%%%%%%%%%%%%%%%%%%%%%

\subsection{A model scalar field  theory }

Let us start with  an illustrative example
of an effective massive
two-derivative field theory  obtained, for example,  by expanding the string
sigma model around some semiclassical solution.
%AAT
\foot{It may have  two possible origins.
First, we  may think of integrating
out all string fields except few  which will appear in the effective action.
Such an  action would
necessarily exhibit divergences which should disappear
once quantum effects  of the modes present in
this  action are also included (the total quantum string theory
is expected to be finite).
This action  should not be interpreted in the Wilsonian
sense, as the remaining fields may be equally massive as those which
have been integrated out at the first stage. Rather, this may be viewed as  a way to
exactly  account for the quantum effects  of some of the fields
while treating others as classical.
Alternatively, we may think of this effective action in the usual 1PI
sense.  Then  all fields are allowed to propagate in the
loops
 so that  this  effective action should be  free of  2d UV divergences.
As usual, the exact quantum 2d S-matrix is then the tree-level S-matrix of such an
 effective
action.}
Such an effective action will  naturally have a  kinetic term with
  two space and two time derivatives and thus, unlike the
LL-type models, will contain  both positive and
negative-energy modes.  As already mentioned, the latter should be  eliminated
in order to bring  it  to the LL form. Below we will describe
how this can be done
and as a result   reproduce
 some characteristic
features of the LL actions  discussed in the previous sections.

One such feature  is the  occurrence of inverse
powers of $e(p)= \sqrt{ 1 + \bl p^2}$ in the quartic vertex.
 While the precise structure  of
the  vertex depends on the details of the effective action,
the  presence  of inverse powers of $e(p)$   appears to be  directly
related to  the elimination of the
 negative-energy modes.

\bigskip

Let us start with  a  generic complex scalar Lagrangian,
%which may arise, e.g., as a
%quantum effective action for one  mode
% of a string sigma model
%(cf. \rf{quad})
\bea \la{sa}
L =- \phi^*(\partial_t^2-\partial_x^2+m^2)\phi
-  \hV_4(\partial^{(i)})\
\phi^*(z_1)\phi^*(z_2)\phi(z_3)\phi(z_4)+ \dots \ .
\eea
Here the ellipsis denote terms involving more than
four fields (which  are presumably fixed by the integrability of the
theory).
$\partial^{(i)}$
collectively denote space and time derivatives acting on the
field at position $z_i=(t_i,x_i)$.\foot{While generically nonlocal, the quartic
interaction term typically can be  expanded in  series of local operators
of increasing dimension. An exception is the case in which
some of the world sheet fields which have been eliminated
are massless around the chosen background.}

Given  the action \rf{sa} we may compute the corresponding
tree-level S-matrix  by solving the corresponding classical equations
with the free in-field  boundary condition.
The free field  can be decomposed into the positive
energy and  negative-energy modes
\be\la{su}
\phi=\phi_+ + \phi_-~ \ ,
\ee
which will then enter also the non-linear solution and thus the resulting S-matrix.
Since we will be interested only in the  2-body S-matrix, i.e. in the quartic vertex,
we may formally use the decomposition \rf{su} directly in the action.

The kinetic operator  can  be
factorized as:
\be\la{usak}
\partial_t^2-\partial_x^2+m^2=
-D_+\  D_- \ee \be
  D_\pm(\partial)  \equiv  i\partial_t \mp   e(i\del_x)  \ ,
  \ \ \ \ \ \ \ \ \ \
e(i\del_x)\equiv  \sqrt{m^2-\partial_x^2} \ ,  \la{defg}
 \ee
so that  $D_+ \p_-=O(\phi^3),\ \ D_- \p_+=O(\phi^3)$.
Suppose we consider diagrams where only $\p_+$ and $\p^*_+$  appear on external lines
and ask which field theory with 1-st order kinetic term  $D_-$   would reproduce
the same S-matrix.
 To arrive at such action from \rf{sa} we may supplement
 the decomposition \rf{su}  by a field  redefinition
\bea\la{jsa}
{\hat\phi}_\pm =
\sqrt{D_\mp(\partial)}\ \phi_\pm \ .
\eea
Then \rf{sa} expressed in terms of ${\hat\phi}_\pm$ becomes
\bea\label{splitpm}
L &=&
{\hat \phi}^*_+ \left(i\partial_t-\sqrt{m^2-\partial_x^2}\right)
{\hat \phi}_+\cr
&- &\frac{\hV_4(\partial^{(i)})}{
{\sqrt{D_-(\partial^{(1)})\  D_-(\partial^{(2)}) \
D_-(\partial^{(3)}) \ D_-(\partial^{(4)})} }}
{\hat \phi}^*_+(z_1){\hat\phi}^*_+(z_2){\hat\phi}_+(z_3){\hat\phi}_+(z_4)\cr
&+&{\rm terms~containing~}{\hat\phi}_-,{\hat\phi}^*_- \ .
\eea
At the tree level ignoring the dependence
on ${\hat\phi}_-,{\hat\phi}^*_-$ means consistently
truncating the S-matrix to the sector
of the positive-energy modes.
The first two terms  in \rf{jsa} then
give a 1-st order action
that resembles the LL actions discussed above (cf. \rf{quad}).
However, there is an obvious difference in that
the interaction term may contain  time derivatives.

To address this issue let us note that
the truncation
to positive energy modes implicitly assumes that
the resulting action is to be used in the low-energy regime
($\omega \ll m $), where the excitations of the field $\phi$ can be
thought of as nonrelativistic.  It is therefore reasonable to
expand the $D_- $  factors in the interaction vertex in \rf{splitpm} in $\del_t  \ov m$.
Then  we can further  eliminate
the time-derivative dependent terms in the vertex using field redefinitions
(as in the relation between the string sigma model and the LL action in
\ci{krt}), i.e.
using that on the free equations of motion
$i\partial_t{\hat \phi}_+  =  e(i\del_x) {\hat \phi}_+$.
Equivalently, we may just replace $i\partial_t$ by $\sqrt{m^2-\partial_x^2}$
in the quartic interaction term.
We then finish with the following
effective  Lagrangian for ${\hat \phi}_+$
%AAT
\be\label{tpm}
L \simeq
{\hat \phi}^*_+ \left[i\partial_t- e(i\partial_x) \right]
{\hat \phi}_+
- V_4  \ ,
\ee
\be  \la{ttpm}
V_4 =
 \frac{\hV_4[  - i e(i\partial^{(i)}_x) ,\partial^{(i)}_x]}{4
{\sqrt{e (i\partial^{(1)}_x) \ e(i\partial^{(2)}_x)
\ e (i\partial^{(3)}_x)  \ e(i\partial^{(4)}_x)}}} \
{\hat \phi}^*_+(z_1){\hat\phi}^*_+(z_2){\hat\phi}_+(z_3){\hat\phi}_+(z_4) \ .
\ee
Comparing this to the on-shell vertex \rf{ex} written in an equivalent
symmetric form
\begin{equation}
\cV(p,p';k,k')=i\bar{\lambda}\frac{pp'+kk'}{\sqrt{e(p)e(p')e(k)e(k')}}
\label{stvert}
\end{equation}
we observe its close similarity  with the vertex  \rf{tpm}
extracted from the BDS S-matrix
provided we choose $V_4$ in a remarkably  simple
local  form
\be \la{veg}
 \hV_4 (\partial^{(i)})\propto
[\partial^{(1)}_x\partial^{(2)}_x
+
\partial^{(3)}_x\partial^{(4)}_x] \prod^4_{i=1} \delta^{(2)}(z_i - z)
%~~~~~~~~m\propto \frac{1}{{\bar\lambda}^{1/2}}
 \ee
There is still
  a difference in the structure of the kinetic terms in
 \rf{tpm} and in \rf{quad}: apart from the replacement of
 $m^2$   by   $\bl^{-1}$
 (and a rescaling of $t$)
 here we are missing the subtraction of $-1$  in the dispersion relation
\rf{dii}.
%Assuming $V_4$ does not contain time derivatives
This can be easily fixed  by applying a field redefinition
${ \phi} = e^{-i m  t} \td\phi $, where
  $\td\phi_+ $ will now be a ``slow'' field. This mimicks
  the ``fast string'' expansion (based on isolating the fast angle variable)
  done in relating the LL action to  string theory
  action in  \ci{krt} (see also \ci{mik,kut}).
  Then
\be     \la{hhgg}
 - \phi^*(\partial_t^2-\partial_x^2+m^2)\phi
  =  2i m  \td\phi^*\partial_t\td\phi  -
\td\phi^*(\partial_t^2-\partial_x^2)\td  \phi \ee
\be  \la{hgg}
{\hat \phi}^*_+ (i\partial_t- \sqrt{m^2 - \del^2_x}  ){\hat \phi}_+
= {\td \phi}^*_+ \bigg[i\partial_t- (\sqrt{m^2 - \del^2_x} -m) \bigg] {\td \phi}_+
 \ee
 and so  the transformation from the relativistic to
  non-relativistic theory can
 be viewed as a  standard non-relativistic expansion.

 %%%%%%%%%%%%%%%%%%%%%%%%%%%%%%%%%%%%%%%%%%%%%%%%5
\subsection{Relation between
 string  sigma model on $R \times S^3$ and the AFS S-matrix}
%%%%%%%%%%%%%%%%%%%%%%%%%%%%%%%%%%%%%%%%%%%

Let us now  turn to string theory
  and explain how the above action appears  in the
  context of the discussion of \ci{krt}.
  There one started  with the classical string action on $R \times S^3$
  with the metric  $ds^2= - dt^2 +  [d \alpha   + C(n)]^2 + d \vn d \vn $,
  performed 2-d duality $\alpha\to \td \alpha $, and then  fixed the ``uniform''
   gauge: $t=\tau, \ \td \alpha = { J \ov \sql}  \sigma$, i.e.
   $p_\alpha = { J \ov \sql}$=const.%VV2
\foot{The choice  of the isometry direction $\alpha$ in fixing the uniform gauge
corresponds to a particular choice of a charge that is assumed to be distributed homogeneously
along the string to match the spin chain picture \ci{krt}. In the  gauge used in \ci{krt,kut}
that isometry direction  corresponded to the total spin $J= J_1 + J_2$ in the $SU(2)$ sector,
while in the uniform gauge used in \ci{AF} the corresponding charge   was single
spin component $J_1$ (i.e. $\alpha$ was the  angle in one of the three rotation planes).}
The resulting action then takes the form \rf{aal};  after
   the redefinition \rf{re}  of the world-sheet coordinate
   $\s \to x= { J \ov 2 \pi } \sigma $
   the string Lagrangian  \ci{krt} takes the  $J$-independent form
   (${\cal S} = \int dt \int dx^J_0 \ L$)
  \be \la{ccc}
\cL=C_t-\sqrt{[1-\frac{1}{4} (\partial_t
\vn)^2][1+\frac{\bar\lambda}{4}(\partial_x \vn)^2]
+\frac{\bar\lambda}{16}(\partial_t \vn\cdot\partial_x
\vn)^2}~~. \ee
Expanding \rf{ccc} near $\vn=(0,0,1)$ and using  \rf{nz}
we get for the terms quartic in fluctuation field
 (cf. \rf{quad},\rf{kj})
\bea L&=&
 i{\phi}^*\partial_t\phi
- \frac{1}{2}{\phi}^*(\partial_t^2-{\bar\lambda}\partial_x^2)\phi
+ \frac{1}{4} \left[
{\phi}^*{}^2 ( \dot \phi^2 - \bl  \phi'^2) + c.c. \right]
\cr
&+& \ \frac{1 }{8}\left(\dot \phi^*{}^2-\bl
{\phi'}^*{}^2\right)
\left(\dot \phi^2-\bl {\phi'}^2\right) + O(\phi^6)
\label{exptest}
\eea
We can now apply
to this action
the procedure from the previous subsection
to read off the quartic vertex in the corresponding ``non-relativistic'' action.

The first step is
to replace
the time derivatives in the quartic interaction
 term in   \rf{exptest}   with their expression following from
the free equations of motion (the result  is the same as doing field
redefinitions and ignoring  higher than quartic terms):
\be \la{joy}
\del_t \phi \ \ \to  \ \   - i [ e(i\del_x) -1 ] \phi \ ,\ \ \ \ \ \
e(i\del_x) \equiv  \sqrt{ 1 - \bl \del^2_x}
 \ee
 The resulting quartic vertex \rf{ine} is then easily found
 from \rf{exptest} in momentum representation:
 \begin{equation}
\cV(p,p';k,k')=
i\frac{\td \cV_4(p,p',k,k')}{\sqrt{e(p)e(p')e(k)e(k')}} \ ,\ \ \ \
\la{ert}
\end{equation}
 \bea
 \la{ddd}
 \td \cV_4 = &&   \bl  (p p' + kk')   -[e(p) -1 ][e(p')-1]
   -[e(k) -1 ][e(k')-1]           \no \\
&& +\  \frac{1 }{2}  \left(\bl p p' -[e(p) -1 ][e(p')-1]  \right)
 \left(\bl k k' -[e(k) -1 ][e(k')-1]  \right)
 \eea
 %AAT
Upon multiplication of this by the energy-momentum  delta-function in
\rf{delta},\rf{ta}  (allowing us to set $k,k'$ to $p,p'$ or $p',p$)
we then reproduce precisely the ``tree-level'' part of the low-energy AFS
S-matrix in \rf{afss},\rf{jpl}:
\be
 \cV (p,p',k,k')\    {  \bl{}^{-1} e(p) e(p') \ov  p e(p') - p' e(p) }
 \  \delta_+(p,p',k,k')
=  (S^{\scriptscriptstyle {\rm SU(2)}}_{\rm string})_{\rm tree}
 \  \delta_+(p,p',k,k')  \la{fero}
\ee
\be \la{zzzz}
(S^{\scriptscriptstyle {\rm SU(2)}}_{\rm string})_{\rm tree}
= { 2 i F(p,p') \ov  p e(p') - p' e(p) }
\ . \ee
Following
the procedure of the previous subsection,
the resulting non-relativistic effective  Lagrangian  corresponding
to \rf{exptest} is (cf. \rf{tpm},\rf{quad})
\be\la{itak}
L = \phi^*\bigg[ i \del_t - (\sqrt{1 - \bl \del^2_x}  -1)\bigg]
 \phi  - V_4(\phi)  +  O(\phi^6) \ , \ee
\bea \la{geq}
V_4 &=&
 \frac{1}{4} \bigg\{
\bigg( { 1 \ov \sqrt { e(i\del_x) }} \phi^* \bigg)^2 \bigg[
\bigg(   {  e(i\del_x) -1  \ov  \sqrt { e(i\del_x) }         }  \phi\bigg)  ^2
 + \bl \bigg( { \del_x \ov \sqrt { e(i\del_x) } } \phi \bigg)^2\bigg] + c.c. \bigg\}
\\ \no
&-&  \frac{1 }{8}\bigg[ \bigg(    {  e(i\del_x) -1  \ov  \sqrt { e(i\del_x) }}
 \phi^*       \bigg)^2
+\bl
   \bigg(     { \del_x \ov \sqrt { e(i\del_x) }}  \phi^*         \bigg)^2\bigg]
\bigg[        \bigg(    {  e(i\del_x) -1  \ov  \sqrt { e(i\del_x) }}  \phi     \bigg)^2
+ \bl
   \bigg(     { \del_x \ov \sqrt { e(i\del_x) } } \phi         \bigg)^2
                     \bigg]
\eea
Expanding  \rf{itak} in powers of spatial derivatives
one can check that it agrees (modulo field redefinitions like the one
 below \rf{vvv})
 with the leading  terms in the
LL vertex \rf{kj}   for the ``string'' value of the coefficient
$b_2$ in \rf{stri}.

%AAT
$V_4$ in \rf{geq}  is  the ``string'' or AFS  analog
of the exact off-shell BDS field theory vertex \rf{vvv}.
To make the  analogy with \rf{geq} more explicit
 we can represent the on-shell BDS vertex \rf{exo}
in a more symmetric form (equivalent on-shell to
 \rf{vvv})
\be \la{seq}
(V_4)_{\rm BDS} =
 \frac{\bl}{4}   \bigg[
\bigg( { 1 \ov \sqrt { e(i\del_x) }} \phi^* \bigg)^2
 \bigg( { \del_x \ov \sqrt { e(i\del_x) } } \phi \bigg)^2 + c.c. \bigg] \ .
 \ee
  This is simply  one of the terms present in \rf{geq}.
  It is then also clear  which is the analog of the
  scalar action \rf{exptest}
  that would  lead to ``non-relativistic'' action
  \rf{itak}   with such a  quartic term:
  \be \la{oy}
  L_{\rm BDS} =
 i{\phi}^*\partial_t\phi
- \frac{1}{2}{\phi}^*(\partial_t^2-{\bar\lambda}\partial_x^2)\phi
-  \frac{\bl }{4}  \left({\phi}^*{}^2  \phi'^2 + c.c. \right)  + O(\phi^6) \ .
\ee
Omitting here  the
  quadratic term with two time derivatives  leads to the
standard (leading order) LL action. Since this  affects
only the quadratic terms, this relation is  perfectly consistent with the fact
that  the BDS ansatz is the minimal (and natural)
generalization of the XXX$_{1/2}$ spin chain,
 affecting only the dispersion relation.
The interaction term in \rf{oy}
 is invariant under
 time-dependent $U(1)$ rotations, so we  can  also put the kinetic term in
 \rf{oy} in the standard relativistic  form by applying
 the rotation
 $\phi = e^{i t} \vp$  that  induces instead a   mass term (cf. \rf{hhgg})
  \be \la{toy}
  L_{\rm BDS} =  - \frac{1}{2}{\vp}^*(\partial_t^2-{\bar\lambda}\partial_x^2-1 )\vp
-  \frac{\bl}{4}   \left({\vp}^*{}^2  \vp'^2 + c.c. \right)  + O(\vp^6) \ .
\ee
This ``BDS'' action \rf{oy} has  obviously quite  different structure from
\rf{exptest} that follows from  string theory.

\bigskip

\subsection{Strings on  $AdS_3\times S^1$, AFS-type S-matrix in the
$SL(2)$ sector and the universality of the dressing phase }
% for bosonic sectors}

%v2
An  important conclusion of the previous  subsection is that the tree-level ``2-magnon''
 S-matrix
following from the string action on $R \times S^3$ is indeed the same
as the low-energy limit of the AFS Bethe ansatz S-matrix \rf{afss} in the $SU(2)$
sector. This provided
%In the previous subsection we described
a direct relation between the
classical string theory
 and the low momentum limit of
the AFS ansatz.
An obvious question is whether this relation can
be extended to other sectors.
The $SL(2)$ sector is of particular
interest:  a successful comparison would
%show conclusively
give a nontrivial check of the suggestion \cite{bs2,beis} that
at the classical level in string theory the dressing
phase $\sigma^2=e^{i\theta_{\rm AFS}}$ is universal to all
%(bosonic)
sectors of the theory.

The construction of the low momentum limit of
the BDS-type scattering matrix in the $SL(2)$ sector proceeds as in
section \ref{lowmomlim} and we postpone the details to Appendix  A.4.
%\ref{vertsl2}.
Assuming that the dressing phase is universal, the
equations \rf{yyy} and  (\ref{tii}) imply that the ``tree-level'' part of the low
energy AFS-type S-matrix in the  $SL(2)$ sector
 (i.e. the counterpart of  \rf{afss},\rf{jpl} in the
$SU(2)$ sector) is
\bea
({\tilde S}^{\scriptscriptstyle {\rm SL(2)}})_{\rm tree} &=& i
\left[(p-p')\
-\frac{p^2+p'^2}{pe(p')-p'e(p)}\right]+i{\tilde\theta}_{\rm AFS}\cr
&=&
 -\frac{i\ pp'}{pe(p')-p'e(p)}\bigg[1 + e(p)e(p') - \bl pp' \bigg]\ .
\label{BStreeSL2}
\eea
In the spirit of the previous subsection, this expression  should be
compared to the world-sheet tree-level scattering matrix of ``magnons''
 (small string fluctuations near the BMN vacuum in the
parameterization \rf{nz}, i.e. the $S^1$ geodesic)
on
$AdS_3\times S^1$. We   shall compute the latter below.

It is worth emphasizing that the  notion that the dressing
factor  $\sigma^2$
is universal has a meaning only
under the assumption that the vacua in  the
various sectors are chosen to be the same.
% (perhaps inherited from the full string
%theory)
\footnote{In multi-component integrable field theories, a change
of vacuum state typically entails a change of the phase of the
scattering matrix. Consequently, changing the vacuum state of only
one sector induces a phase change of the scattering matrix of only
that sector and thus a relative phase compared to the other sectors.}
This means, in particular, that to
extract  the relevant vertex it is necessary to use the  uniform
gauge \ci{krt,AF} as in our discussion of the $SU(2)$ sector
(i.e.
$t=\tau, $ $ p_\alpha=$const, or   $ \td \alpha = { J \ov \sql}
 \sigma = {  x \ov \sqrt {\bl}}$,
 where $\alpha$ now is the coordinate  of $S^1$ from  $S^5$)
 and to consider small string fluctuations near the
 $S^1$ geodesic.

The corresponding gauge-fixed string
action  on $AdS_3 \times S^1$, i.e. the counterpart of \rf{ccc},
   was constructed in  \cite{ptt}
(see sect. 2 there).
Using slightly different parametrization than in \ci{ptt} (see \ci{st})
 the  metric of  $AdS_3 \times \td S^1$  (after 2d duality $\alpha \to \td \alpha$)
 may be written
 as
$ds^2= - [d t   + B(\vl)]^2 + d \vl d \vl + d \td \alpha^2 $.
Here
$ \vl $ is a pseudo-unit 3-vector
\be
\ell^i \ell^j\eta_{ij}=-1~,   \ \ \ \ \ \   ~~~~\eta_{ij}={\rm diag}(-1, 1, 1)~,
\ee
 with a parameterization in terms of a single complex scalar
  convenient for   an
expansion near the vacuum ${\vl}=(1,0,0)$ being
\be {\vl}=\left(1+2|\phi|^2,\,-i(\phi-\phi^*)\sqrt{1+|\phi|^2}, \,
(\phi+\phi^*)\sqrt{1+|\phi|^2}\right)~.
\ee
The connection 1-form $B$ (the analog of $C$ in \rf{aal},\rf{ccc})
projected on the world sheet has components
\be
B_a=-\frac{1}{2}\int d\xi \ \epsilon_{ijk}\ell^i\partial_\xi \ell^j \partial_a \ell^k \,
~~~~~~~~ B_a=\frac{\ell^2\partial_a \ell^3 -
\ell^3\partial_a \ell^2}{2(1+\ell^1)} =-\frac{1}{2}(\phi^*\partial_a\phi -
\phi\partial_a\phi^*) \ .
\ee
Then the gauge-fixed string action on $AdS_3 \times S^1$
       \ci{ptt} becomes
\bea
\label{Lsl2}
{\cal S}&=&-  \int   dt \int^{J}_0 dx\  L \ ,\ \ \ \ \ \ \ L= \sqrt{-h} ~ ,   ~\\
h &=&
\left[-(1+B_t)^2 +  \frac{1}{4}(\partial_t {\vl})^2\right]
\left[ 1-\bl B_x^2+\frac{\bl}{4}(\partial_x {\vl})^2\right] -{\bl}
\bigg[B_x(1+B_t)
-  { 1 \ov 4} \partial_t {\vl}\cdot \partial_x {\vl} \bigg]^2~.
\nonumber
\eea
 This action has a form of the  Nambu action in a static gauge,  but,  in contrast to \rf{ccc},
without  a WZ-type term (the analog of the latter, i.e. $B_t$, here
comes out of the square root term  upon ``fast-string''  expansion
leading to the LL action \ci{ptt}).

With these preliminaries, we are ready to extract the four-point
vertex following the same steps as in section 7.1.
 Expanding
(\ref{Lsl2}) to quartic order in the fluctuation field $\phi$ gives
\bea L&=&
 i{\phi}^*\partial_t\phi -
\frac{1}{2}{\phi}^*(\partial_t^2-{\bar\lambda}\partial_x^2)\phi\cr
&-& \frac{1}{4} \left[ {\phi}^*{}^2 \left(\dot \phi^2 - \bl
\phi'^2\right) + c.c. \right]  +  \frac{1}{4}\left[ i \phi^*{\dot\phi}{}^*
\left(\dot \phi{}^2-\bl {\phi'}{}^2\right) + c.c. \right] \cr
&+&\  \frac{1 }{8} \left(\dot
\phi^*{}^2-\bl {\phi'}^*{}^2\right) \left(\dot \phi^*{}^2-\bl
{\phi'}^*{}^2\right)
 + O(\phi^6) \ .  \la{ghk}
\eea
The difference compared to
the $R_t \times S^3$ case \rf{exptest} is in the  change of sign of the
second-derivative quartic term (that has to do with the opposite sign
of the curvature of $AdS_3$ compared to $S^3$) and also in the presence of the
3-derivative term.
  From here it
follows immediately that the quartic vertex for ``magnons'' with 1-st order dispersion
relation, i.e. the analog of \rf{ert},\rf{ddd},
 is
\bea\la{ny}
&&{\cal V}^{\scriptscriptstyle {\rm SL(2)}}(p,p'; k,k')
= i\frac{{\tilde {\cal V}}^{\scriptscriptstyle {\rm
SL(2)}}_4}{\sqrt{e(p)e(p')e(k)e(k')}}\  , \\
&&{\tilde {\cal V}}^{\scriptscriptstyle {\rm SL(2)}}_4 =-\bl(pp'+kk')
+[e(p)-1][e(p')-1]+
[e(k)-1][e(k')-1]\no \\
&&~~~~~~~~~~\vphantom{\frac{1}{2|}}
+\frac{1}{2} \left(\bl pp' -
[e(p)-1][e(p')-1]\right)\left(\bl kk' -
[e(k)-1][e(k')-1]\right)\cr
&&~~~~~~~~~~
-\frac{1}{2}\Big[([e(p)-1]+[e(p')-1])\left(\bl kk' -
[e(k)-1][e(k')-1]\right)\cr
&&~~~~~~~~~~~~~~~~~~
+(p,p')\leftrightarrow (k,k')
\Big] \la{my}
%\nonumber
%\cr &&~~~~~~~~ \left.+
%([e(k)-1]+[e(k')-1]) \left(\bl pp' -
%[e(p)-1][e(p')-1]\right)\right]
\eea
The vertex ${\cal V}^{\rm SL(2)}$ simplifies considerably
upon multiplication by the delta-function (\ref{delta})
enforcing the  energy-momentum conservation.  It follows then that  the
tree-level S-matrix of this $SL(2)$ sigma model, i.e.
the analog of \rf{fero},\rf{zzzz} in the $SU(2)$ case,  is
\bea\la{mero}
 &&{{\cal V}}^{\scriptscriptstyle {\rm SL(2)}}
\frac{{\bar\lambda}^{-1}e(p)e(p')}{pe(p')-p'e(p)}
\delta_+(p,p',k,k')
 =
  (S^{\scriptscriptstyle {\rm SL(2)}}_{\rm string})_{\rm tree} \   \delta_+(p,p',k,k') \ ,
\cr
&& (S^{\scriptscriptstyle {\rm SL(2)}}_{\rm string})_{\rm tree}  = i\ pp'
\frac{\bar{\lambda}pp'-e(p)e(p')-1}{pe(p')-p'e(p)} ~~ .
\eea
Remarkably, this is indeed the same
as  the ``tree-level'' part  \rf{BStreeSL2}  of the low
energy AFS-type S-matrix  in the $SL(2)$ sector.\foot{
%VV2
As in the $SU(2)$ sector, this conclusion about the relation of the low-energy
AFS S-matrix and {\it classical} string sigma model S-matrix is of course
consistent with the ``derivation'' of the $SL(2)$ dressing
 phase  \ci{s} from the {\it classical} string model on $AdS_3 \times S^1$
 by discretizing the integral equation \ci{KZ}
describing classical
solutions of $AdS_3 \times S^1$ string sigma model.}

We have therefore shown, in a  low energy  and strong coupling (classical string)
approximation, that the
dressing phase relating the gauge theory Bethe ansatz
and the world sheet S-matrix  is the {\it same}
 in the $SU(2)$ and the $SL(2)$
sectors. This provides a  non-trivial test of the proposed
 \ci{s,bs2,beis}  generalization of the $SU(2)$    AFS ansatz  to
other sectors.

 \bigskip

\subsection{Fermionic $SU(1|1)$ truncation of the superstring action
%further tests of universality
 and  AFS-type S-matrix }
%%%%%%%%%%%%%%%%%%%%%%%%%%%%%%%%%%%%%%%%%%%%%%%%%%%%%%%%%%%%%%%%%%%%%%

As  a further  test of the  universality of the dressing phase
we will now discuss the tree-level S-matrix \ci{kz}
of the $SU(1|1)$ truncation \ci{aaf}
of the \adss superstring action
and its comparison with  the
corresponding phase shift  in the AFS-type Bethe ansatz  in
\cite{s,bs2}.
The matching of  the two S-matrices in the low-energy limit was already
noted     in \ci{kz}, but since our perspective is somewhat different
  and also to clarify
some conceptual issues that seem to
have more general  importance  we shall
discuss this case in  detail below.

%We will encounter a subtlety pointed out above, related to the choice of vacuum.

Using the low-energy limit  of the BDS-type S-matrix  in the $SU(1|1)$
sector  from Appendix A.3 \rf{jlo}   and assuming
the universality of the AFS phase we find as in \rf{BStreeSL2}
\bea
({\tilde S}^{\scriptscriptstyle {\rm SU(1|1)}})_{\rm tree} &=&
\frac{i}{2} \ (p-p') \left[1  -\frac{p-p'}{ p \ e(p') - p' \  e(p)}\right]
+i{\tilde\theta}_{\rm AFS}\cr
&=&
\frac{i}{2}\bigg(p[e(p')-1] -p'[e(p)-1]\bigg)  \ .
\label{suuu}
\eea
This is the expression we would, by analogy with the bosonic sector cases,
  expect to get
as a  tree-level S-matrix in the corresponding sector
 of the superstring theory.

%Let us begin with analyzing the world sheet theory.
Similarly to the truncation of the string  theory to
the $SU(2)$  and $SL(2)$ sectors, the \adss world sheet sigma model
may be consistently truncated  \cite{st1,aaf}
% (consistently \ci{aaf} at the classical level)
  to a fermionic
model \cite{aaf}  containing the $AdS_5$ time
coordinate $t$, two fermionic components $\Psi_{1},\Psi_2$
 and a boson $\alpha$  parameterizing an $S^1$  direction in
$S^5$. One may  then  fix the  uniform gauge condition
%which in  \ci{aaf} was that  of
\ci{AF,aaf}, i.e.
$t=\tau, \ p_\alpha = { J \ov \sql}$.
% or, equivalently, after
%dualizing $\alpha \to \td \alpha$, by fixing the static gauge
%$t=\tau, \ \td \alpha ={J \ov \sql} \sigma$ in the
%corresponding Nambu-type action.
\foot{In the more general
sector of string theory, the analogous
gauge may be fixed by picking the isometric direction $\alpha$ on $S^5$
corresponding to the large R-charge, rescaling all other coordinates
by $e^{iq_j\alpha}$ such that they become neutral under the
$U(1)$  transformation, dualizing $\alpha \to \td \alpha$
 and setting $t= \tau, \ {\tilde\alpha}={J \ov \sql} \sigma$.}

%This is not choice of vac -- vac is same; this is choice of length
%
%

Note that in contrast to the previous  two cases of the bosonic sectors where
$J$  in the uniform  gauge  corresponded to the {\it length}
of the spin chain on the  gauge theory side ($J$ was total spin $J_1+J_2$ in the
$SU(2)$ case in \rf{ccc}  and the $U(1)$ R-charge in the $SL(2)$  case in \rf{Lsl2})
here $J$  is the  bosonic R-charge  while
the length of the chain is  $L= J + \ha M $   where $M$ is
the number of fermionic impurities \ci{bei2,sw} (the
corresponding operators are Tr$(Z^{L-M} \psi^M)$).
This suggests a subtlety in
the identification of  the spin chain and the world-sheet S-matrices
 in this case
(which was indeed already mentioned in \ci{s,kz}).\foot{Let us note also that the
 dependence of the form of the
``string'' Bethe ansatz  on a choice of world-sheet gauge was
emphasized in \ci{af2,afz}.} Indeed, the S-matrix  on the spin chain
side has a meaning only in the context of a specific choice of a
ground state
 %lowest energy state
 and its quantum numbers
%
%nnnew
%Bethe ansatz,
so the identification of the length on the l.h.s. of the Bethe equations \rf{ook}
with the length of the  world-sheet spatial direction is
important.\foot{In general,  the gauge choice is
 also related to the  issue  of identification of the vacua; the choices of vacua
 in the three rank-one sectors are  formally  the same -- the vacuum
 is the  BMN one;  however,  its
  embedding in the full  string theory may appear to be  different.}

Using the consistent truncation of  \cite{aaf},
T-dualizing the coordinate $\alpha\to \td \alpha$
%introduced in the truncation of \cite{aaf}
%following the discussion above
and fixing the  gauge
$t=\tau$ and ${\tilde\alpha}={J\ov \sql}\sigma = {x \ov \sqrt{\bl} }$
%or using the result \cite{aaf} of the phase space gauge fixing,
%the world sheet Lagrangian is (up to quartic order in fields)
%is
one finds the  fermionic analog of the bosonic
actions in \rf{ccc} and \rf{Lsl2}, i.e.
the action of \ci{aaf} with each spatial derivative $\del_x$
having an additional factor of $\sqrt{\bl}$
\bea\la{aru}
L=&-&i{\bar\Psi}(\rho^0\partial_t+\sqrt{{\bar\lambda}}\rho^1\partial_x)\Psi
+{\bar\Psi}{\Psi} -\frac{1}{4} \sqrt{\bar\lambda} \
\epsilon^{ab} \ ({\bar\Psi}\partial_a\Psi\
{\bar\Psi}\rho^3
%_{\scriptscriptstyle -1}
\partial_b\Psi
- \partial_a{\bar\Psi}\Psi\
\partial_b{\bar\Psi}\rho^3
%_{\scriptscriptstyle -1}
\Psi) \cr
&+&
{ 1 \ov 8} \sqrt{\bar\lambda}\  \ep^{ab}\ (\bar \Psi \Psi)^2
 \del_a \bar \Psi  \rho^3 \del_b \Psi \ .
%\dots~~
\eea
Here $\Psi$ is a two-component spinor (formed from  2 components of the original
fermions of the \adss superstring),
${\bar\Psi}=\Psi^\dagger\rho^0$ and
\bea
\rho^0=
\pmatrix{
-1 & 0\cr
0& 1} \ ,
~~~~~~
\rho^1=
\pmatrix{
0 & i\cr
i& 0}\ ,
~~~~~~    \rho^3
%_{\scriptscriptstyle -1}
=\rho^0\rho^1
~~.
\eea
Like  the bosonic actions \rf{ccc} and \rf{Lsl2}, this non-linear fermionic  action
is classically  integrable \ci{aaf}
 but should  not be expected to be meaningful
at the quantum level (in particular, it is not renormalizable, cf. \ci{kz}):
 to  compute  quantum corrections  one is to include couplings to all other
 superstring modes.
 As in the previous bosonic  cases, here  we will be interested only in the  tree-level
 2-particle S-matrix  corresponding to \rf{aru} (where
  the 6-point interaction term
  may thus  be dropped out).

As was pointed out   in \cite{st1}, to compare the truncation of the
superstring action  to the
spin chain side,
% for the comparison of the  coherent state
%continuum limit of the gauge theory dilatation operator with the world
%sheet theory,
 the $SU(1|1)$ spin chain fermionic magnon should be
 identified  with one
 %(``light'')
  of the two
components of the fermion field  $\Psi=\pmatrix{
\Psi_1 \cr  \Psi_2} $.  Our aim is thus
to  compute the tree S-matrix for this component.

Written explicitly  in terms of
$\Psi_1$ and $\Psi_2$,  the Lagrangian
\rf{aru} is, up to the relevant fourth order in the  fields,
\bea\la{ddss}
L= &-&
\Psi_1^*(i\partial_t + 1 )\Psi_1 -   \Psi_2^*(i\partial_t-1)\Psi_2 \cr
&+&
  \sqrt{\bl} \bigg[
   \Psi_2^* \del_x \Psi_1  - \ha
    \Psi_1^*\Psi_2^*\left(i \partial_t\Psi_1
\partial_x\Psi_1 + i\partial_t\Psi_2 \partial_x\Psi_2\right) + h.c. \bigg]
 + O(\Psi^6)  \eea
Since $\Psi_1$ and $\Psi_2$ have opposite signs of mass terms,
applying the  redefinition like the one in \rf{hhgg} or the one  relating
\rf{toy} and \rf{oy}, i.e.
\be \la{kjj}
 \Psi_1 = e^{ i t} \z  \ , \ \ \ \ \ \ \ \  \ \ \ \Psi_2 = e^{ i t} \chi \ ,
  \ee
we make  $\z$    ``massless''
while $\chi$ more massive, and thus  can
 naturally  integrate out the latter.
 This rotation of the fluctuation
 fields is necessary in order to relate the  truncated string action to a
 ``non-relativistic'' LL-type action  for the fermionic ``magnons'' \ci{st1}
 (cf. \rf{fer}).
 Similar rotation was automatically incorporated in the choice of
 gauge fixing and field parametrization in the $SU(2)$   \rf{ccc} and
 $SL(2)$ \rf{Lsl2} sectors
 where   \rf{exptest}   and  \rf{ghk}    contained  linear in time derivative
 ``friction'' term.
  For comparison, such term would be absent if one would start
  with  the BMN-type action for bosonic fields.\foot{For example, in the
  case of $R_t \times S^3$   we gauge-fixed the ``fast'' coordinate
  which was the combined angle in the two rotational planes
  (corresponding  to the total spin  $J_1+J_2$).
  Had we fixed, as in  the BMN fluctuation  case,
   the angle  in only one rotation
   plane, we would need  to apply an extra time-dependent rotation
   to the fluctuation fields.}

  Solving the equation for $\chi$ gives
\be
\chi =  -  \frac{ \sqrt{\bl} \del_x }{2- i\partial_t  } \z  + O(\z^2) \ee
and thus the effective Lagrangian for $\z$  becomes\foot{An equivalent
 (up to change of notation) quadratic part of the action appeared
 in the same context in  \ci{st1}.}
\bea
&&L(\z) = - \z^*\bigg(i\partial_t-
 \frac{\bl\partial_x^2}{2- i\partial_t}\bigg)\z   \cr
&& -
  \ha \bl  \bigg[
    \z^*    \frac{  \del_x }{2 + i\partial_t  } \z^*\bigg(
    (1 - i \partial_t ) \z \partial_x\z
 + \bl  \frac{ (1-i\partial_t) \del_x }{2- i\partial_t  } \z
        \frac{  \del^2_x }{2- i\partial_t  } \z
       \bigg) + h.c. \bigg]
 +... \eea
 The dispersion relation  for $\zeta$ is thus
%
% signs changed
%
 $i\del_t \zeta = (1 \pm \sqrt{ 1 - \bl \del_x^2}) \zeta + O(\z^3)$. Using
 that
 \be \la{pee} i\partial_t-
 \frac{\bl\partial_x^2}{2- i\partial_t} =
 \left(  i \del_t + \sqrt{1 - \bl \del_x^2}- 1  \right) \    P^2 \ , \ \ \ \
 \ \ \   P^2 \equiv
%
% signs changed
%
  { 1 +  \sqrt{1 - \bl \del_x^2} - i \del_t   \ov 2- i\partial_t}
 \ , \ee
 we can then find the effective Lagrangian
 for the positive-energy part of $\zeta$
 with the expected dispersion relation by redefining
 \be
 \zeta = P^{-1} \psi  \ee
 getting (cf.  \rf{fer} and \rf{itak})
\bea\la{loo}
\cL(\psi) &=& -  \psi^*\left(i\partial_t+ \sqrt{ 1 - \bl \del^2_x} -1 \right) \psi
\cr
&& -\
  \ha \bl  \bigg[
    { \bar P}^{\scriptscriptstyle -1}\psi^*
    \frac{ {\bar  P}^{\scriptscriptstyle -1} \del_x }{2 + i\partial_t  }
    \psi^*\bigg(
    (1 - i \partial_t )P^{\scriptscriptstyle -1}
     \psi \partial_xP^{\scriptscriptstyle -1}\psi
    \cr
    &&~~~~~~~
 + \bl  \frac{ (1-i\partial_t) \del_x P^{\scriptscriptstyle -1
 }}{2- i\partial_t  } \psi
        \frac{  \del^2_x P^{\scriptscriptstyle -1}}{2- i\partial_t  } \psi
       \bigg) + {\rm h.c.} \bigg]
 +...
 \eea
where $\bar P= P^\dagger$.

The on-shell 4-vertex corresponding to this Lagrangian (found by
 eliminating the
 time derivatives
in the quartic term using the free equation of motion for $\psi$, so that
$  i \del_t \to  1 - e(i\del_x), \       P^{-1} \to  { \sqrt{ 1 +   e(i\del_x)}
\ov \sqrt{ 2   e(i\del_x)}}   $)
 is (cf. \rf{ert},\rf{ddd} and \rf{ny},\rf{my}):
\bea
\label{ve}
&&{\cal V}^{\scriptscriptstyle\rm SU(1|1)}(p, p';k,k')= { i \ov 2}
\frac{{\tilde {\cal V}}^{\scriptscriptstyle\rm SU(1|1)}(p,p';k,k')+
{\tilde {\cal V}}^{\scriptscriptstyle\rm SU(1|1)}(k,k';p,p')}
{\sqrt{e(p)e(p')e(k)e(k')}} \ ,  \\
&&{{\tilde {\cal V}}^{\scriptscriptstyle\rm SU(1|1)}}(p,p';k,k')=
-\frac{\bl}{4}\ A(k,k')\frac{[p'-p -A(p,p')] \big[
(1+e(k))(1+e(k'))-\bl kk'\big]
}{\sqrt{(1+e(p))(1+e(p'))(1+e(k))(1+e(k'))}} \nonumber \eea where
$A(k,k')\equiv ke(k')-k'e(k)$. Multiplying this vertex (\ref{ve})
by the kinematic factor leads as in \rf{fero},\rf{mero} (with
$\delta_+ \to \delta_-$ as in \ci{kz})
  to a very simple result for the S-matrix
\bea
&& {{\cal V}}^{\scriptscriptstyle {\rm SU(1|1)}}
\frac{{\bar\lambda}^{-1}e(p)e(p')}{pe(p')-p'e(p)}
\delta_-(p,p',k,k')
 =
  (S^{\scriptscriptstyle {\rm SU(1|1)}}_{\rm string})_{\rm tree}(p',p) \   \delta_-(p,p',k,k') \ ,
\cr
&&(S^{\scriptscriptstyle {\rm SU(1|1)}}_{\rm string})_{\rm tree}  =
\frac{i}{2}\left[  p\ e(p')-p'\ e(p)  \right] ~ .
\label{u11}
\eea
% quadratic term in the  familiar 1st order form \foot{
%
% nnnew
%
%As in the
%case of the model field theory in section 7.1, the energy differs by a
%constant shift compared to that of magnons. As in that case, this may
%be accounted for by performing a time-dependent rephasing of
%$\psi$. This  transformation may be done at the level of the action
%\rf{laga} or after truncating away the negative energy
%modes of $\psi$.  We have checked that final results for the
%tree-level scattering matrix is independent of when this rephasing
%is done. We will choose to do it explicitely at the end, when its
%effect on the truncated action is minimal.}
This  is the same tree-level S-matrix as one  finds   from \ci{kz}
 by interpreting their  result  for the S-matrix of the model of \ci{aaf}
 in terms of a
``non-relativistic'' single-component fermionic field theory.

The string theory result  \rf{u11} is  different
from the low momentum limit of the AFS-type  S-matrix  for the $SU(1|1)$
sector \rf{suuu}   by  a $\l$-indepedent term
 $\frac{i}{2} (p'-p)$.
 The difference    is  actually  the
necessary correction to the scattering phase appearing from expressing
the Bethe equations in terms of the R-charge $J$  rather than the length
$L= J + \ha M $ of the chain \cite{s,kz}.
On the string theory side, this shift may be attributed to a
choice of the uniform gauge that fixed $J$ instead of $L$. \foot{An
alternative  interpretation
of this  change in the scattering phase is that it is due to
a  difference between the choice of the  vacua in  the $SU(1|1)$ and
the rank one bosonic sectors.}
We conclude that the direct string-theory computations of the ``magnon''
S-matrix confirm that
the non-trivial $\l$-dependent
dressing phase relating the ``gauge'' and ``string'' Bethe
ans\"atze is universal for  all the rank one sectors.

\bigskip

\section{Outlook}

One  important lesson of the present paper is the following.
To compare
the spin chain Bethe ansatz phase shift for magnons  near the {\it
ferromagnetic}
vacuum  which have ``non-relativistic'' first-order dispersion relation
to string theory
one should  re-organize the string-theory  S-matrix for BMN-type modes
(which originally have  relativistic dispersion relation)
into the S-matrix for an  effective field theory of the
positive-energy modes
as discussed  above in section 7.1.

A  potential application of the   relation between  the
low-energy limit of the AFS-type   S-matrix and the ``non-relativistic''
form  of the
classical string action we have investigated in this paper
%on $R \times S^3$ and on $AdS_3\times S^1$
is a possibility to shed light on the  connection  between the structure
of string $\alpha'\sim { 1 \ov \sql} $   corrections   and subleading
terms in the string  phase in \rf{sss},\rf{doa}.

The are several  open  issues that require analyzing the complete
world sheet theory at the quantum level.
 One is relation to quantum corrections within the Landau-Lifshitz framework,
 e.g., whether (part of) higher-order terms in \rf{jpp}
may  be interpreted as quantum loop corrections in  the LL model
\rf{itak},\rf{geq}.
%is an indication of that.
Still, the ``non-causal'' loops of the LL model  do not
involve  negative-energy modes which are present in the full string loop
contributions (where not only quartic but also higher-order vertices
will be contributing to 2-particle S-matrix),  so to account for the
latter  one needs to go beyond the specific low-energy
approximation to the AFS  scattering matrix we  considered in section 4.

%%%%%%%%%%%
%%%%%
%%%%%  Original ending
%%%%%
%%%%%%%%%%%%

\iffalse
????????????????????????
By the way,  one lesson of sect 7.2.   is basically how we can  go
back again
to general superstring, fix similar gauge, get quartic action
generalizing that scalar one
and start computing S-matrix without nonrelativ approx.
may be we need a note like that at the end.
 wonder what will happen if we start with  7.14 -- string action on
S3 to quartic order -- and just extract tree level S-matrix -- it will
be same
as we got in 7.16 I guess.  Difference will appear in the loop  since
now propagator   is causal; we will get true log divergences; that
will be related to zamolodchikovs s-matrix
 and Kazakov et al story (modulo we are in a different gauge).
Then we can think of supplementing this action by fermions -- there is
indeed
susy version of O(n) model of Witten et al but we need string theory
version of it (something like Faddeev-Reshetikhin but I think we are
after S-matrix of elementary excitations). then we will need to go
back to superstring, at least ads3 x s3 version of it...
\fi

\bigskip
\bigskip

\section*{Acknowledgments }
%%%%%%%%%%%%%%%%%%%%%%%%%%%

We are  grateful to   G. Arutyunov, S. Frolov, V. Kazakov, J. Minahan,
J. Russo and K. Zarembo for useful discussions.
 The work of A.T. and A.A.T.
  was supported  in part by the DOE grant DE-FG02-91ER40690.  A.A.T.
acknowledges also the support of
% PPARC PPA/G/O/2002/00474,
 INTAS  03-51-6346, EC MRTN-CT-2004-005104   and the RS Wolfson award.

\renewcommand{\theequation}{A.\arabic{equation}}
\renewcommand{\thesection}{A}
 \setcounter{equation}{0}
\setcounter{section}{1} \setcounter{subsection}{0}

 \section*{Appendix A:   Effective field theory
 vertices  in  $SL(2)$ and $SU(1|1)$ sectors       }

In this Appendix we shall generalize the  construction of the effective 2d field
 theory vertex from BDS   $SU(2)$
 S-matrix to the other two rank 1 sectors.
The same can be repeated also
 for the AFS-type case (using the general expressions in  \ci{bs2}).
 %as discussed for the $SU(2)$ case  in the main part of the paper.

\subsection{Bethe Ansatz equations }

Let us  start with   recalling the $S$-matrices that  enter the BDS-type
 Bethe
ansatze for $SU(2)$, $SU(1|1)$ and $SL(2)$ sectors \ci{bs2,bds,s}.
The Bethe ansatz equations can be written in the form \cite{bs2}
\begin{eqnarray}\la{ses}
e^{i p_j L} =\prod_{k\ne j}^M  S_\eta(p_j,p_k) \ ,\ \ \ \ \ \ \
 S_\eta(p_j,p_k) = \left(\frac{x_j^+-x_k^-}{x_j^--x_k^+}\right)^\eta
\ \frac{1-{ \bl \ov 4x_j^+x_k^-}  }{1-{\bl \ov 4x_j^-x_k^+}}\ ,
\end{eqnarray}
where $\eta=-1,0,1$ for the $SL(2)$, $SU(1|1)$ and $SU(2)$ sectors.
Here
\be
x^\pm_k = {e^{ \pm { i \ov 2} p_k} \ov 4 \sin {p_k \ov 2}    }
\sqrt{ 1 + 4 \bl \sin^2 {p_k \ov 2}}   \ ,
%x^\pm_k = x(u_k\pm \textstyle{\frac{i}{2}}) \ , ~~~~
%e^{ip}=\frac{x^+}{x^-} \ , ~~~~ x(u) =
%\frac{1}{2} u \left(1+\sqrt{1 - {\bl \ov u^2}}\right) \ ,
 \ ~~~ ~~~~
 ~~~~ \bl \equiv  \frac{\lambda}{(2\pi)^2} \ .
 \ee
 An  important property of the S-matrix in  \rf{ses} is that it can be
written as
\begin{eqnarray}\la{sws}
S_\eta(p_j,p_k)=
%&=&\frac{f_\eta(p_j,p_k,z)}{f_\eta(p_j,p_k,z)}\Big|_{z=1}
%\no \\
%&=&
\frac{A_\eta(p_j,p_k)+B_\eta(p_j,p_k)}{A_\eta(p_j,p_k)-B_\eta(p_j,p_k)}=
\frac{1+B_\eta(p_j,p_k)/A_\eta(p_j,p_k)}{1-B_\eta(p_j,p_k)/A_\eta(p_j,p_k)}
\end{eqnarray}
where $B_\eta$ is purely imaginary and
%In terms of the numerator and denominator of $S_\eta$, the
%functions $A_\eta$ and $B_\eta$ are simply
\begin{eqnarray}
A_\eta(p_j,p_k)&=&\frac{1}{2}\left[
(x_j^+-x_k^-)^\eta(1-{\bl  \ov 4x_j^+x_k^-})+
(x_j^--x_k^+)^\eta(1-{\bl \ov 4x_j^-x_k^+}) \right]\cr
B_\eta(p_j,p_k)&=&\frac{1}{2}\left[
(x_j^+-x_k^-)^\eta(1-    {\bl  \ov 4x_j^+x_k^-}  )-
(x_j^--x_k^+)^\eta(   1-{\bl \ov 4x_j^-x_k^+}   ) \right]
\la{klp}\end{eqnarray}

%%%%%%%%%%%%%%%%%%%%%%%%%%%%%%%%%%%%%%%%%%%%%%%
\subsection{4-point vertex from  S-matrix}
%%%%%%%%%%%%%%%%%%%%%%%%%%%%%%%%%%%%%%%%%%%%%%%%%%%%%%%%

Given a field theory of the type
\rf{quad}, it is a simple exercise to find its tree-level
S-matrix. It is related to the 4-point vertex $\cV(p, p';k, k')$ in \rf{ine}
by multiplication by the kinematic factor \rf{ta}
%
%Assuming again that  the corresponding field theory
%is a non-relativistic one with propagator
%$ D(\omega, q) = \frac{i}{\omega-\left(e(q) -1\right)+i\epsilon} $
%and a 4-point vertex  of the form
%\begin{eqnarray}
%\cV(p, p';k, k')= \frac{1}{2} \bigg[ \cV(p, p')+ \cV(k, k')\bigg] \ ,
%\end{eqnarray}
%we find, following the discussion
% in the $SU(2)$ sector, that
% the tree-level   S-matrix is
 \be \la{trr}
S_{\rm tree} =
%(-1)^{[f]}
 \ \cV(p, p';k,k')\  \frac{\bl{}^{-1} e(p)\ e(p')}{p'\ e(p)-p\
e(p')} \ . \ee
Here we are free to use the on-shell condition for the momenta
$p,p',k$ and $k'$ as well as momentum conservation. Making use of
this freedom (which, as discussed above, implies
 that $p,p'$ equals $k,k'$ or $k',k$) it is always possible to
put the vertex in the form
\begin{eqnarray}
\cV(p, p';k, k')= \frac{1}{2} \bigg[ \cV(p, p')+ \cV(k, k')\bigg] \ .
\label{vertgen}
\end{eqnarray}
This freedom brings in  the issue of reconstructing the off-shell
vertex $\cV$ from the
knowledge of the tree-level S-matrix. This issue is particularly
relevant  for  rank 1 fermion sector. The S-matrix depends on  two
incoming momenta and as such the vertex will have the structure
(\ref{vertgen}). But then it may seem  impossible to write
a nontrivial Lagrangian $L$ for two anticommuting
fields $\psi, \bps$  since  (\ref{vertgen}) implies that the
corresponding terms in $L$ will contain either $\bps^2=0$ or
$\psi^2=0$. This is,  however,  an illusion stemming from a naive
use of free equations of motion as well as  momentum conservation.

To identify a way to undo
the use of momentum conservation
we are to
take  into account the
 statistics of the scattered particles.
 In particular, the
vertex should be either symmetric or antisymmetric
depending whether
the scattered particles are bosons or fermions.\foot{This is
implicitly taken into account by the fact that the relative sign
between the two terms in $\delta_+$ in \rf{delta} is positive for bosons and
negative for fermions (see, e.g.,  \ci{kz}).}
 Below we will  use the
momentum conservation constraint on the vertex which is
extracted from the
scattering matrix in such a way that the required symmetry properties are manifest.

Up to
divergent terms that we discard,
  the 1-loop contribution to the
resulting field-theory S-matrix is   (cf. \rf{quy}):
\be \la{orr}
 S_{\rm 1-loop } = I_{\rm 1}(p,p')\  \frac{\bl{}^{-1} e(p)\ e(p')}{p'\ e(p)-p\
e(p')} \ ,\   \ee
\be \la{prr}
 I_{ 1}(p,p') =(-1)^{[f]} [ \cV(p, p')]^2 \frac{\bl{}^{-1} e(p)\ e(p')}{p'\ e(p)-p\
e(p')}  \ , \ \ \ \ \ \ \ \ \ \ \
S_{\rm tree} =  {I_{ 1}(p, p') \ov \cV(p, p') }  \ .
\ee
Then the  all-loop S-matrix is:
\begin{eqnarray}
S_{\rm all-loop} = \frac{1+ \ha S_{\rm tree} }{1-\ha S_{\rm tree} }
\label{it}
\end{eqnarray}
By comparing this with the S-matrix extracted from the Bethe
equations \rf{sws}       it follows  quite generally  that
\be\la{kik}
S_{\rm tree}=  2 \frac{B_\eta(p,p')}{A_\eta(p,p')}  \ , \ee \bea
\cV(p, p')&=& 2  \bl \frac{p'\ e(p)-p\ e(p')}{e(p) \ e(p') }
 \ \frac{B_\eta(p,p')}{A_\eta(p,p')}
 \\
&=&2  \bl \frac{p'e(p)-p e(p')}{e(p)e(p')}\
\frac{(x_1^+-x_2^-)^\eta(1- { \bl \ov 4x_1^+x_2^-} )-
(x_1^--x_2^+)^\eta(1-{ \bl\ov  4x_1^-x_2^+} )}
{(x_1^+-x_2^-)^\eta(1-{\bl\ov  4x_1^+x_2^-} )+
(x_1^--x_2^+)^\eta(1-{\bl \ov 4x_1^-x_2^+} )} \nonumber
\end{eqnarray}
where $x^\pm_1 = x^\pm (p), \ x^\pm_2 = x^\pm (p')$.

\subsection{4-vertex in the $SU(1|1)$ sector  \label{su11BDSlim}}

As explained in the $SU(2)$ case, to  compare with the field theory
S-matrix we need to take
a specific low-momentum  limit in the Bethe ansatz S-matrix, in which
$p \to 0$ and one keeps  only the  leading term in $p$
at each order in $\l$.
This amounts to the replacement:
 \be \la{zlo}
x^\pm (p)  \ \ \to \ \ \  {e(p) \ov 2 p  } \ ,\ \ \ \ \ \ \ \ \ \ \ \ \
 e(p)=\sqrt{1+\bl  p^2} \ .
\ee
Then
\begin{eqnarray}
&& (S_{su(1|1)})_{\rm tree} =
2 \frac{(1-{ \bl \ov 4x_1^+x_2^-})- (1-   { \bl\ov  4x_1^-x_2^+}         )}
{(1-   { \bl \ov 4x_1^+x_2^-}      )+ (1-    { \bl\ov  4x_1^-x_2^+}        )}
\cr
&&
\  \to\   (\td S_{su(1|1)})_{\rm tree} =
\frac{i}{2} \ (p-p') \left[1  -\frac{p-p'}{ p \ e(p') - p' \  e(p)}\right]
\la{jlo} \\
&&~=\frac{i{\bl}}{4}p p'(p-p')
-\frac{i{\bl}^2}{16}pp'(p^3-2p^2p'+2p p'^2-p'^3)\cr
 &&~
+\frac{i{\bl}^3}{32}p p'(p^5-2p^4p'+2p^3p'^2-2p^2p'^3
+2p p'^4-p'^5)\cr &&~ -\frac{5i{\bl}^4}{256}p
p'(p^7-2p^6p'+2p^5p'^2-2p^4p'^3 +2p^3p'^4 -2p^2p'^5+2pp'^6
-p'^7) + \ldots  \no
\end{eqnarray}
Thus, after dividing by the kinematic factor,
\begin{eqnarray}\la{llo}
\cV_{su(1|1)}(p,p') =\frac{i}{2} \bl (p-p')
\left[\frac{p\ e(p')-p'\ e(p)}{e(p) \ e(p')}-\frac{p-p'}{e(p) \ e(p')}\right]
\end{eqnarray}
i.e.
\begin{eqnarray}\la{slo}
 \cV_{su(1|1)}(p,p') &=&-\frac{i{\bl^2}}{4}pp'(p-p')^2 +
\frac{i{\bl}^3}{16} pp'(p-p')^2(3p^2+pp'+3p'^3)\cr
&-&\frac{i{\bl}^4}{32}pp'(p-p')^2(5p^4+2p^3p'+5p^2p'^2
+2pp'^3+5p'^5)\cr &+&\!\!
\frac{i{\bl}^5}{256}pp'(p-p')^2(35p^6+15p^5p'+35p^4p'^2\cr
& & \ \ \ \   +  17p^3p'^3 +35p^2p'^4 +15pp'^5+35p'^6 )  +       \ldots
\end{eqnarray}
%The exact in $\bl$ form of the  vertex
%to leading order in $a$
%is:
%\begin{eqnarray}\la{llo}
%\cV_{su(1|1)}(p,p') =\frac{i}{2} \bl (p-p')
%\left[\frac{p\ e(p')-p'\ e(p)}{e(p) \ e(p')}-\frac{p-p'}{e(p) \ e(p')}\right]
%\end{eqnarray}
Note that this vertex starts with the 2-loop order $\l^2$ term,
in agreement with the fact that the leading-order LL action
in the $su(1|1)$ sector is free \ci{s,bei2,hl1,st1,cal2}.

Clearly, the vertex \rf{llo} does not have the symmetry properties
necessary to arise from a fermionic action, as it is symmetric under
$p\leftrightarrow p'$.
%As emphasized above, this vertex
%results from the one derived from a Lagrangian through
% the use of the
%momentum conservation.
 The required antisymmetry is restored by using the
momentum conservation to express the overall factor $(p-p')$ in
terms of the outgoing momenta:
\begin{eqnarray}\la{mllo}
\td \cV_{su(1|1)}(k,k';p,p') &=&\frac{i}{2} \bl (k-k') \left[\frac{p\
e(p')-p'\ e(p)}{e(p) \ e(p')}-\frac{p-p'}{e(p) \ e(p')} \right]
\nonumber\\
&& \ \ = -\frac{i{\bl^2}}{4} pp'(p-p') (k-k')  +
O(\bar{\lambda}^3) \ \ .
\end{eqnarray}
This  vertex allows us to determine the leading
part  in the quartic interaction term
of the resulting fermionic coherent state
action  corresponding to the BDS-type Bethe ansatz
in the $SU(1|1)$ sector (cf. \rf{quad},\rf{fer})
\be  \la{er}
\cS=\int dt \int_{0}^{J}dx\ \bigg\{
  -\bps  \bigg[i \del_t + (\sqrt{1- \bl  \partial_x^2}-1)\bigg]\psi
  \   - \  V(\psi,\bps ) \bigg\}\  , \label{uad}  \ee
\be \la{amm}
V= V_{4} + V_6 + ... , \ \ \ \ \ \ \
V_4 =\frac{\bar{\lambda}^2}{4} \big(
 \bps\bps{}'\psi' \psi'' +{\rm h.c.}\big)+O(\bar{\lambda}^3)
 \ ,  \ee
where $\psi'= \del_x \psi$  and $\psi$
is a complex   anticommuting field.\foot{As discussed in \ci{st1}, to
find a similar  quadratic term  on the string theory side
where one starts  with a
relativistic massive fermion action one is to solve for  one
of the two fermionic components in terms of the other.}
The exact form of $V_4$ that follows
from \rf{mllo} is (cf. \rf{vvv})
\begin{equation}
V_{4} =\frac{\bar{\lambda}}{8}~\bps\partial_x\bps
\bigg[\psi\ \frac{\partial_{x}}{\sqrt{1-
\bar{\lambda}\partial_{x}^2}} \psi
-  \bigg(\frac{1}{\sqrt{1-
\bar{\lambda}\partial_{x}^2}} \psi\bigg)
\bigg(\frac{\partial_{x}}{\sqrt{1-
\bar{\lambda}\partial_{x}^2}} \psi\bigg)\bigg]+h.c. \la{ds}
\label{su11lag}
\end{equation}
%%R
%
This  interaction term
was constructed  using the BDS-type  S-matrix.\foot{It is  of course
  different from the  relativistic  model   of \ci{aaf}
  (obtained  by a particular truncation of the classical
  string action of \ci{mt})
for which  the quantum  S-matrix was computed in \ci{kz}.}
Including a proper ``string'' phase in the S-matrix one
should be able  to reconstruct the effective action that should  be more
closely   related to string theory.\foot{Let us note again that  there is a freedom of
off-shell extension that
% noted that our construction does not
%specify whether
allows,e.g., to write  the denominators here
%are indeed as written in \rf{su11lag}
%or
in a more symmetric form, as suggested by the string theory considerations
of section 7 (cf.
\rf{stvert}).}

%%%%%%%%%%%%%%%%%%%%%%%%%%%%%%%%%%%%%%%%%%%%%%%%%%%%%%%%%%%%%%%%%%%%%%%%%%%%
\subsection{4-vertex in the $SL(2)$ sector
%\label{vertsl2}
}
%%%%%%%%%%%%%%%%%%%%%%%%%%%%%%%%%%%%%%%%%%%%%%%%%%%%%%%%%%%%%%%

%%new
Similarly, in the  $SL(2)$
sector one finds for  the low momentum limit of the BDS-type S-matrix
(using \rf{kik},\rf{zlo}, etc.)
%Use $p\rightarrow a p$ and $h a^2$ is kept fixed (keep leading
%terms in momenta at each order in $h$).
%Similarly, in the  $SL(2)$ sector we find
\begin{eqnarray}
&&  (S_{sl(2)})_{\rm tree} = 2\frac{(x_1^+-x_2^-)^{-1}(1-{ \bl  \ov 4x_1^+x_2^-}
)-
(x_1^--x_2^+)^{-1}(1-  { \bl  \ov 4x_1^-x_2^+}  )}
{(x_1^+-x_2^-)^{-1}(1-  { \bl  \ov 4x_1^+x_2^-} )+
(x_1^--x_2^+)^{-1}(1-   { \bl  \ov 4x_2^+x_1^-}   )}\cr
&&\ \to \ \  (\td  S_{sl(2)})_{\rm tree}
= i  \left[(p-p')\
-\frac{p^2+p'^2}{pe(p')-p'e(p)}\right]
\la{yyy}
\\
&&=\frac{2ipp' }{p-p'}
 +\frac{i{\bl}}{2}\frac{pp'
}{p-p'}(p^2+p'^2)\cr &&~~~~
-\frac{i{\bl}^2}{8}\frac{pp' }{p-p'}
(p^4-p^3p'+2p^2p'^2-2pp'^3+p'^4)\cr
&&~~~~
+\frac{i{\bl}^3}{16}\frac{pp' }{p-p'}
(p^6-p^5p'+2p^4p'^2-2p^3p'^3 +2p^2p'^4-pp'^5+p'^6) + \dots \no
%\cr &&~~~~
%-\frac{5i{\bl}^4}{128}\frac{pp' }{p-p'}
%(p^8-p^7p'+2p^6p'^2 -2p^5p'^3+2p^4p'^4 -2p^3p'^5 +2p^2p'^6-pp'^7
%+p'^8) +    \ldots \no
\end{eqnarray}
Multiplying by kinematic factor,
the exact expression for the
vertex summing up the leading small  momentum terms at each order in   $\l$
 expansion is (cf. \rf{ex},\rf{llo})
\begin{eqnarray}\la{kuy}
\cV_{sl(2)}(p,p') = i \bl
\left[(p-p')\ \frac{p\ e(p')-p'\ e(p)}{e(p) \ e(p')}
-\frac{p^2+p'^2}{e(p)\ e(p')}\right]
\end{eqnarray}
i.e.
\begin{eqnarray}\la{uy}
 \cV_{sl(2)}(p,p') &=&-2i\bl pp'  +  \frac{i{\bl^2}}{2}pp'(p^2+p'^2)
-\frac{i{\bl}^3}{8}pp' (3p^4+5p^3p'+5pp'^3+3p'^4)\cr
 &+&\frac{i{\bl}^4}{16}pp' (5p^6+8p^5p'+6p^3p'^3+8pp'^5+5p'^6) + \dots
 %\cr
%&-&\frac{i{\bl}^5}{128}pp' (35p^8+55p^7p'+38p^5p'^3
%+38p^3p'^5 +55pp'^7 +35p'^8)\cr &&  + \ldots
\end{eqnarray}
%The exact expression for the
%vertex summing up the leading small  momentum terms at each order in   $\l$
% expansion is (cf. \rf{ex},\rf{llo})
%\begin{eqnarray}\la{kuy}
%\cV_{sl(2)}(p,p') = i \bl
%\left[(p-p')\ \frac{p\ e(p')-p'\ e(p)}{e(p) \ e(p')}
%-\frac{p^2+p'^2}{e(p)\ e(p')}\right]
%\end{eqnarray}
%\subsection{$SU(2)$}
%I hope it agrees :)
%\begin{eqnarray}
%\frac{1}{a}V(D_1,D_2) &=&iD_1D_2\cr
%&-&\frac{i{\bl}}{2}D_1D_2(D_1^2+D_2^2)\cr
%&+&\frac{i{\bl}^2}{8}D_1D_2 (3D_1^4+2D_1^2D_2+3D_2^4)\cr
% &-&\frac{i{\bl}^3}{16}D_1D_2 (D_1^2+D_2^2)(5D_1^4-2D_1^2D_2^2+5D_2^4)\cr
%&&\ldots
%\end{eqnarray}
%The resummed vertex to leading order in $a$ is:
%\begin{eqnarray}
%\frac{1}{a}V_{SU(2)}(k_1,k_2)={i}\frac{k_1k_2}{e(k_1)e(k_2)}
%\end{eqnarray}
%which agrees with your calculations.

%Multiplying this by the kinematic factor (\ref{ta}) leads to the first
%term in equation (\ref{BStreeSL2}).

The leading-order LL action in the $SL(2)$ sector
(depending on  a pseudo-unit vector  parametrizing $AdS_3$)  was
constructed in
\ci{st,bell} (see also \ci{KZ}); the  higher-order corrections to spin-chain LL action were
not computed before (corrections to string LL action
can be found from the expressions in \ci{ptt}).
 The S-matrix approach  avoids the problem of
complicated explicit form of the dilatation operator in  this
sector and gives one  an efficient method of reconstructing the
 higher order terms in the
 effective  action.
 We thus get the following quartic term in the
$SL(2)$ effective action (cf.  \rf{vvv} or  \rf{ds})
%%
%We get for the quartic term in the $SL(2)$ effective
%action the following analog of \rf{vvv} (cf. \rf{ds}):
\begin{equation}
V_{4}=\frac{\bar{\lambda}}{4}\phi^{*}\phi^{*}\bigg[\bigg(\frac{1}{\sqrt{1-
\bar{\lambda}\partial_{x}^2}}\phi\bigg)
\frac{\partial_{x}^2}{\sqrt{1-
\bar{\lambda}\partial_{x}^2}} \phi+ \del_x \phi
\frac{\partial_{x}} {\sqrt{1-
\bar{\lambda}\partial_{x}^2}} \phi
-\phi \frac{\partial_{x}^2}{\sqrt{1-
\bar{\lambda}\partial_{x}^2}} \phi\bigg]+c.c.
\end{equation}

%\iffalse

\bigskip

As was observed  in \ci{s,bs2}, the BDS S-matrices in
the three sectors are formally related by  $S_{su(2)} S_{sl(2)} = [S_{su(1|1)}]^2$,
implying that
\bea\la{relp}
&& (S_{su(2)})_{\rm tree} + (S_{sl(2)})_{\rm tree} = 2 (S_{  su(1|1)        })_{\rm tree}
\ , \\
&& \cV_{su(2)} + \cV_{sl(2)}     = 2 \cV_{su(1|1)}
\la{jka} \ ,   \eea
which provides a simple check on the expressions in \rf{ex}, \rf{kuy}  and \rf{llo}.
The  relation \rf{relp} is true also  after the inclusion of the
AFS phase  contribution \rf{tii} which is the same for all sectors.
%Thus one should  get also the  same relation between the tree-level S-matrices
%on the string theory side (see  section 7).

\bigskip

%%%%%%%%%%%%%%%%%%%%%%%%%%%%%%%%%%%%%%%%%%%%%%%%%%%%%%%%%%%%%
\subsection{Deformation of the BDS  S-matrix and  the vertex}

Let us now  comment on the impact  of the extra
phase  that relates  the BDS S-matrix $S_g$
and the S-matrix $S_s$ (called, respectively,  $S_1$  and $S$ in \rf{sss})
 entering
the string Bethe equations
on  the structure of the resulting field theory vertex.
If  we start with
\begin{eqnarray}\la{hhi}
S_s(p, p') =S_g(p, p')\ [\sigma(p, p')]^2~ , \ \ \ \ \ \ \
\sigma^2= e^{ i \theta} = \frac{a+b }{a -b} \ ,
\end{eqnarray}
where $B$  in \rf{sws} and $b$ in \rf{hhi} are assumed to be purely imaginary,
then it is not hard to trace this deformation through the steps made
above and find the deformed (string)  theory analogs  of the quantities $A$ and
$B$  in \rf{klp}
\begin{eqnarray}\la{llk}
S_s&=&S_g\ \sigma^2 = \frac{A_g-B_g}{A_g +B_g }\ \frac{a+b}{a-b}\cr
&=&\frac{(A_g a - B_g b)-(A_g b-B_g a)}{(A_g a - B_g b)+(A_g b-B_g
a)}\equiv \frac{A_s-B_s}{A_s +B_s }
\end{eqnarray}
We can then reconstruct the
4-point vertex (tree S-matrix)  for
the ``string'' LL model:
\begin{eqnarray}
V_s \equiv  \frac{B_s}{A_s}=\frac{A_g b-B_g a}{ A_g a - B_g
b}=\frac{v_\theta-V_g}{1-V_g v_\theta}
\ , \ \ \ \ \ \
V_g=\frac{B_g}{A_g}
\ , \ \ \ \ \ \
v_\theta = \frac{b}{a} = \tan \frac{\theta }{2}\ ,
\end{eqnarray}
where $\theta$ is the dressing phase relating the gauge and string theory Bethe
Ans\"atze.
To relate this to the discussion in section 4
  one should apply the
small momentum expansion to simplify the entries in \rf{llk}.

 %\fi

\renewcommand{\theequation}{B.\arabic{equation}}
 \setcounter{equation}{0}
\setcounter{section}{1} \setcounter{subsection}{0}

 \section*{Appendix B:  Momentum expansion of the  leading strong
coupling correction  to the AFS  phase     }

Here  we shall present some details of the small
  momentum expansion of the  leading quantum correction
   to the AFS phase  which we used in section 4
and show that
it modifies  the small momentum expansion in (\ref{tida}) by terms
nonanalytic in ${\bar\lambda}$ leading to further deviations from (\ref{faa}).

The dressing phase in \rf{sss},\rf{doa}  is defined by its large
${\bar\lambda}$ expansion
\begin{eqnarray}
\theta=\theta_{\rm AFS}+
%\frac{1}{\sqrt{\bar\lambda}}
\theta_{\rm HL}
+{\cal O}\bigg(\textstyle{\frac{1}{(\sqrt{\bar\lambda})^2}}\bigg)\ ,
\end{eqnarray}
where $\theta_{\rm HL}$ is the first correction in \rf{doa},\rf{cvc}
given by \cite{hl}
\begin{eqnarray}
\theta_{\rm HL}=
%\sqrt{\bar\lambda}
\sum_{r=2}^\infty\sum_{s=r+1}^\infty a_{rs}
\left(\frac{\bar\lambda}{4}\right)^{\frac{r-1}{2}+\frac{s-1}{2}}
\left[q_s(p)q_r(p')-q_s(p')q_r(p)\right]
\end{eqnarray}
with $a_{rs}={ 4 \ov \pi} \frac{(r-1)(s-1)}{(r-1)^2-(s-1)^2}$ for odd $r+s$
and 0 otherwise.
Using the small momentum limit (\ref{lii}) of the charge densities $q_r$, it is easy
to see that $\theta_{\rm HL}$ may be written as
\begin{eqnarray}
\label{thetaHL}
\theta_{\rm HL}=
%\sqrt{\bar\lambda}
pp'g(p)g(p')
\sum_{r=2}^\infty\sum_{s=r+1}^\infty
a_{rs}\left[\frac{1}{g(p)^rg(p')^s}- \frac{1}{g(p')^rg(p)^s}\right]
%~~~
%{\rm where}
%~~~
%g(p)=\frac{\sqrt{\bar\lambda}p}{e(p)-1}
%~~.
\end{eqnarray}
where $$g(p)=\frac{\sqrt{\bar\lambda}\ p}{e(p)-1}$$
 is fixed in the
small momentum limit.
After changing the summation index $s$ to $s=2n +r+1$ to take into
account the vanishing of $a_{rs}$ for even $r+s$,
%
%
\iffalse
%
\begin{eqnarray}
\theta_{\rm HL}=\sqrt{\bar\lambda}
pp'g(p)g(p')
\sum_{r=2}^\infty\sum_{n=0}^\infty
a_{r,2n+r+1}\left[\frac{1}{g(p)^rg(p')^{2n+r+1}}-
\frac{1}{g(p')^rg(p)^{2n+r+1}}\right]
\label{thetaHL}
\end{eqnarray}
%
\fi
%
the double sum in (\ref{thetaHL}) equals the second mixed derivative
of a double sum  computed in  \cite{af}:
\begin{eqnarray}
\chi_1(x,y)&=&\sum_{r=2}^\infty\sum_{n=0}^\infty
\frac{a_{r,2n+r+1}}{(r-1)(2n+r)}\frac{1}{x^{r-1}y^{2n+r}}\cr
&=& {2 \ov \pi}
\Big[\,
\log\frac{y-1}{y+1} \log\frac{x-\frac{1}{y}}{x-y}\\\nonumber
&&+{\rm
Li}_2\frac{\sqrt{y}-{\frac{1}{\sqrt{y\vphantom{A}}}}}{\sqrt{y}-\sqrt{x}}-{\rm
Li}_2\frac{\sqrt{y} +   {\frac{1}{\sqrt{y\vphantom{A}}}} }{\sqrt{y}-\sqrt{x}}+{\rm
Li}_2\frac{\sqrt{y}-{\frac{1}{\sqrt{y\vphantom{A}}}}
%\sqrt{\frac{1}{y}}
}{\sqrt{y}+\sqrt{x}}-{\rm
Li}_2\frac{\sqrt{y}+
{\frac{1}{\sqrt{y\vphantom{A}}}}
%\sqrt{\frac{1}{y}}
}{\sqrt{y}+\sqrt{x}}\, \Big]~~.
\end{eqnarray}
Some algebra then leads to
\begin{eqnarray}
\theta_{\rm HL}&=&
\frac{2
%\sqrt{\bar\lambda}
~pp'g(p)g(p')}{\pi (g(p)-g(p'))(1-g(p)g(p'))}  \\
&\times &\bigg[  1 - \ha \frac{(1-g(p)g(p'))^2+(g(p)-g(p'))^2}
{(g(p)-g(p'))(1-g(p)g(p'))}
\ln\frac{(1+g(p))(1-g(p'))}{(1-g(p))(1+g(p'))}\bigg]
\nonumber
\end{eqnarray}
As expected, in the small momentum limit $p\rightarrow 0$ and
$\lambda p^2$=fixed, the correction $\theta_{\rm HL}$
to $\theta_{\rm AFS}$ scales
as $p^2$; however, its  dependence on  $\bar\lambda$ is nonanalytic:
\begin{eqnarray}
%\frac{1}{\sqrt{\bar\lambda}}
\theta_{\rm HL}\propto p^2 f(\sqrt{\bar\lambda}\ p)
\end{eqnarray}
Including this additional phase in the (\ref{tida}) will modify its
low momentum expansion by terms nonanalytic $\bar\lambda$ starting with
\begin{eqnarray}\la{hll}
\delta {\tilde S}_{\rm AFS}=
-\frac{i}{3 \pi } {\bar\lambda}^{3/2} p^2 p'^2 (p-p') + ...
\end{eqnarray}
This implies further deviations from the naive expectation (\ref{faa})  (expected  by
analogy with the BDS case) for the
low-energy limit of the scattering matrix of the ``string'' ansatz.
It is  possible in principle  that this  nonanalytic dependence on ${\bar\lambda}$
may change once all higher order corrections  to $\theta$ are
resummed. This may be expected
 on the grounds that the dressing phase
should have an analytic expansion at small ${\bar\lambda}$
if it is eventually to agree
with perturbative gauge theory.

%
%\begin{eqnarray}
%\ ,
%\end{eqnarray}
%

%%%%%%%%%%%%%%%%%%%%%%%%%%%%%%%%%%%%%%%%%%%%%%%%%%%%%%%%%%%%%%%%%%%%%%%%%%%


\begin{thebibliography}{20}

%%%%%%%%%%%%%%%%%%%%%%%%%%%%%%%%%%%%%%%%%%%%%%%%%%%%%%%%%%%%%%%%%%%%



\bibitem{mz}
 J.~A.~Minahan and K.~Zarembo,
 ``The Bethe-ansatz for N = 4 super Yang-Mills,''
 JHEP {\bf 0303}, 013 (2003)
 [hep-th/0212208].
 %%CITATION = HEP-TH 0212208;%%


\bi{bks}
 N.~Beisert, C.~Kristjansen and M.~Staudacher, ``The dilatation
operator of N = 4 super Yang-Mills theory,'' Nucl.\ Phys.\ B {\bf
 664}, 131 (2003) [hep-th/0303060].
%%CITATION = HEP-TH 0303060;%%

\bibitem{bds}
  N.~Beisert, V.~Dippel and M.~Staudacher,
  ``A novel long range spin chain and planar N = 4 super Yang-Mills,''
  JHEP {\bf 0407}, 075 (2004)
  [hep-th/0405001].
  %%CITATION = HEP-TH 0405001;%%


\bibitem{s}
  M.~Staudacher,
  ``The factorized S-matrix of CFT/AdS,''
  JHEP {\bf 0505}, 054 (2005)
  [hep-th/0412188].
  %%CITATION = HEP-TH 0412188;%%


\bibitem{bs2}
  N.~Beisert and M.~Staudacher,
  ``Long-range PSU(2,2$|$4) Bethe ansaetze for gauge theory and strings,''
  Nucl.\ Phys.\ B {\bf 727}, 1 (2005)
  [hep-th/0504190].
  %%CITATION = HEP-TH 0504190;%%




\bibitem{afs}
  G.~Arutyunov, S.~Frolov and M.~Staudacher,
  ``Bethe ansatz for quantum strings,''
  JHEP {\bf 0410}, 016 (2004)
  [hep-th/0406256].
  %%CITATION = HEP-TH 0406256;%%




\bibitem{manpol}
  N.~Mann and J.~Polchinski,
  ``Bethe ansatz for a quantum supercoset sigma model,''
  Phys.\ Rev.\ D {\bf 72}, 086002 (2005)
  [hep-th/0508232].
  %%CITATION = HEP-TH 0508232;%%

\bibitem{kz}
  T.~Klose and K.~Zarembo,
  ``Bethe ansatz in stringy sigma models,''
  hep-th/0603039.
  %%CITATION = HEP-TH 0603039;%%


\bibitem{kaz}
  N.~Gromov, V.~Kazakov, K.~Sakai and P.~Vieira,
  ``Strings as multi-particle states of quantum sigma-models,''
  hep-th/0603043.
  %%CITATION = HEP-TH 0603043;%%


\bi{beis}
N.~Beisert,
  ``The su(2$|$2) dynamic S-matrix,''
  hep-th/0511082.
  %%CITATION = HEP-TH 0511082;%%
  ``An SU(1$|$1)-invariant S-matrix with dynamic representations,''
  hep-th/0511013.
  %%CITATION = HEP-TH 0511013;%%


\bibitem{beiklos}
  N.~Beisert and T.~Klose,
  ``Long-range gl(n) integrable spin chains and plane-wave matrix theory,''
  hep-th/0510124.
  %%CITATION = HEP-TH 0510124;%%


\bibitem{afz}
  S.~Frolov, J.~Plefka and M.~Zamaklar,
  ``The AdS(5) x S5  superstring in light-cone gauge and its Bethe
  equations,''
  hep-th/0603008.
  %%CITATION = HEP-TH 0603008;%%



\bibitem{cal}
  C.~G.~Callan, H.~K.~Lee, T.~McLoughlin, J.~H.~Schwarz, I.~J.~Swanson and X.~Wu,
  ``Quantizing string theory in AdS(5) x S5 : Beyond the pp-wave,''
  Nucl.\ Phys.\ B {\bf 673}, 3 (2003)
  [hep-th/0307032].
  %%CITATION = HEP-TH 0307032;%%
C.~G.~Callan, T.~McLoughlin and I.~J.~Swanson,
  ``Holography beyond the Penrose limit,''
  Nucl.\ Phys.\ B {\bf 694}, 115 (2004)
  [hep-th/0404007].
  %%CITATION = HEP-TH 0404007;%%


\bibitem{ss}
  D.~Serban and M.~Staudacher,
  ``Planar N = 4 gauge theory and the Inozemtsev long range spin chain,''
  JHEP {\bf 0406}, 001 (2004)
  [hep-th/0401057].
  %%CITATION = HEP-TH 0401057;%%

\bibitem{bt}
  N.~Beisert and A.~A.~Tseytlin,
  ``On quantum corrections to spinning strings and Bethe equations,''
  Phys.\ Lett.\ B {\bf 629}, 102 (2005)
   [hep-th/0509084]
  %%CITATION = HEP-TH 0509084;%%

\bi{szz} S.~Schafer-Nameki, M.~Zamaklar and K.~Zarembo,
  ``Quantum corrections to spinning strings in AdS(5) x S5 and Bethe ansatz:
  A comparative study,''
  JHEP {\bf 0509}, 051 (2005)
  [hep-th/0507189].
  %%CITATION = HEP-TH 0507189;%%



\bi{sz} S.~Schafer-Nameki and M.~Zamaklar,
  ``Stringy sums and corrections to the quantum string Bethe ansatz,''
  JHEP {\bf 0510}, 044 (2005)
  [hep-th/0509096].
  %%CITATION = HEP-TH 0509096;%%


\bibitem{janik}
  R.~A.~Janik,
  ``The AdS(5) x S5  superstring worldsheet S-matrix and crossing symmetry,''
  hep-th/0603038.
  %%CITATION = HEP-TH 0603038;%%



\bibitem{hl}
  R.~Hernandez and E.~Lopez,
  ``Quantum corrections to the string Bethe ansatz,''
  hep-th/0603204.
  %%CITATION = HEP-TH 0603204;%%




\bibitem{af}
  G.~Arutyunov and S.~Frolov,
  ``On AdS(5) x S5  string S-matrix,''
  hep-th/0604043.
  %%CITATION = HEP-TH 0604043;%%


\bibitem{fk}
  L.~Freyhult and C.~Kristjansen,
  ``A universality test of the quantum string Bethe ansatz,''
  hep-th/0604069.
  %%CITATION = HEP-TH 0604069;%%


\bibitem{mt}
R.~R.~Metsaev and A.~A.~Tseytlin,
  ``Type IIB superstring action in AdS(5) x S(5) background,''
  Nucl.\ Phys.\ B {\bf 533}, 109 (1998)
  [hep-th/9805028].
  %%CITATION = HEP-TH 9805028;%%


\bibitem{mald}
  D.~M.~Hofman and J.~Maldacena,
  ``Giant magnons,''
  hep-th/0604135.
  %%CITATION = HEP-TH 0604135;%%

  \bi{dor}
N.~Dorey,
  ``Magnon Bound States and the AdS/CFT Correspondence,''
 hep-th/0604175.
  %%CITATION = HEP-TH 0604175;%%


\bi{bmn} D.~Berenstein, J.~M.~Maldacena and H.~Nastase, ``Strings
in flat space and pp waves {}from N =4 super Yang Mills,'' JHEP
{\bf 0204}, 013 (2002) [hep-th/0202021].
%%CITATION = HEP-TH 0202021;%%

\bibitem{gmr}
  D.~J.~Gross, A.~Mikhailov and R.~Roiban,
  ``Operators with large R charge in N = 4 Yang-Mills theory,''
  Annals Phys.\  {\bf 301}, 31 (2002)
  [hep-th/0205066].
  %%CITATION = HEP-TH 0205066;%%


\bibitem{saz}
  A.~Santambrogio and D.~Zanon,
  ``Exact anomalous dimensions of N = 4 Yang-Mills operators with
large R charge,''
  Phys.\ Lett.\ B {\bf 545}, 425 (2002)
  [hep-th/0206079].
  %%CITATION = HEP-TH 0206079;%%


\bibitem{bcv}
  D.~Berenstein, D.~H.~Correa and S.~E.~Vazquez,
  ``All loop BMN state energies from matrices,''
  JHEP {\bf 0602}, 048 (2006)
  [hep-th/0509015].
  %%CITATION = HEP-TH 0509015;%%



\bi{fradkin} E.~H.~Fradkin,
  ``Field theories of condensed matter systems,''
 Redwood City, USA: Addison-Wesley (1991) 350 p. (Frontiers in physics, 82).



\bibitem{kru}
  M.~Kruczenski,
  ``Spin chains and string theory,''
  Phys.\ Rev.\ Lett.\  {\bf 93}, 161602 (2004)
  [hep-th/0311203].
  %%CITATION = HEP-TH 0311203;%%


\bibitem{krt}
  M.~Kruczenski, A.~V.~Ryzhov and A.~A.~Tseytlin,
  ``Large spin limit of AdS(5) x  S5 string theory and low energy  expansion
  of ferromagnetic spin chains,''
  Nucl.\ Phys.\ B {\bf 692}, 3 (2004)
  [hep-th/0403120].
  %%CITATION = HEP-TH 0403120;%%


\bibitem{kmmz}
  V.~A.~Kazakov, A.~Marshakov, J.~A.~Minahan and K.~Zarembo,
  ``Classical / quantum integrability in AdS/CFT,''
  JHEP {\bf 0405}, 024 (2004)
  [hep-th/0402207].
  %%CITATION = HEP-TH 0402207;%%




\bibitem{rt}
  A.~V.~Ryzhov and A.~A.~Tseytlin,
  ``Towards the exact dilatation operator of N = 4 super Yang-Mills theory,''
  Nucl.\ Phys.\ B {\bf 698}, 132 (2004)
  [hep-th/0404215].
  %%CITATION = HEP-TH 0404215;%%


\bibitem{tsecarg}
  A.~A.~Tseytlin,
  ``Semiclassical strings and AdS/CFT,''
   in: Cargese 2004, String theory: From gauge interactions to cosmology, p.
    265-290.
  [hep-th/0409296]
  %%CITATION = HEP-TH 0409296;%%


\bibitem{mtt} J.A. Minahan, A. Tirziu and A.A. Tseytlin,
  ``$1/J$ corrections to semiclassical AdS/CFT states from quantum
  Landau-Lifshitz model,''
  Nucl.\ Phys.\ B {\bf 735}, 127 (2006)
  [hep-th/0509071].
%%CITATION = HEP-TH 0509071;%%


\bibitem{mtt1}
  J.~A.~Minahan, A.~Tirziu and A.~A.~Tseytlin,
  $``1/J^2$ corrections to BMN energies from the quantum long range
  Landau-Lifshitz model,''
  JHEP {\bf 0511}, 031 (2005)
  [hep-th/0510080].
  %%CITATION = HEP-TH 0510080;%%


\bibitem{t}
  A.~Tirziu,
  ``Quantum Landau-Lifshitz model at four loops: 1/J and 1/J2 corrections to
  BMN energies,''
  hep-th/0601139.
  %%CITATION = HEP-TH 0601139;%%

\bi{rss}
A.~Rej, D.~Serban and M.~Staudacher,
  %``Planar N = 4 gauge theory and the Hubbard model,''
  JHEP {\bf 0603}, 018 (2006)
  [hep-th/0512077].
  %%CITATION = HEP-TH 0512077;%%

\bi{za}
K.~Zarembo,
  ``Antiferromagnetic operators in N = 4 supersymmetric Yang-Mills theory,''
  Phys.\ Lett.\ B {\bf 634}, 552 (2006)
  [hep-th/0512079].
  %%CITATION = HEP-TH 0512079;%%

\bi{at}
G.~Arutyunov and A.~A.~Tseytlin,
  ``On highest-energy state in the su(1$|$1) sector of N = 4 super Yang-Mills
  theory,''
 hep-th/0603113.
  %%CITATION = HEP-TH 0603113;%%





\bibitem{beisert}
  N.~Beisert,
  ``The dilatation operator of N = 4 super Yang-Mills theory and
  integrability,''
  Phys.\ Rept.\  {\bf 405}, 1 (2005)
  [hep-th/0407277].
  %%CITATION = HEP-TH 0407277;%%

 \bi{btz}
 N.~Beisert, A.~A.~Tseytlin and K.~Zarembo,
  ``Matching quantum strings to quantum spins: One-loop vs. finite-size
  corrections,''
  Nucl.\ Phys.\ B {\bf 715}, 190 (2005)
  [hep-th/0502173].
  %%CITATION = HEP-TH 0502173;%%

\bi{kruznew}
  M.~Kruczenski,
  ``Renormalization group flow of reduced string actions,''
  hep-th/0509178.
  %%CITATION = HEP-TH 0509178;%%



\bi{bei2}
N.~Beisert,
  ``The su(2$|$3) dynamic spin chain,''
  Nucl.\ Phys.\ B {\bf 682}, 487 (2004)
  [hep-th/0310252].
  %%CITATION = HEP-TH 0310252;%%




\bibitem{hl1}
  R.~Hernandez and E.~Lopez,
  ``Spin chain sigma models with fermions,''
  JHEP {\bf 0411}, 079 (2004)
  [hep-th/0410022].
  %%CITATION = HEP-TH 0410022;%%

\bibitem{st1}
  B.~Stefanski and A.~A.~Tseytlin,
  ``Super spin chain coherent state actions and AdS(5) x S5  superstring,''
  Nucl.\ Phys.\ B {\bf 718}, 83 (2005)
  [hep-th/0503185].
  %%CITATION = HEP-TH 0503185;%%


\bi{mik}
A.~Mikhailov,
  ``Speeding strings,''
  JHEP {\bf 0312}, 058 (2003)
  [hep-th/0311019].
  %%CITATION = HEP-TH 0311019;%%
``Slow evolution of nearly-degenerate extremal surfaces,''
  J.\ Geom.\ Phys.\  {\bf 54}, 228 (2005)
  [hep-th/0402067].
  %%CITATION = HEP-TH 0402067;%%

  \bi{kut}
M.~Kruczenski and A.~A.~Tseytlin,
  ``Semiclassical relativistic strings in S5 and long coherent operators  in
  N = 4 SYM theory,''
  JHEP {\bf 0409}, 038 (2004)
  [hep-th/0406189].
  %%CITATION = HEP-TH 0406189;%%

\bi{mets}
  R.~R.~Metsaev,
  ``Type IIB Green-Schwarz superstring in plane wave Ramond-Ramond
  background,''
  Nucl.\ Phys.\ B {\bf 625}, 70 (2002)
  [hep-th/0112044].
  %%CITATION = HEP-TH 0112044;%%

\bibitem{ft1}
S.~Frolov and A.~A.~Tseytlin, ``Semiclassical quantization of
rotating superstring in AdS(5) x  S(5),'' JHEP {\bf 0206}, 007
(2002) [hep-th/0204226].
%%CITATION = HEP-TH 0204226;%%


\bi{parn}
  A.~Parnachev and A.~V.~Ryzhov,
  ``Strings in the near plane wave background and AdS/CFT,''
  JHEP {\bf 0210}, 066 (2002)
  [hep-th/0208010].
  %%CITATION = HEP-TH 0208010;%%

\bibitem{cal2}
  C.~G.~Callan, J.~Heckman, T.~McLoughlin and I.~J.~Swanson,
  ``Lattice super Yang-Mills: A virial approach to operator dimensions,''
  Nucl.\ Phys.\ B {\bf 701}, 180 (2004)
  [hep-th/0407096].
  %%CITATION = HEP-TH 0407096;%%


\bibitem{aaf}
  L.~F.~Alday, G.~Arutyunov and S.~Frolov,
  ``New integrable system of 2dim fermions from strings on AdS(5) x S5,''
  JHEP {\bf 0601}, 078 (2006)
  [hep-th/0508140].
  %%CITATION = HEP-TH 0508140;%%


\bi{st}
B.~J.~Stefanski and A.~A.~Tseytlin,
  ``Large spin limits of AdS/CFT and generalized Landau-Lifshitz equations,''
  JHEP {\bf 0405}, 042 (2004)
  [hep-th/0404133].
  %%CITATION = HEP-TH 0404133;%%



\bibitem{bell}
  S.~Bellucci, P.~Y.~Casteill, J.~F.~Morales and C.~Sochichiu,
  ``SL(2) spin chain and spinning strings on AdS(5) x S5 ,''
  Nucl.\ Phys.\ B {\bf 707}, 303 (2005)
  [hep-th/0409086].
  %%CITATION = HEP-TH 0409086;%%

\bi{ptt}
 I.~Y.~Park, A.~Tirziu and A.~A.~Tseytlin,
  ``Spinning strings in AdS(5) x S5: One-loop correction to energy in  SL(2)
  sector,''
  JHEP {\bf 0503}, 013 (2005)
  [hep-th/0501203].
  %%CITATION = HEP-TH 0501203;%%


\bibitem{KZ}
  V.~A.~Kazakov and K.~Zarembo,
  ``Classical / quantum integrability in non-compact sector of AdS/CFT,''
  JHEP {\bf 0410}, 060 (2004)
  [hep-th/0410105].
  %%CITATION = HEP-TH 0410105;%%

\bibitem{AF}
  G.~Arutyunov and S.~Frolov,
  ``Integrable Hamiltonian for classical strings on AdS(5) x S5,''
  JHEP {\bf 0502}, 059 (2005)
  [hep-th/0411089].
  %%CITATION = HEP-TH 0411089;%%

\bi{sw}
C.~G.~.~Callan, J.~Heckman, T.~McLoughlin and I.~J.~Swanson,
  ``Lattice super Yang-Mills: A virial approach to operator dimensions,''
  Nucl.\ Phys.\ B {\bf 701}, 180 (2004)
  [hep-th/0407096].
  %%CITATION = HEP-TH 0407096;%%

\bi{af2}
G.~Arutyunov and S.~Frolov,
  ``Uniform light-cone gauge for strings in AdS(5) x S5: Solving su(1$|$1)
  sector,''
  JHEP {\bf 0601}, 055 (2006)
  [hep-th/0510208].
  %%CITATION = HEP-TH 0510208;%%



\end{thebibliography}
\end{document}